\documentclass[reprint,amsmath,amssymb,aps,]{revtex4-2}

\usepackage{tikz}
\usepackage{quantikz}
\usepackage{multirow}
\usepackage{graphicx}
\usepackage{dcolumn}
\usepackage{caption}
\usepackage{booktabs}
\usepackage{placeins}  
\usepackage{bm}
\usepackage{bbold}
\usepackage{float}  
\usepackage{mathtools}
\usepackage{xcolor}
\usepackage{multirow}
\usepackage{graphicx}  
\usepackage{array}
\usepackage{colortbl} 
\usepackage{nicematrix}
\usepackage{siunitx}
\usepackage{algpseudocode}
\usepackage[T1]{fontenc}
\usepackage[caption=false]{subfig}
\usetikzlibrary{positioning, quotes}
\usetikzlibrary{shadows, shapes.geometric}

\usepackage{hyperref}

\begin{document}
\preprint{APS/123-QED}

\title{Qubit-Scalable CVRP via Lagrangian Knapsack Decomposition and Noise-Aware Quantum Execution}

\author{Monit Sharma$^{1}$}
\author{Hoong Chuin Lau$^{1,2}$}
\email{Corresponding author email: hclau@smu.edu.sg}
\address{$^1$School of Computing and Information Systems,  Singapore Management University, Singapore}
\address{$^2$Institute of High Performance Computing, A*STAR, Singapore}

\begin{abstract}

Hybrid quantum optimization for vehicle routing faces a practical bottleneck: (i) direct QUBO encodings of CVRP quickly exceeds near-term qubit and gate budgets; and (ii) near-term quantum evaluations are expensive, noise-limited, and highly sensitive to backend and circuit configuration. We address this gap with an end-to-end decomposition-based pipeline that converts CVRP into a sequence of bounded-width quantum subproblems and treats quantum execution itself as a decision problem within the optimization loop. Starting from a Fisher--Jaikumar assignment linearization, we apply Lagrangian relaxation to dualize customer-assignment couplers, yielding independent, per-vehicle knapsack subproblems that admit QUBO/Ising evaluation. To replace brittle subgradient tuning in the outer loop, we learn a multiplier-update controller using a two-stage protocol (expert-guided pretraining followed by reinforcement learning fine-tuning) with rewards defined by execution-realized progress and route reconstruction. Furthermore, we introduce constrained contextual bandit as a hardware-aware execution layer that selects backend and circuit configuration with feasibility screening, thereby enabling safe adaptation across heterogeneous noisy resources and parallel multi-QPU scheduling.
Computational results on multiple CVRPLIB families demonstrate that (i) the decomposition yields stable bounded-width subproblems across instance sizes, (ii) learned multiplier updates improve end-to-end routing quality relative to classical subgradient control under matched budgets, and
(iii) hardware-aware configuration on real devices reduces median optimality gaps compared with static hardware execution, improving the utility of each costly quantum evaluation. Our goal is not to claim quantum advantage, but together, these
components provide a practical end-to-end framework for scaling hybrid quantum CVRP optimization through OR decomposition, AI (learning-augmented dual control), and adaptive hardware-aware execution.

\end{abstract}

\maketitle

\section{\label{sec:level_introduction}Introduction}

Near-term quantum processors impose strict constraints on logical problem size and effective circuit budget due to noise, sparse connectivity, and compilation overhead~\cite{Preskill_2018,sharma2025comparative,sharma2025cutting}. In hybrid quantum--classical optimization, these constraints are coupled with operational heterogeneity: multiple backends may be available, but the effective cost and quality of an evaluation depend on device-specific compilation, queueing, and calibration drift~\cite{sharma2025comparative}. These realities motivate algorithmic designs that (i) expose quantum-executable subproblems and (ii) allocate/configure quantum evaluations in a hardware-aware manner.

This paper develops and evaluates such a design for the capacitated vehicle routing problem (CVRP)~\cite{toth2002vehicle,dantzig1959truck,clarke1964scheduling}. Our framework combines (i) a Lagrangian decomposition that converts each instance into a stream of bounded-width 0--1 subproblems with an end-to-end connection to routing quality, (ii) a learning-augmented controller for multiplier updates in the resulting dual loop, and (iii) a hardware-aware execution layer that selects backend and circuit configurations across heterogeneous QPUs. The resulting perspective is not ``quantum CVRP'' in isolation, but a hybrid OR system in which optimization structure and hardware constraints are jointly modeled and exploited.

\paragraph{Lagrangian CVRP decomposition into quantum-sized subproblems.}
Direct encodings of CVRP into QUBO form quickly exceed effective circuit budget~\cite{sharma2025comparative}. 
We instead adopt a decomposition pathway that produces a stream of smaller subproblems while 
retaining an end-to-end connection to routing quality. Starting from a linearization that 
yields a generalized assignment structure~\cite{fisher1981generalized}, we apply Lagrangian 
relaxation to dualize customer assignment constraints and obtain a separable form that 
decomposes into knapsack-structured subproblems (one per vehicle) coupled only through multipliers. 
Each subproblem can be encoded as a QUBO and solved by a variational routine, yielding sampled 
bitstrings that are decoded into candidate assignments; a classical reconstruction step then 
forms feasible routes and evaluates the corresponding CVRP objective. This design is explicitly 
motivated by qubit scalability: it replaces a monolithic encoding with repeated executions of 
bounded-width subproblems whose sizes can be screened against available devices.

\paragraph{Learning multiplier updates via pretraining and fine-tuning.}
\label{sec:para_learned_multipliers}
The decomposition induces an outer-loop optimization over Lagrange multipliers. Classical subgradient updates can be sensitive to stepsize schedules and may oscillate, slowing progress and degrading decoded primal solutions~\cite{bollapragada2023adaptive,xiao2022class}. We learn a multiplier-update policy using a two-stage protocol: expert-guided pretraining to initialize a stable controller, followed by reinforcement-learning fine-tuning with a curriculum that combines dense proxy signals with periodic end-to-end routing feedback. Importantly, the learning signal is based on \emph{execution-realized} outcomes after decoding/repair and reconstruction, enabling robustness to sampling variability and quantum execution noise. The learned policy is a drop-in replacement for hand-tuned multiplier rules and shapes both the difficulty and the quality of the generated subproblems.

\paragraph{Hardware-aware quantum execution across heterogeneous backends.}
Even with bounded-width subproblems, execution quality depends strongly on backend choice and circuit configuration. 
Sparse coupling graphs can induce SWAP overhead; deeper or more entangling
ans\"atze may violate coherence-limited budgets; and calibration drift can make static qubit-selection heuristics unreliable~\cite{bhartikishor2022noisy,Preskill_2018,fang2025caliqec,gaye2025model,proctor2020detecting}.
We therefore treat configuration as an explicit decision problem conditioned on the joint problem--hardware context. Specifically, we train a constrained contextual bandit in a trace-driven offline phase, using backend descriptors together with required logical width to select qubit placement, entanglement pattern, and depth subject to feasibility safeguards (gate-budget screening). At deployment, the learned policy routes quantum evaluations across heterogeneous backends online, and performance is assessed using execution-realized end-to-end outcomes after decoding/repair and reconstruction (gap to BKS), together with robustness metrics (runtime, latency, failures).

Our objective is not to claim quantum advantage over state-of-the-art classical CVRP solvers. 
Rather, we study a hybrid regime in which quantum evaluations are expensive and imperfect, and study the question of how decomposition and control mechanisms can make such evaluations 
\emph{useful} within an end-to-end OR pipeline. 

We make the following contributions.
\begin{enumerate}
  \item \textbf{Bounded-width Lagrangian decomposition with end-to-end primal recovery.}
    We develop a Lagrangian relaxation pipeline that yields per-vehicle knapsack QUBO subproblems with bounded logical width, together with decoding/repair and route reconstruction that map subproblem samples to feasible CVRP routes.

  \item \textbf{Learning-augmented multiplier control robust to noisy subproblem solutions.}
    We introduce a two-stage learning framework for multiplier updates using expert-guided pretraining followed by reinforcement-learning fine-tuning, with execution-realized progress signals to improve stability and reduce reliance on hand-designed stepsize schedules.

  \item \textbf{Hardware-aware backend and circuit configuration via constrained contextual bandits.}
    We formulate backend selection, circuit configuration, and qubit placement as a constrained decision problem and propose a contextual bandit policy that selects entanglement pattern, depth, and physical-qubit subsets conditional on backend descriptors and subproblem width, with explicit feasibility safeguards.

  \item \textbf{System-level evaluation under heterogeneous backends and fixed quantum budgets.}
    We provide a reproducible experimental protocol and implementation supporting multiple backend models, 
    demonstrating scaling behavior as instance size grows under fixed quantum-evaluation budgets and heterogeneous backend orchestration.
\end{enumerate}

This paper proceeds as follows. Section~\ref{sec:related_work} reviews the decomposition and hybrid-quantum ingredients. Section~\ref{sec:formulation} presents the Fisher--Jaikumar linearization and the
Lagrangian relaxation that converts CVRP into a stream of bounded-width knapsack subproblems.
Section~\ref{sec:primal_recovery} then specifies the decoding/repair and route-reconstruction steps
used to obtain feasible routes and defines the evaluation metrics and learning signals used throughout.
Section~\ref{sec:learned_multipliers_section} introduces the learned multiplier-update controller and
its two-stage training protocol, while Section~\ref{sec:hardware_aware} formulates hardware-aware
quantum execution as a constrained contextual bandit and describes feasibility screening and
heterogeneous-backend orchestration. Section~\ref{sec:experiments} describes our experimental design, and Section~\ref{sec:results} reports the experimental results for end-to-end comparison followed by
component ablations and execution-level analyses.

\section{Background and Related Work}
\label{sec:related_work}

We position our framework around four threads that determine whether hybrid quantum--classical
routing pipelines are practically useful: (i) OR decompositions for CVRP that separate assignment
from sequencing, (ii) learning-based optimization and control for dual updates, (iii) bounded-width
QUBO/Ising optimization primitives and variational routines, and (iv) hardware-aware execution under
heterogeneous, time-varying backends. We intentionally focus on the specific ingredients needed to
understand our decomposition into knapsack subproblems, our learned multiplier-update controller,
and our backend-aware configuration policy.

\subsection{OR decompositions for CVRP: separating assignment and sequencing}
\label{sec:cvrp_assignment_decomp}

CVRP couples \emph{assignment} (which vehicle serves each customer) with \emph{sequencing}
(the visit order within each route), and this coupling is a major barrier to scalable solution
methods \cite{toth2002vehicle,dantzig1959truck,clarke1964scheduling}. A long-standing approach is
to decouple these decisions using ``cluster-first, route-second'' pipelines: first partition
customers into capacity-feasible sets, then solve a routing subproblem within each set
\cite{fisher1981generalized,gillett1974heuristic}. Fisher and Jaikumar's generalized-assignment
heuristic is particularly influential because it makes the assignment stage explicit through a
GAP linearization based on seed locations and insertion-cost proxies \cite{fisher1981generalized}.
Subsequent refinements improve seed selection and local improvement, but the core modeling idea
remains: isolating assignment exposes separability across vehicles and defers full routing
complexity to a reconstruction phase \cite{baker1999extensions,toth2002vehicle}. We adopt this
view to obtain an assignment-driven formulation that is compatible with repeated bounded-width
subproblem solves.

\subsection{Learning to optimize: stabilizing multiplier updates in Lagrangian pipelines}
\label{sec:lr_gap_knapsack}

Lagrangian relaxation is a standard mechanism for handling a small set of coupling constraints in
decomposition schemes by moving them into the objective with multipliers, thereby yielding
separable subproblems and a concave (typically nonsmooth) dual \cite{geoffrion1974lr,fisher1981lagrangian}.
In GAP-like assignment models, dualizing the ``each customer assigned once'' constraints produces
independent per-vehicle subproblems that reduce to knapsack structure under capacity limits
\cite{fisher1981lagrangian,geoffrion1974lr}. The classical challenge is the outer-loop multiplier
optimization: subgradient methods are scalable but sensitive to stepsize schedules and stabilization,
and poor tuning can lead to oscillation or slow progress, especially when primal recovery depends on
noisy or approximate subproblem solutions \cite{polyak1977subgradient,fisher1981lagrangian}.
These sensitivities motivate treating multiplier updates as a control problem. In our setting, the
need is amplified because subproblem feedback is sampling-based and can be execution-noisy; we
therefore learn an update policy that replaces hand-tuned schedules while preserving explicit
feasibility safeguards and reconstruction.

\subsection{Quantum optimization for bounded-width routing subproblems}
\label{sec:qubo_vqe_qaoa}

QUBO/Ising modeling provides a canonical interface between discrete optimization and quantum
evaluation \cite{lucas2014ising,kochenberger2014unconstrained}. For near-term gate-based devices,
however, usefulness is governed by \emph{logical width} and \emph{effective circuit cost}: monolithic
routing encodings often introduce dense penalty couplings and additional variables that exceed
width/depth limits even before noise is considered. This motivates architectures that repeatedly
solve \emph{bounded-width} subproblems and integrate their outputs into a higher-level classical
procedure. Variational routines such as VQE and QAOA provide a flexible mechanism to approximately
optimize bounded-width Ising/QUBO objectives via sampling from parameterized circuits
\cite{peruzzo2014variational,farhi2014quantum}. In hybrid pipelines, the decoding step is
algorithmic rather than cosmetic: measurement samples must be mapped to feasible subproblem decisions,
and the resulting decoded quality determines both reported metrics and learning signals.

\subsection{Hardware execution: configuration, variability, and online adaptation}
\label{sec:hardware_aware_exec}

Even at fixed logical width, realized optimization utility varies substantially across backends and
circuit configurations due to compilation overhead and hardware variability. Limited connectivity
induces routing via SWAP insertion, inflating depth and two-qubit gates and degrading performance on
noisy devices \cite{li2019tackling,sivarajah2021t}. In addition, calibration parameters drift and are
heterogeneous across qubits and couplers, making static ``best qubits'' heuristics unreliable
\cite{nation2023suppressing,kurniawan2024use}. These effects suggest that backend choice, qubit
placement, ansatz family, and depth should be treated as decisions that adapt online to observed
problem and hardware context. Contextual bandits provide a lightweight framework for this setting:
actions are discrete, feedback is costly and noisy, and safety constraints (e.g., gate-budget
feasibility) can be enforced via filtering \cite{li2010contextual,chu2011contextual}. We adopt this
perspective and drive adaptation using execution-realized rewards computed from decoded outcomes,
rather than device metadata alone.

\subsection{Quantum approaches to the capacitated vehicle routing problem}
\label{sec:quantum_cvrp_lit}

Table~\ref{tab:quantum_cvrp_lit} summarizes recent representative quantum and hybrid quantum--classical
approaches to the CVRP.
A consistent pattern emerges: direct quantum formulations remain limited to very small instances,
while larger studies typically rely on hybridization, decomposition, or the use of quantum methods
only within a restricted subroutine. This literature context motivates our focus on bounded-width
subproblems, learned outer-loop control, and hardware-aware execution within a full end-to-end
pipeline.

\begin{table*}[t]
\centering
\caption{Representative quantum and hybrid quantum--classical approaches for the CVRP. The table highlights the dominant design pattern in prior work: either direct formulations evaluated only on very small instances, or hybrid approaches in which quantum computation is applied only to restricted subproblems or local improvement stages.}
\label{tab:quantum_cvrp_lit}
\scriptsize
\setlength{\tabcolsep}{3pt}
\renewcommand{\arraystretch}{1.15}
\begin{tabular*}{\textwidth}{@{\extracolsep{\fill}}lllllll}
\hline
\textbf{Reference} & \textbf{CVRP strategy} & \textbf{Quantum role} & \textbf{Platform} & \textbf{Scale} & \textbf{Main outcome} & \textbf{Main limitation} \\
\hline

Feld et al.\ \cite{feld2019hybrid} &
\parbox[t]{2.0cm}{Two-phase hybrid CVRP\\ (clustering + routing)} &
\parbox[t]{1.9cm}{QA on route-level\\ subproblems} &
\parbox[t]{1.7cm}{D-Wave\\ annealer} &
\parbox[t]{1.5cm}{Benchmark\\ CVRP sets} &
\parbox[t]{3.0cm}{Early hybrid quantum--classical CVRP method showing that partitioning is necessary for annealer-based execution} &
\parbox[t]{3.1cm}{Quantum component is only one phase of the pipeline; no learned multiplier control or hardware-aware execution} \\

Irie et al.\ \cite{irie2019quantum} &
\parbox[t]{2.0cm}{Direct QUBO with time, state, and capacity} &
\parbox[t]{1.9cm}{QA on direct\\ CVRP formulation} &
\parbox[t]{1.7cm}{D-Wave\\ 2000Q} &
\parbox[t]{1.5cm}{Very small\\ embedded instances} &
\parbox[t]{3.0cm}{Introduces a richer direct CVRP QUBO with explicit time/state/capacity modeling} &
\parbox[t]{3.1cm}{Resource growth is severe; practical scaling remains limited by embedding and qubit requirements} \\

Palackal et al.\ \cite{palackal2023quantum} &
\parbox[t]{2.0cm}{CVRP reduced to clustering + route-level TSPs} &
\parbox[t]{1.9cm}{QAOA/VQE on reduced\\ subproblems} &
\parbox[t]{1.7cm}{Gate-model\\ simulation} &
\parbox[t]{1.5cm}{Toy-scale reduced\\ instances} &
\parbox[t]{3.0cm}{Shows decomposition is needed under current hardware limits; VQE outperforms QAOA on the reduced route subproblems} &
\parbox[t]{3.1cm}{Not an end-to-end large-scale CVRP solver; quantum optimization is applied only after reduction to smaller routing pieces} \\

Xie et al.\ \cite{xie2024feasibility} &
\parbox[t]{2.0cm}{Feasibility-preserving\\ CVRP encoding} &
\parbox[t]{1.9cm}{AOA / constrained\\ QAOA-style solver} &
\parbox[t]{1.7cm}{Gate-model\\ study} &
\parbox[t]{1.5cm}{Small illustrative\\ instances} &
\parbox[t]{3.0cm}{Preserves feasibility and improves probability of sampling optimal solutions relative to penalty-based QAOA} &
\parbox[t]{3.1cm}{Scale remains small; contribution is mainly encoding and mixer design rather than end-to-end benchmark performance} \\

Holliday et al.\ \cite{holliday2024hybrid} &
\parbox[t]{2.0cm}{Classical tabu search with quantum route refinement} &
\parbox[t]{1.9cm}{QA inside a hybrid\\ metaheuristic} &
\parbox[t]{1.7cm}{D-Wave hybrid /\\ annealing-based} &
\parbox[t]{1.5cm}{CMT instances up\\ to 199 nodes} &
\parbox[t]{3.0cm}{Finds the best-known solution on CMT1 and reports competitive results on larger instances} &
\parbox[t]{3.1cm}{Quantum module is a local improvement component, not a decomposition-driven end-to-end optimization framework} \\

Mario et al.~\cite{mario2024qa_hybrid} &
\parbox[t]{2.0cm}{Hybrid real-time route optimization strategies for CVRP} &
\parbox[t]{1.9cm}{QA-assisted hybrid phases} &
\parbox[t]{1.7cm}{Annealing-based hybrid workflow} &
\parbox[t]{1.5cm}{CVRP-focused study} &
\parbox[t]{3.0cm}{Reports promising hybrid performance for route-optimization variants} &
\parbox[t]{3.1cm}{Evidence is more proof-of-concept than a broad benchmark study with standardized large-scale CVRPLIB-style reporting} \\

Huang et al.\ \cite{huang2025cvrp_qaoa_cg} &
\parbox[t]{2.0cm}{Column generation\\ for CVRP} &
\parbox[t]{1.9cm}{QAOAnsatz for pricing / route-generation subproblems} &
\parbox[t]{1.7cm}{Gate-model\\ hybrid algorithm} &
\parbox[t]{1.5cm}{Up to 6\\ customers} &
\parbox[t]{3.0cm}{Combines column generation with a one-hot mixer and AL-inspired capacity handling; faster convergence than standard QAOA on small cases} &
\parbox[t]{3.1cm}{Very small scale; still far from realistic CVRPLIB-sized instances} \\

Onah and Michielsen\ \cite{onah2025requirements} &
\parbox[t]{2.0cm}{Resource-feasibility analysis for CVRP encodings} &
\parbox[t]{1.9cm}{No solver; scaling and feasibility study} &
\parbox[t]{1.7cm}{Analytic resource\\ model} &
\parbox[t]{1.5cm}{Encoding-level\\ scaling analysis} &
\parbox[t]{3.0cm}{Shows that direct NISQ CVRP remains unlikely even under efficient encodings; motivates decomposition-based strategies} &
\parbox[t]{3.1cm}{Not a solver paper; value is positioning and feasibility analysis rather than algorithmic performance} \\

\hline
\end{tabular*}
\end{table*}

Taken together, prior quantum-CVRP work is dominated by two regimes: \emph{direct formulations}
that remain limited to very small problem sizes, and \emph{hybrid methods} in which quantum
computation is used only within a narrow subroutine such as local routing improvement, route-level
TSP solving, or pricing in a decomposition method. Recent resource analyses further suggest that
direct CVRP encodings are unlikely to be practical on near-term hardware at realistic instance sizes.
This creates a clear opening for decomposition-centric pipelines that reduce logical width, learn how
to steer the outer loop, and allocate scarce quantum evaluations in a hardware-aware manner.

\subsection{Summary: gaps and how we address them}
\label{sec:gaps_summary}



Prior work provides the main ingredients needed for hybrid quantum routing, but not their integration within a single end-to-end pipeline. In particular, existing work does not jointly address three issues
that become coupled in practice: generating bounded-width subproblems, stabilizing the multiplier updates that govern those subproblems, and adapting execution to heterogeneous noisy hardware. In contrast, this work develop this integrated formulation and evaluate its effect on end-to-end routing performance.

\section{Decomposition and Quantum-Sized Subproblems for CVRP}
\label{sec:formulation}

We consider the CVRP on a complete graph with depot 
node $0$ and customer set $V=\{1,\dots,n\}$. We use \emph{customer} to denote a non-depot location and reserve \emph{node} for the underlying graph
representation; thus an instance with DIMENSION $N$ contains $n=N-1$ customers plus a depot. Each customer $i\in V$ has demand $d_i\ge 0$ 
and travel costs $c_{ij}\ge 0$ are given for all $(i,j)\in \{0\}\cup V$. The goal is to 
construct up to $K$ depot-starting routes that serve each customer exactly once, satisfy 
vehicle-capacity constraints, and minimize total travel cost. This section establishes notation and a reference arc-flow formulation for CVRP, then introduces an
assignment-based linearization in the spirit of Fisher--Jaikumar. The resulting generalized assignment
structure is the starting point for the Lagrangian decomposition used to produce bounded-width
subproblems in subsequent sections.

\subsection{CVRP notation and baseline arc-flow formulation}
\label{sec:cvrp_baseline}

Let $N=\{0\}\cup V$ denote all nodes including the depot. We use binary decision variables 
$x_{ij}\in\{0,1\}$ indicating whether any vehicle traverses directed arc $(i,j)$, and 
auxiliary variables to enforce capacity feasibility. A standard arc-flow formulation minimizes 
total travel cost
\begin{equation}
\min \sum_{i\in N}\sum_{\substack{j\in N\\ j\neq i}} c_{ij}\,x_{ij},
\end{equation}
subject to degree constraints ensuring each customer is entered and exited exactly once,
\begin{align}
\sum_{\substack{j\in N\\ j\neq i}} x_{ij} &= 1 && \forall i\in V,\\
\sum_{\substack{j\in N\\ j\neq i}} x_{ji} &= 1 && \forall i\in V,
\end{align}
and depot flow constraints that limit the number of routes (vehicles),
\begin{align}
\sum_{j\in V} x_{0j} \le K,\qquad \sum_{i\in V} x_{i0} \le K.
\end{align}
To enforce vehicle capacity, one may use standard single-commodity or multi-commodity flow 
constructions (or equivalently a family of capacity cuts); modern exact algorithms typically 
adopt a route-based (set-partitioning) master formulation with pricing and cutting planes 
rather than a pure arc-flow model \cite{toth2002vehicle,baldacci2008exact}. 
We present the arc-flow view here only as a reference, it makes explicit the global coupling 
induced by routing decisions and clarifies why direct QUBO encodings of CVRP tend to become 
wide and dense.

In contrast, we decouple \emph{assignment} from \emph{sequencing}. A Fisher--Jaikumar-style
linearization exposes a generalized assignment structure; we then apply Lagrangian relaxation,
dualizing the coupling constraints that enforce customer coverage, to obtain independent
knapsack-structured subproblems (one per vehicle). Sequencing is reintroduced only in the
final reconstruction stage.

\subsection{Fisher--Jaikumar assignment linearization (GAP construction)}
\label{sec:fj_gap}

We adopt an assignment-based linearization in the spirit of Fisher and Jaikumar
\cite{fisher1981generalized} to separate \emph{which vehicle serves which customers} from the
\emph{within-route sequencing} decisions. For each vehicle $k\in\{1,\dots,K\}$ we select a
representative \emph{seed} customer $s_k$ and define an insertion-cost surrogate
$a_{ik}$ that approximates the marginal routing cost of assigning customer $i$ to vehicle $k$.

\paragraph{Seed selection.}
When customer coordinates are available, we select seeds via an angle-based sector rule:
we compute the polar angle of each customer relative to the depot, sort customers by angle,
partition the ordered list into $K$ contiguous sectors, and choose as $s_k$ the customer farthest
from the depot within sector $k$. This produces geographically distributed seeds and encourages
compact clusters. If coordinates are unavailable, we fall back to selecting the $K$ customers farthest
from the depot as seeds. 

\paragraph{Insertion-cost surrogate.}
Given seeds, we use the Fisher--Jaikumar-style surrogate
\begin{equation}
a_{ik} \;=\; c_{0i} + c_{i s_k} - c_{0 s_k},
\label{eq:fj_cost}
\end{equation}
where $c_{uv}$ denotes the metric travel cost between nodes $u$ and $v$ and $0$ denotes the depot.
Intuitively, \eqref{eq:fj_cost} measures the marginal cost of serving $i$ on a seed-anchored route
for vehicle $k$.

Let $y_{ik}\in\{0,1\}$ indicate whether customer $i$ is assigned to vehicle $k$. With demands
$d_i$ and vehicle capacity $Q$, the assignment stage becomes a generalized assignment problem (GAP):
\begin{align}
\min \quad & \sum_{i\in V}\sum_{k=1}^K a_{ik}\, y_{ik} \label{eq:gap_obj}\\
\text{s.t.}\quad
& \sum_{k=1}^K y_{ik} = 1 && \forall i\in V \label{eq:gap_assign_once}\\
& \sum_{i\in V} d_i\, y_{ik} \le Q && \forall k\in\{1,\dots,K\} \label{eq:gap_cap}\\
& y_{ik}\in\{0,1\} && \forall i\in V,\; k\in\{1,\dots,K\}. \label{eq:gap_bin}
\end{align}

For our purposes, \eqref{eq:gap_obj}--\eqref{eq:gap_bin} is not merely a heuristic 
approximation: it is the structured intermediate model on which we apply Lagrangian relaxation. 
In particular, dualizing the coupling constraints \eqref{eq:gap_assign_once} yields a 
separable Lagrangian subproblem across vehicles that reduces to per-vehicle knapsack-type 
optimization (Section~\ref{sec:lagrangian_relaxation}), enabling repeated bounded-width 
subproblem solves and subsequent reconstruction into feasible routes.

\subsection{Lagrangian relaxation of the GAP assignment couplers}
\label{sec:lagrangian_relaxation}

We apply Lagrangian relaxation to the GAP formulation \eqref{eq:gap_obj}--\eqref{eq:gap_bin} by
dualizing the customer-assignment coupling constraints \eqref{eq:gap_assign_once}. Following the standard LR 
construction for 0--1 integer programs, we associate a multiplier $\lambda_i\in\mathbb{R}$ 
with each customer-assignment constraint $\sum_{k=1}^K y_{ik}=1$ and move these constraints 
into the objective \cite{geoffrion1974lr,fisher1981lagrangian}. For fixed multipliers 
$\lambda=(\lambda_1,\dots,\lambda_n)$, the Lagrangian relaxation is

\begin{equation}
\begin{split}
L(\lambda) = \min_{y} \quad & \sum_{i\in V}\sum_{k=1}^K a_{ik} y_{ik} \\
& + \sum_{i\in V}\lambda_i\Big(1-\sum_{k=1}^K y_{ik}\Big) \\
\text{s.t.} \quad & \sum_{i\in V} d_i y_{ik} \le Q, \quad \forall k \\
& y_{ik}\in\{0,1\}, \quad \forall i,k
\end{split}
\label{eq:lagrangian_def}
\end{equation}
Rearranging \eqref{eq:lagrangian_def} yields

\begin{equation}
\begin{split}
L(\lambda) = \sum_{i\in V}\lambda_i 
+ \sum_{k=1}^K \min_{y_{\cdot k}} \bigg\{ & \sum_{i\in V}(a_{ik}-\lambda_i)y_{ik} : \\
& \sum_{i\in V} d_i y_{ik}\le Q, \; y_{ik}\in\{0,1\} \; \forall i \bigg\}
\end{split}
\label{eq:lagrangian_separable}
\end{equation}

which is \emph{separable by vehicle}. Each inner minimization in \eqref{eq:lagrangian_separable}
is a 0--1 knapsack-structured subproblem (one per vehicle) with adjusted item costs 
$(a_{ik}-\lambda_i)$ and weights $d_i$ \cite{fisher1981lagrangian}. This is the structural 
property we exploit: LR converts the globally coupled assignment constraints into a collection 
of independent subproblems coupled only through the multiplier vector $\lambda$.

For minimization problems, $L(\lambda)$ provides a valid lower bound on the optimal value of 
the primal GAP (and hence on the quality of any feasible assignment under the same 
linearization), and the best such bound is obtained by solving the Lagrangian dual
\begin{equation}
\max_{\lambda\in\mathbb{R}^{|V|}} \; L(\lambda),
\label{eq:lagrangian_dual}
\end{equation}
a concave (generally nonsmooth) optimization problem \cite{geoffrion1974lr,fisher1981lagrangian}. Classical approaches optimize \eqref{eq:lagrangian_dual} via subgradient ascent, where the subgradient component at customer $i$ is the violation of the relaxed constraint, i.e.,
\begin{equation}
g_i(\lambda) = 1 - \sum_{k=1}^K y_{ik}^\star(\lambda),
\label{eq:subgrad}
\end{equation}
with $y^\star(\lambda)$ denoting an optimal (or approximately optimal) solution of the 
Lagrangian subproblem \eqref{eq:lagrangian_separable} at $\lambda$ 
\cite{held1974validation,fisher1981lagrangian}. In our setting, the subproblems are solved 
approximately (via variational sampling), and we therefore treat multiplier updates as a 
control problem under stochastic subproblem feedback; Section~\ref{sec:learned_multipliers_section} 
replaces hand-tuned subgradient stepsizes with a learned update policy while preserving the 
LR separability that enables bounded-width quantum subproblem solves.

Finally, because \eqref{eq:lagrangian_separable} relaxes the ``assign exactly once'' 
constraints, the vehicle-wise subproblem solutions generally do not define a feasible 
assignment. We therefore apply a decoding/repair step that converts sampled subproblem 
bitstrings into a feasible assignment (satisfying \eqref{eq:gap_assign_once} and 
\eqref{eq:gap_cap}) and then reconstruct routes to evaluate the end-to-end CVRP objective. 
This dual--primal interplay (dual updates guided by subproblem solutions plus explicit primal 
recovery) is standard in practical LR-based algorithms and is the mechanism by which our 
pipeline retains an end-to-end routing connection \cite{fisher1981lagrangian}.

\paragraph{Candidate restriction for bounded-width subproblems.}
In the Lagrangian subproblem for vehicle $k$, customer $i$ has reduced cost $(a_{ik}-\lambda_i)$.
We restrict the candidate set for each vehicle to customers with positive reduced profit,
$\lambda_i-a_{ik} > 0$, since items with nonpositive profit cannot improve the subproblem objective.
This dynamic filtering reduces the number of active binary decisions per subproblem and supports
bounded-width quantum evaluations in the subsequent QUBO mapping.

\subsection{Per-vehicle knapsack subproblem and QUBO encoding}
\label{sec:knapsack_qubo}

Equation~\eqref{eq:lagrangian_separable} shows that, for fixed multipliers $\lambda$, the
Lagrangian subproblem decomposes into $K$ independent 0--1 knapsack--structured problems.
For each vehicle $k$, define adjusted costs $\tilde a_{ik}(\lambda):=a_{ik}-\lambda_i$, and let
$y_{ik}\in\{0,1\}$ indicate whether customer $i$ is tentatively selected for vehicle $k$ in the
relaxed subproblem. The per-vehicle subproblem is
\begin{align}
\min_{y_{\cdot k}}\quad & \sum_{i\in V}\tilde a_{ik}(\lambda)\,y_{ik} \label{eq:knap_lin}\\
\text{s.t.}\quad & \sum_{i\in V} d_i\,y_{ik} \le Q,\qquad y_{ik}\in\{0,1\}\ \forall i\in V.\nonumber
\end{align}
This formulation is knapsack-structured in that it selects a subset of items (customers) under a
single capacity constraint, with costs shifted by the multipliers. It is convenient for our purposes
because (i) it is separable across vehicles and (ii) it admits a compact QUBO/Ising representation
of bounded logical width.

\paragraph{ QUBO/Ising representation and logical width.}
A quadratic unconstrained binary optimization (QUBO) model has the form
\begin{equation}
\min_{z\in\{0,1\}^m}\; z^\top Q z + q^\top z + c,
\label{eq:qubo_def}
\end{equation}
and is equivalent (via $s=2z-1$) to minimizing an Ising Hamiltonian over spins
\cite{lucas2014ising,kochenberger2014unconstrained}. Variational routines such as QAOA and VQE
operate naturally on such Ising/QUBO objectives \cite{farhi2014quantum,peruzzo2014variational}.
In our setting, the \emph{logical width} of a subproblem is the number of binary variables appearing
in its QUBO encoding, i.e., the number of qubits required to represent the objective before
compilation to a hardware coupling graph.

\paragraph{ Capacity-constraint encoding via a tilted quadratic penalty.}
To convert \eqref{eq:knap_lin} into an unconstrained QUBO without introducing slack variables, we
penalize violations of the knapsack inequality $\sum_i d_i y_{ik}\le Q$ using a \emph{tilted quadratic}
penalty. Dropping the vehicle index $k$ for readability, define
\[
W(y):=\sum_{i\in V} d_i y_i,\qquad r(y):=W(y)-Q.
\]
We minimize an energy of the form
\begin{equation}
E_{\mathrm{Tilt}}(y)
=
\sum_{i\in V}\tilde a_i(\lambda)\,y_i
+
\rho\Big(r(y)^2 + s\,r(y)\Big),
\label{eq:tilt_energy_main}
\end{equation}
where $\rho>0$ controls penalty strength and $s\ge 0$ controls the degree of \emph{headroom}
encouraged below capacity. This is a standard penalty-based QUBO modeling approach \cite{glover2019tutorial},
with the additional linear ``tilt'' term shaping the feasible region.

The tilt has a simple interpretation: $r(y)^2+s\,r(y)$ is minimized at $r^\star=-s/2$, i.e.,
at an intended load $W^\star = Q - s/2$, and it induces a preferred band $W\in[Q-s,Q]$.
In the present setting, subproblems are solved approximately (via sampling under noise), so biasing
toward slightly under-capacity selections improves the yield of feasible candidates after decoding/repair
and stabilizes downstream reconstruction. A full coefficient expansion of \eqref{eq:tilt_energy_main},
together with comparisons to a Taylor-style quadratic penalty baseline, is provided in
Appendix~\ref{app:penalties}. Related unbalanced/tilted penalty encodings for inequality constraints
in QUBO models have also been advocated to improve feasibility behavior under approximate optimization.

\paragraph{Implication for scalability.}
Under the tilted slack-free encoding \eqref{eq:tilt_energy_main}, the logical width of a vehicle
subproblem is the number of active customer-selection variables $y_i$ (i.e., the size of the candidate
set considered for that vehicle at the current LR iteration). In our pipeline, this candidate set is
typically much smaller than $|V|$ because LR and decoding restrict attention to a subset of promising
customers per vehicle, and we can screen subproblem widths against backend limits. This yields a stream
of bounded-width QUBOs that can be solved repeatedly by variational routines and integrated into the
outer-loop multiplier update and final reconstruction.

Given \eqref{eq:tilt_energy_main}, we instantiate a binary quadratic objective and evaluate it via an
Ising-form measurement primitive; sampled bitstrings are then decoded (and repaired if needed) into
feasible assignments used in downstream reconstruction.

\paragraph{Why a tilted penalty.}
Penalty-based QUBO encodings of one-sided capacity inequalities can be brittle under 
sampling noise: near-feasible solutions that ``hug'' the boundary may flip to slight overloads, 
reducing the yield of feasible candidates passed to decoding/repair. We therefore adopt a 
\emph{tilted} quadratic penalty that induces a preferred band just below capacity (controlled
headroom), improving feasibility rates under noisy variational optimization. 
Appendix~\ref{app:penalties:microbench} provides a controlled micro-benchmark on random 
knapsack instances solved on a noisy simulator, showing that the tilted encoding maintains the 
feasible-sample yield and improves the probability of recovering the exact optimum, especially
when combined with CVaR-based objectives \cite{glover2019tutorial,montanez2024unbalanced}.

\section{Primal recovery and evaluation metrics}
\label{sec:primal_recovery}

Lagrangian relaxation produces high-quality subproblem solutions but does not, in general, 
return a feasible primal assignment because the ``assign exactly once'' constraints are 
dualized. Moreover, in our setting each per-vehicle knapsack subproblem is solved approximately 
via sampling from a variational quantum routine. We therefore treat \emph{primal recovery},
decoding sampled bitstrings into feasible assignments and reconstructing feasible CVRP routes
, as a first-class component of the pipeline, and we define learning signals and evaluation 
metrics in terms of the best repaired solution obtained from measurement samples, rather than 
raw energy estimates. This dual--primal interplay is standard in practical Lagrangian-relaxation 
algorithms and is often essential for obtaining competitive primal solutions 
\cite{fisher1981lagrangian}.

\subsection{Decoding rule and best-of-samples selection}
\label{sec:decode_rule}

For a fixed multiplier vector $\lambda$ and vehicle $k$, the quantum routine produces a set of 
measurement samples over the knapsack selection variables $\{y_{ik}\}_{i\in V}$ (and, if 
present, any auxiliary bits from the penalty encoding). Each sample induces a tentative 
selection set
$
S_k=\{i\in V: y_{ik}=1\}.
$
Because sampling is stochastic and the relaxed solutions need not define a feasible global 
assignment, we apply a deterministic repair procedure (Section~\ref{sec:assignment_repair})
to each sampled candidate and select the \emph{best repaired} outcome according to a 
downstream score (subproblem proxy or end-to-end routing cost when evaluated). Concretely, 
given a batch of samples $\mathcal{B}$, we define
\begin{equation}
\hat y(\lambda) \in \arg\min_{y \in \mathcal{R}(\mathcal{B})} \; \mathrm{Score}(y),
\label{eq:best_of_samples}
\end{equation}
where $\mathcal{R}(\mathcal{B})$ denotes the set of repaired feasible assignments 
obtained by applying the repair map to all samples in $\mathcal{B}$. This ``best-of-samples 
after repair'' rule is used consistently: it defines the reward for multiplier learning and 
hardware-aware configuration, and it is the unit of comparison in our subproblem-level 
evaluation (Section~\ref{sec:metrics}).

\subsection{Assignment repair}
\label{sec:assignment_repair}

The repair step converts tentative per-vehicle selections $\{S_k\}_{k=1}^K$ into a feasible 
assignment $\hat y$ that satisfies the GAP constraints (each customer assigned exactly once 
and per-vehicle capacity). Because the underlying structure is a generalized assignment model, 
we adopt a lightweight, deterministic repair strategy inspired by standard GAP heuristics and 
Lagrangian heuristics \cite{fisher1981lagrangian,cattrysse1992survey,osman1995heuristics}. 
The repair has two stages:

\paragraph{Stage 1: conflict resolution (duplicate selections).}
If a customer $i$ appears in multiple sets $S_k$, we keep $i$ in a single vehicle and remove it
from others. We assign $i$ to the vehicle $k$ that minimizes a local criterion (e.g., smallest
surrogate cost $a_{ik}$ or smallest reduced cost $\tilde a_{ik}(\lambda)$), while respecting
remaining capacity whenever possible. Ties are broken deterministically.

\paragraph{Stage 2: completion (unassigned customers).}
If a customer $i$ is not selected by any vehicle after conflict resolution, we insert it into 
a vehicle with sufficient remaining capacity using a greedy rule (e.g., minimal incremental 
surrogate cost), and if no such vehicle exists we attempt a small number of local exchanges 
(moving one or two customers) to free capacity. If completion fails, we mark the candidate 
infeasible and exclude it from $\mathcal{R}(\mathcal{B})$.

This repair procedure is intentionally simple: it preserves the bounded-width quantum 
primitive and shifts feasibility enforcement to a transparent, auditable classical step. 
Full pseudo-code (including tie-breaking and exchange rules) is provided in 
Appendix~\ref{app:repair_pseudocode}.

\subsection{Route reconstruction}
\label{sec:reconstruction}

Given a repaired feasible assignment $\hat y$, we construct CVRP routes by solving 
(approximately) a routing problem within each vehicle's assigned customer set. Specifically, 
for each vehicle $k$ we form the cluster
$
V_k=\{i\in V:\hat y_{ik}=1\}
$
and construct a depot-starting tour that visits all nodes in $V_k$ and respects vehicle 
capacity by construction. In practice, this can be implemented using a standard 
``cluster-first, route-second'' approach: route construction and improvement within each 
cluster can be performed by TSP heuristics (e.g., savings-based initialization plus local 
search) or by calls to a classical VRP/TSP solver for the cluster sizes encountered in our 
experiments \cite{toth2002vehicle}. The reconstruction step outputs (i) a set of feasible routes, 
(ii) the corresponding end-to-end CVRP cost, and (iii) feasibility flags if any cluster-level 
construction fails (rare when assignment repair succeeds). Implementation details and parameter 
settings are given in Appendix~\ref{app:reconstruction_details}.

\subsection{Evaluation metrics}
\label{sec:metrics}

We report results at three levels to isolate where performance gains arise and under what 
conditions they persist:

\paragraph{Circuit-level feasibility and overhead.}
For each backend and circuit configuration, we report compilation and execution statistics 
that influence reliability and cost, including logical width, physical qubit count after 
mapping, circuit depth, two-qubit gate counts, and feasibility with respect to a 
hardware-imposed gate-budget constraint (Section~\ref{sec:bandit_actions_screening}). These metrics quantify 
routing overhead and the frequency with which a configuration must be rejected or overridden.

\paragraph{Outer-loop sample efficiency and robustness.}
Because quantum evaluations are costly and imperfect in our hybrid setting, we evaluate the
outer loop in terms of (i) the quantum-evaluation budget (total circuit executions or quantum
jobs invoked per instance), (ii) wall-clock cost and its decomposition into quantum-execution
time and classical processing time, and (iii) execution robustness, quantified by job-level
latency statistics and the fraction of failed submissions. These metrics capture how effectively
the control mechanism converts limited quantum resources into end-to-end improvement in
reconstructed CVRP solutions, and they support cost--quality comparisons across backends.

\paragraph{End-to-end CVRP performance.}
After assignment repair and route reconstruction, we report the total CVRP cost and 
(when applicable) the optimality gap relative to best-known solutions under a fixed 
sampling/computation 
budget. This end-to-end view is the primary measure of ``routing utility'' for our pipeline.

Together, these metrics support a transparent assessment of how decomposition, multiplier 
control, and hardware-aware execution affect final routing quality.

\section{Learned multiplier update policy}
\label{sec:learned_multipliers_section}

The Lagrangian decomposition induces an outer-loop optimization problem over multipliers 
$\lambda$ that control the per-vehicle subproblem stream. Classical approaches optimize 
the Lagrangian dual using subgradient ascent with carefully designed stepsize rules and 
stabilization heuristics \cite{fisher1981lagrangian}. In our setting, however, each subproblem 
is solved approximately via sampling and is followed by decoding/repair and route 
reconstruction (Section~\ref{sec:primal_recovery}), yielding a noisy and delayed feedback 
signal. We therefore learn a multiplier update policy that replaces hand-tuned stepsize 
schedules with data-driven updates designed to improve stability and end-to-end routing 
utility under stochastic subproblem feedback.

\subsection{Subgradient baseline and motivation for learning}
\label{sec:subgrad_motivation}

The Lagrangian dual \eqref{eq:lagrangian_dual} is concave but generally nonsmooth, and the 
classical workhorse for updating multipliers is subgradient ascent 
\cite{fisher1981lagrangian,held1974validation}. Let $y^\star(\lambda)$ denote an 
(approximately) optimal solution of the Lagrangian subproblem at multipliers $\lambda$. 
A subgradient of $L(\lambda)$ at customer $i$ is the violation of the relaxed assignment 
constraint,
\begin{equation}
g_i(\lambda) \;=\; 1 - \sum_{k=1}^K y^\star_{ik}(\lambda),
\label{eq:subgrad_repeat}
\end{equation}
and a generic subgradient update has the form
\begin{equation}
\lambda^{t+1}_i \;=\; \lambda^t_i + \eta_t\, g_i(\lambda^t),
\label{eq:subgrad_update}
\end{equation}
where $\eta_t>0$ is a stepsize. In practical LR algorithms, the quality of the resulting dual 
bound and (crucially) the quality of recovered primal solutions can depend strongly on how 
$\eta_t$ is selected and stabilized across iterations \cite{fisher1981lagrangian,polyak1977subgradient}.

\paragraph{Stepsize sensitivity and oscillations.}
Subgradient methods do not enjoy the smooth contraction properties of gradient methods; their 
behavior is dominated by stepsize choice and by the scale of subgradients 
\cite{polyak1977subgradient}. Diminishing stepsizes guarantee convergence in convex settings 
but can be slow in practice, while Polyak-type stepsizes can accelerate progress but require 
knowledge (or reliable estimates) of the optimal dual value and can be unstable when that 
estimate is inaccurate. In LR pipelines, this sensitivity is amplified because multiplier 
updates steer which items appear attractive in each subproblem: overly aggressive steps can 
induce cycling (large swings in assignments across vehicles), whereas conservative steps can 
stall progress and yield weak primal recoveries \cite{fisher1981lagrangian,held1974validation}.

\paragraph{Approximate/noisy subgradients in the hybrid setting.}
Our setting introduces additional sources of inexactness beyond classical LR. Each per-vehicle
subproblem is solved approximately via a sampling-based variational routine, and the resulting
bitstrings are passed through decoding/repair (Section~\ref{sec:primal_recovery}). 
Consequently, the effective subgradient signal \eqref{eq:subgrad_repeat} is noisy: it 
depends on finite-shot sampling, backend noise, and the nonlinearity introduced by repair. 
More broadly, the subgradient literature recognizes that inexact subgradients (arising 
from noise or approximation) can degrade stability and require additional care in stepsize 
selection and averaging \cite{nedic2010effect,kiwiel2004convergence}. In our pipeline, this 
manifests as high variance in the dual ascent direction and as sensitivity of end-to-end 
routing quality to multiplier dynamics.

\paragraph{Dual progress versus primal routing utility.}
A second practical issue is that subgradient ascent targets the dual objective $L(\lambda)$, 
whereas our ultimate goal is end-to-end routing performance after reconstruction. In classical 
LR practice, strong algorithms typically couple dual ascent with explicit primal recovery 
heuristics, and tuning is guided as much by primal quality as by the dual bound 
\cite{fisher1981lagrangian}. In the hybrid quantum setting, this dual--primal coupling 
becomes even more pronounced: small changes in multipliers can significantly affect the 
distribution of sampled candidates (and thus the feasible yield after repair), even if the 
dual bound changes little.

\paragraph{Motivation for learning.}
These considerations motivate replacing hand-designed stepsize schedules with a learned 
update rule that maps observed signals (subgradient components, feasibility/yield statistics 
from decoding, and periodic end-to-end routing feedback) to multiplier updates. The learned 
policy is trained to stabilize the multiplier dynamics under stochastic subproblem feedback 
and to maximize realized routing utility rather than relying on globally tuned stepsize 
heuristics.

\subsection{Multiplier updates as a controlled decision process}
\label{sec:mdp_formulation}

We model multiplier control as a sequential decision problem in which the state summarizes the
current LR outer-loop context and the action specifies the next multiplier update. This framing
is natural because (i) subproblem solves are approximate and sampling-based, (ii) decoding/repair
induces a nonlinear mapping from samples to feasible assignments, and (iii) routing feedback is
delayed and expensive. We therefore learn a policy that maps observed signals to multiplier
updates in order to maximize a cumulative notion of routing utility. We use standard
reinforcement-learning terminology for Markov decision processes (MDPs) and policy optimization
\cite{sutton1998reinforcement}.

\paragraph{State.}
At LR iteration $t$, the environment state $s_t$ aggregates signals available after solving
the current set of per-vehicle subproblems and applying decoding/repair (Section~\ref{sec:primal_recovery}).
Our implementation uses a structured state composed of:
(i) the current multipliers $\lambda_t$ (excluding the depot index),
(ii) the subgradient proxy $g_t$ from \eqref{eq:subgrad_repeat},
(iii) statistics of the best-of-samples-after-repair candidates (e.g., feasibility flags, feasible-yield
proxies, and best repaired score), and
(iv) optional periodic end-to-end routing feedback (e.g., reconstructed CVRP cost/gap) when such
evaluations are performed.
This state design is deliberately ``control-oriented'', it exposes both constraint-violation signals
($g_t$) and the realized utility of the sampled candidates that the multipliers induce.

\paragraph{Action.}
The action $a_t$ specifies an update to the multipliers. We parameterize the update as
\begin{equation}
\lambda_{t+1} \;=\; \Pi_{\Lambda}\!\left(\lambda_t + \Delta\lambda_t\right),
\label{eq:lambda_update_policy}
\end{equation}
where $\Delta\lambda_t=\pi_\theta(s_t)$ is produced by a learned policy $\pi_\theta$ and
$\Pi_{\Lambda}$ is a projection/clipping operator enforcing simple bounds (to prevent numerical
instability). This form is a direct drop-in replacement for subgradient updates
\eqref{eq:subgrad_update}, subgradient methods correspond to the special case
$\Delta\lambda_t=\eta_t g_t$ with a hand-designed stepsize $\eta_t$.

\paragraph{Reward and credit assignment.}
The learning signal is defined in terms of the repaired candidates produced from measurement samples
(Section~\ref{sec:decode_rule}). Specifically, at each iteration we compute a normalized progress proxy
from the \emph{best repaired} solution among samples (best-of-samples after repair), and we optionally
augment it with a periodic end-to-end routing score when reconstruction is executed. This design yields
a dense signal that is aligned with the outer-loop objective (improving the quality and feasibility of
candidate assignments) while still tethering learning to end-to-end routing performance at a controlled
frequency. The exact reward instantiation used in our experiments is reported in
Section~\ref{sec:training_protocol}.

\paragraph{Policy optimization.}
We optimize $\pi_\theta$ using a policy-gradient method. In particular, we adopt Proximal Policy
Optimization (PPO) due to its empirical stability and its ability to reuse on-policy data through
multiple minibatch epochs while controlling update magnitude via a clipped surrogate objective
\cite{schulman2017proximal}.
Because multiplier dynamics can exhibit oscillations reminiscent of classical Lagrangian methods,
we also view the learned update as a stabilizing controller; this is consistent with prior observations
that naive Lagrangian-style updates can overshoot and oscillate, motivating more responsive update rules
in constrained RL settings \cite{stooke2020responsive}.

\paragraph{Two-stage training preview.}
Direct RL training from scratch can be unstable because the environment is nonstationary (the induced
subproblem stream changes as $\lambda$ changes) and rewards are noisy due to sampling and repair.
We therefore use a two-stage protocol: an expert-guided pretraining phase initializes $\pi_\theta$ to
a reasonable update behavior, followed by PPO fine-tuning under a curriculum that gradually increases
problem difficulty and introduces periodic routing feedback. Full details are given in
Section~\ref{sec:training_protocol}.

\subsection{Two-stage training protocol: expert-guided pretraining and PPO fine-tuning}
\label{sec:training_protocol}

We train the multiplier-update policy using a two-stage protocol. The motivation is that direct
policy-gradient training from scratch is unstable in our setting due to (i) noisy subproblem feedback
from finite-shot sampling and backend noise, (ii) a nonstationary induced subproblem stream as
$\lambda$ changes, and (iii) delayed end-to-end routing feedback (Section~\ref{sec:primal_recovery}).
We therefore first initialize the policy by imitation from expert updates, then fine-tune with
reinforcement learning.

\paragraph{Stage 1: Expert-guided pretraining (behavior cloning).}
We construct a dataset of expert multiplier updates by running a classical baseline controller
(e.g., stabilized subgradient updates with carefully tuned stepsizes) and recording tuples
$(s_t, \Delta\lambda_t^{\mathrm{exp}})$ along trajectories. We then pretrain a policy
$\pi_\theta$ by supervised learning to predict the expert update:
\begin{equation}
\min_{\theta}\; \mathbb{E}_{(s,\Delta\lambda^{\mathrm{exp}})\sim\mathcal{D}}
\left[\;\|\pi_\theta(s)-\Delta\lambda^{\mathrm{exp}}\|_2^2\;\right],
\label{eq:bc_loss}
\end{equation}
where $\mathcal{D}$ denotes the aggregated demonstration dataset. This stage provides a stable
initial controller and mitigates the distribution-shift problem that arises when a policy visits
states that are underrepresented in offline expert data. To further reduce compounding error, we
iteratively aggregated data from rollouts under the current policy and relabel visited states
with expert actions (DAgger-style dataset aggregation) \cite{ross2011reduction}. In our implementation,
we use expert-guided pretraining as a robust initializer and treat any dataset aggregation as an enhancement.

\paragraph{Stage 2: Reinforcement-learning fine-tuning (PPO).}
Starting from the pretrained parameters, we fine-tune $\pi_\theta$ using Proximal Policy Optimization
(PPO) \cite{schulman2017proximal}. PPO maximizes a clipped surrogate objective that constrains the change
in policy likelihood ratios between successive updates, improving stability under noisy returns:
\begin{equation}
\max_{\theta}\; \mathbb{E}_t\Big[\min\big(r_t(\theta)\hat A_t,\;\mathrm{clip}(r_t(\theta),1-\epsilon,1+\epsilon)\hat A_t\big)\Big],
\label{eq:ppo_obj}
\end{equation}
where $r_t(\theta)$ is the policy likelihood ratio and $\hat A_t$ is an advantage estimate computed
from rollout returns. We implement PPO with standard actor--critic components, and we define rewards
using the best-of-samples-after-repair primitive from Section~\ref{sec:decode_rule}, augmented by
periodic end-to-end routing feedback when reconstruction is executed.

\paragraph{Curriculum and noise-aware training.}
To improve learning stability and generalization, we employ a curriculum over instance difficulty
(e.g., increasing $n$, tightening capacity ratios, or increasing candidate-set sizes) and gradually
introduce delayed routing feedback at a fixed frequency. Curriculum learning is a standard technique
for stabilizing training in nonconvex settings by presenting easier instances before harder ones
\cite{bengio2009curriculum}. In addition, we train and evaluate under both classical and noisy quantum
settings, which helps the policy adapt to sampling variability and execution noise that affect the
effective dynamics of the outer loop.

\paragraph{Outputs and logging.}
At each LR iteration $t$ we log: the state features $s_t$, the chosen update $\Delta\lambda_t$,
subgradient proxy statistics, feasibility/yield statistics from decoding, and (when computed) the
reconstructed routing cost. This logging supports the ablations reported in Section~\ref{sec:results}
and provides an auditable record of how learned updates influence the subproblem stream.

\paragraph{Implementation mapping.}
The above protocol corresponds directly to our code pipeline: expert demonstrations are generated and
stored, a policy is pretrained on these demonstrations, and PPO fine-tuning is performed using the
environment defined by the LR outer loop with decoding/repair and reconstruction in the loop.

\subsection{Policy parameterization and architecture}
\label{sec:policy_architecture}

We represent multiplier control with a graph-based actor--critic policy that outputs a continuous
update for each customer multiplier. The default model used in our two-stage training pipeline is
\texttt{DiagonalPrecondPolicy} (a preconditioned GAT policy) \cite{velivckovic2017graph,brody2021attentive}. 
The policy operates on a heterogeneous graph observation (PyTorch Geometric \texttt{HeteroData}),
batched.

\paragraph{Graph encoder (GATv2 message passing).}
Let $h_i^t$ denote the embedding of customer node $i$ at outer-loop iteration $t$. The policy encoder
computes node embeddings using stacked GATv2 convolution layers with edge features 
(\texttt{edge\_dim}=1 in our implementation), which allows the attention mechanism to incorporate
pairwise edge attributes in message passing \cite{brody2021attentive}. We denote the resulting
node embedding matrix by $H^t=[h_1^t,\dots,h_n^t]^\top$.

\paragraph{Diagonal preconditioned update with residual drift.}
The actor predicts a per-customer Gaussian policy over multiplier increments. Rather than predicting
$\Delta\lambda$ with an unconstrained head, we parameterize the mean update as a \emph{diagonally
preconditioned} subgradient step plus a residual drift term:
\begin{equation}
\mu_i^t \;=\; \underbrace{\alpha_i^t\, g_i^t}_{\text{preconditioned subgradient}} \;+\;
\underbrace{\beta_i^t}_{\text{residual drift}},
\qquad i\in V,
\label{eq:diag_precond_update}
\end{equation}
where $g_i^t$ is the subgradient proxy from \eqref{eq:subgrad_repeat}, and $(\alpha_i^t,\beta_i^t)$ are
outputs of two light-weight node-wise heads applied to $h_i^t$. In the implementation, $\alpha_i^t$ is
constrained to be positive via a \texttt{softplus} transformation (with a small floor for numerical
stability), yielding a diagonal scaling of the subgradient direction. The residual
term $\beta_i^t$ is included to avoid a ``zero-gradient trap'': when $g_i^t\approx 0$ (e.g., the relaxed
solution is locally assignment-feasible), a pure preconditioned update would freeze. The residual head
permits continued exploration and fine adjustment of multipliers even when subgradient magnitudes vanish.

\paragraph{Stochastic policy and action bounds.}
We model the policy as a factorized Gaussian over customers,
\begin{equation}
\Delta\lambda^t \sim \mathcal{N}(\mu^t,\,\Sigma),
\label{eq:gaussian_policy}
\end{equation}
with diagonal covariance $\Sigma=\mathrm{diag}(\sigma^2)$ parameterized by a learnable log-standard
deviation vector that is shared across states (state-independent \texttt{log\_std}). This is a standard
and stable parameterization used in many PPO implementations to ensure positivity of $\sigma$ and to
control exploration \cite{schulman2017proximal}. The resulting update is applied through the
environment projection/clipping operator in \eqref{eq:lambda_update_policy} to enforce bounded multiplier
magnitudes and prevent numerical instability.

\paragraph{Value function.}
The critic predicts a scalar state value $V(s_t)$ from the node embeddings using global mean pooling
followed by an MLP head. This provides a permutation-invariant aggregation of the customer-wise features
and supports advantage estimation for PPO \cite{schulman2017proximal}.

\paragraph{Hyperparameters.}
Layer widths, number of GATv2 layers/heads, and dropout rates are fixed across experiments and reported
in Appendix~\ref{app:training_hparams} to keep the main description focused on the control
parameterization \eqref{eq:diag_precond_update}.

\subsection{Reward definition and credit assignment}
\label{sec:reward_definition}

A central design choice is the learning signal used to train the multiplier-update policy. In our
pipeline, multipliers $\lambda$ influence the \emph{distribution} of sampled knapsack candidates and,
after decoding/repair and reconstruction (Section~\ref{sec:primal_recovery}), determine realized routing
utility. We therefore define rewards using \emph{execution-realized} outcomes (best repaired candidates,
and periodic routing evaluations) rather than raw energy estimates. To stabilize training, we employ a
curriculum reward shaper that progressively introduces more informative but more delayed objectives
(curriculum learning; reward shaping) \cite{bengio2009curriculum,ng1999policy}.

\paragraph{Logged signals.}
At LR iteration $t$, after applying the policy action and resolving the relaxed subproblem, the
environment computes and logs:
(i) a subgradient proxy $g_t$ from \eqref{eq:subgrad_repeat},
(ii) a violation magnitude $V_t := \|g_t\|_1$,
(iii) a proxy assignment correctness count
\begin{equation}
n_t^{\mathrm{corr}} := \big|\{i\in V:\ |g_{t,i}|<\tau\}\big|,\qquad \tau=0.1,
\end{equation}
(iv) the current Lagrangian lower bound $LB_t$,
(v) the best known upper bound $UB_t$ (tracked by the solver), and
(vi) periodic end-to-end routing evaluations: reconstructed CVRP cost $C_t$ and (when available)
optimality gap $\mathrm{gap}_t$ (Section~\ref{sec:metrics}).

\paragraph{Curriculum phases.}
We compute rewards using three phases (feasibility $\rightarrow$ dual progress $\rightarrow$ primal
quality), implemented by \texttt{CurriculumRewardShaper}. Let $\Delta(\cdot)$ denote the one-step
improvement, e.g., $\Delta V_t := V_{t-1}-V_t$ and $\Delta LB_t := LB_t-LB_{t-1}$. Phase transitions
use smoothed metrics over a fixed window and occur when (i) the assignment-correctness ratio exceeds a
threshold (default: $0.95$), and (ii) the gap falls below a threshold (default: $0.15$), with the
smoothing window and thresholds reported in Appendix~\ref{app:training_hparams}.

\paragraph{Phase 1: feasibility shaping.}
The first phase emphasizes feasibility and stable constraint handling. The reward combines
subgradient-alignment, reduction in violation magnitude, and assignment-correctness bonuses:

\begin{equation}
\begin{split}
r_t^{(1)} = \;& w_{\mathrm{align}}\cdot \mathrm{clip}\!\left(\langle a_t, g_{t-1}\rangle,\,[-3,3]\right) \\
& + w_V \cdot \Delta V_t \\
& + w_{\mathrm{corr}}\cdot n_t^{\mathrm{corr}} \\
& + w_{\mathrm{full}}\cdot \mathbb{I}\{n_t^{\mathrm{corr}} \approx |V|\}
\end{split}
\label{eq:reward_phase1}
\end{equation}

where $a_t$ is the policy action (either $\Delta\lambda_t$ directly or a preconditioning scale,
depending on action mode), and $\mathrm{clip}(\cdot)$ denotes scalar clipping. This stage encourages
updates that are directionally consistent with the subgradient signal while reducing violation mass.

\paragraph{Phase 2: dual progress shaping.}
Once feasibility behavior stabilizes, we add dual improvement and gap reduction:

\begin{equation}
\begin{split}
r_t^{(2)} = \;& r_t^{(1)} \\
& + w_{LB}\cdot \mathrm{clip}\!\left(\widetilde{\Delta LB_t},\,[-3,3]\right) \\
& + w_{\mathrm{gap}}\cdot \mathrm{clip}\!\left(\Delta \mathrm{gap}_t,\,[-5,5]\right) \\
& + w_{LB+}\cdot \mathbb{I}\{LB_t>0\}
\end{split}
\label{eq:reward_phase2}
\end{equation}

where $\widetilde{\Delta LB_t}$ is a normalized lower-bound improvement used to stabilize scaling
across instances.
This phase aligns learning with the classical LR objective (improve the dual bound) while retaining
feasibility pressure.

\paragraph{Phase 3: primal routing utility.}
In the final phase, we incorporate end-to-end routing cost improvements based on reconstructed routes:

\begin{equation}
\begin{split}
r_t^{(3)} = \;& r_t^{(2)} \\
& + w_C\cdot \mathrm{clip}\!\left(\widetilde{\Delta C_t},\,[-5,5]\right) \\
& + w_{\mathrm{rec}}\cdot \mathbb{I}\{C_t < C_{\min}\} \\
& + w_{\mathrm{tier}}\cdot \mathrm{Tier}(\mathrm{gap}_t) \\
& + w_{\mathrm{assign}}\cdot \frac{n_t^{\mathrm{corr}}}{|V|}
\end{split}
\label{eq:reward_phase3}
\end{equation}

where $\widetilde{\Delta C_t}$ is a normalized cost improvement, $C_{\min}$ is the best cost observed
so far in the episode, and $\mathrm{Tier}(\cdot)$ denotes progressive gap-tier bonuses (e.g., bonuses
when $\mathrm{gap}_t$ falls below $20\%,15\%,10\%,5\%$), as implemented in our shaper.

This stage explicitly optimizes end-to-end routing performance while retaining a small amount of dual
guidance to prevent the policy from sacrificing assignment quality. Finally, to prevent pathological
updates, we apply penalties for extreme negative lower bounds and for high violation regimes, and clip
the total reward to a fixed range.

\paragraph{Why this credit assignment.}
The curriculum implements a practical compromise between dense and delayed rewards, feasibility and
dual progress provide informative step-wise signals early in training, while end-to-end routing cost is
introduced once the policy reliably produces near-feasible assignments. This staged shaping is
consistent with curriculum-learning principles \cite{bengio2009curriculum} and with reward-shaping
practice for improving learning efficiency without changing the underlying task objective
\cite{ng1999policy}. In our experiments, this design yields stable training under stochastic
subproblem feedback and allows the learned controller to optimize realized routing utility under a
fixed quantum evaluation budget.

\subsection{Implementation and computational overhead}
\label{sec:impl_overhead}

The learned multiplier controller is designed to add negligible overhead relative to subproblem
solving. Each LR outer-loop iteration performs: (i) a forward pass of the graph-based actor--critic on
a batched PyTorch Geometric graph observation,
(ii) a small number of PPO update epochs per batch of collected rollouts, and (iii) logging of
constraint-violation and primal-recovery statistics. Graph batching is implemented using the standard
PyG batching mechanism, which merges a list of \texttt{Data}/\texttt{HeteroData} objects into a single
disconnected graph with a batch-assignment vector. PPO updates follow the standard
loop of alternating rollout collection and repeated minibatch gradient updates on the clipped surrogate
objective.

In contrast, the dominant runtime cost in hybrid settings is the subproblem evaluation (classical or
quantum), including compilation and execution when hardware backends are used. To support circuit-level
evaluation and feasibility safeguards, we log compilation-time overhead metrics from the transpiled
circuits: depth and gate counts (including two-qubit gates and SWAPs) using standard Qiskit interfaces
and analysis passes (e.g., \texttt{CountOps}). These
metrics are used both for reporting (Section~\ref{sec:metrics}) and for enforcing gate-budget screening
in the hardware-aware execution policy (Section~\ref{sec:bandit_actions_screening}). Overall, the learned multiplier
controller introduces modest incremental computation compared with the cost of repeated subproblem
solves and provides an auditable log of how multiplier dynamics influence feasibility yield and
end-to-end routing utility.

\section{Hardware-aware quantum execution across heterogeneous backends}
\label{sec:hardware_aware}

Even after Lagrangian decomposition yields bounded-width knapsack QUBOs, quantum execution quality 
depends strongly on the choice of backend and circuit configuration. Sparse coupling graphs can 
induce substantial routing overhead (SWAP insertion), deep ans\"atze can exceed coherence-limited 
budgets, and calibration drift can invalidate static ``best-qubits'' heuristics. To make repeated 
quantum evaluations useful inside the LR outer loop, we treat \emph{backend selection and circuit 
configuration} as an online decision problem: at each knapsack call, we must choose (i) a backend 
(possibly from multiple providers) and (ii) an execution configuration (physical-qubit subset, 
entanglement pattern, depth) that balances solution quality against hardware constraints and 
time-window availability.

Our implementation provides a multi-QPU orchestration layer with automatic failover and parallel 
dispatch across heterogeneous devices, including IBM Quantum and AWS Braket QPUs, with a local
noisy simulator as a final fallback. Within each chosen backend, we learn a hardware-aware 
configuration policy using a contextual bandit (LinUCB) that maps a compact backend/problem 
descriptor to one of a small set of discrete circuit ``arms'' (27 configurations). Critically, 
we enforce an explicit \emph{gate-budget feasibility safeguard} by screening candidate configurations 
using a conservative compiled-gate estimator, rejecting actions predicted to violate a fixed gate 
limit. This yields a transparent mechanism for allocating quantum evaluations across heterogeneous, 
noisy resources while controlling routing overhead introduced during compilation and routing.

The remainder of this section formalizes the contextual bandit (context, actions, reward, and
constraints), introduces our hardware descriptors for heterogeneous QPUs, and describes feasibility
screening and integration with multi-QPU scheduling for parallel execution.

\subsection{Contextual bandit formulation for backend-aware configuration}
\label{sec:bandit_formulation}

We cast circuit configuration as a contextual bandit problem. At each invocation of the quantum
knapsack primitive (Section~\ref{sec:knapsack_qubo}), the solver observes a \emph{context} summarizing
the current subproblem size and the candidate backend, selects one configuration from a small discrete
set of circuit ``arms,'' and receives an execution-realized reward. This is a natural abstraction in
our setting because (i) each decision is made online as the LR outer loop generates a stream of
subproblems, (ii) the reward is observed after execution (sampling) and decoding/repair, and (iii) the
decision must be repeated many times under calibration drift and heterogeneous hardware constraints.
We adopt a linear contextual bandit (LinUCB) because it is simple, data-efficient, and supports fast
per-decision updates \cite{li2010contextual,auer2002finite}.

\paragraph{Context (problem--hardware descriptor).}
For each backend, we construct a lightweight \texttt{HardwareDescriptor} containing node-level
(calibration-like) quantities and a coupling graph representation; we summarize the coupling topology
by the average degree (connectivity) and the diameter (swap pressure proxy). For each knapsack call,
we assemble a \texttt{GlobalState} feature vector consisting of: (i) a problem-complexity proxy,
(ii) average hardware error, (iii) connectivity, (iv) diameter, and (v) logical width (problem size),
with explicit normalization constants to stabilize learning. The
multi-QPU evaluator uses this descriptor interface uniformly across providers (IBM and Braket), with
backend snapshots (JSON) and optional live QPU extraction.

\paragraph{Actions (27-armed configuration set).}
The action space is the Cartesian product of three discrete design choices:

\[
\begin{aligned}
\texttt{placement} &\in \{\textsf{dense}, \textsf{quality}, \textsf{random}\}, \\
\texttt{entanglement} &\in \{\textsf{linear}, \textsf{circular}, \textsf{full}\}, \\
\texttt{depth} &\in \{1, 2, 3\}.
\end{aligned}
\]

yielding 27 configuration arms. Placement selects a physical-qubit subset (dense cluster, highest
quality qubits, or random) using the coupling graph and calibration features; entanglement selects
the logical interaction pattern; and depth controls the number of entangling layers. 
This discrete action design intentionally limits policy complexity while still
capturing the main hardware-sensitive tradeoffs (routing overhead vs expressibility vs noise).

\paragraph{Feasibility safeguard via gate-budget screening.}
A key practical constraint is that the compiled circuit may be infeasible in practice due to routing
(SWAP insertion) and basis translation inflating depth and two-qubit gate count, especially on sparse,
high-diameter coupling graphs. Qiskit’s transpiler explicitly separates \emph{layout} (qubit selection)
from \emph{routing} (SWAP insertion to satisfy coupling constraints), which is the mechanism behind
this overhead. We therefore enforce a conservative gate-budget constraint prior to
execution. Given $(N,\texttt{placement},\texttt{entanglement},\texttt{depth})$ and topology summaries
(connectivity, diameter), we estimate a compiled gate count and reject actions predicted to exceed a
fixed limit (20k gates by default for AWS), falling back to a safe default configuration when needed. 
This turns configuration selection into a \emph{constrained} contextual
bandit: maximize reward subject to a deterministic feasibility filter.

\paragraph{Training reward (trace-driven proxy).}
For trace-driven bandit training, we use a bounded proxy reward $r_{\text{proxy}}\in[0,10]$
that approximates the quality--overhead tradeoff using backend trace features and a circuit
overhead model; actions predicted to violate the gate budget receive zero reward.
Section~\ref{sec:bandit_training} defines $r_{\text{proxy}}$ explicitly.

\paragraph{Execution-realized utility (evaluation and solver integration).}
At deployment time, the bandit selects a configuration (placement, entanglement, depth) that is
executed through the same quantum pipeline as the main solver. Measurement samples are decoded into
candidate knapsack selections and integrated into the LR outer loop, and end-to-end performance is
assessed by the reconstructed CVRP objective (gap to BKS) together with cost metrics (circuits/jobs,
runtime, latency, failures). QPU failures trigger failover and, if necessary, a noisy local simulation
fallback to preserve progress.

\paragraph{LinUCB update rule.}
We maintain a separate ridge-regression model for each arm. For context feature vector $x$, LinUCB
computes an upper-confidence score $\hat\mu_a(x) + \alpha\sqrt{x^\top A_a^{-1}x}$ for each arm $a$,
selects the maximizing arm, and updates the chosen arm’s statistics with a rank-1 update. To support
fast online learning, we store $A_a^{-1}$ explicitly and update it using the Sherman--Morrison formula,
matching the implementation in \texttt{ConfigurationBandit} \cite{li2010contextual}.

\subsection{Hardware descriptors and context feature construction}
\label{sec:hardware_descriptors}

To support backend-agnostic configuration, we represent each candidate device using a common
\emph{hardware descriptor} interface and derive a compact context vector for the bandit. The design
goal is pragmatic: expose only the hardware signals that most strongly predict compilation overhead
and execution reliability, while remaining portable across providers.

\paragraph{Unified device abstraction.}
We use a \texttt{HardwareDescriptor} object that stores (i) per-qubit features
(e.g., $T_1$, $T_2$, single-qubit error proxy, readout error, and an ``available'' flag) and
(ii) per-coupler features (two-qubit error proxy, duration, availability), together with a coupling
graph representation. For IBM devices, these fields are extracted from backend
\texttt{configuration} and \texttt{properties} (including qubit coherence, readout error, and native
two-qubit gate error/length). This is consistent with Qiskit's backend model, where
\texttt{BackendProperties} exposes qubit and gate calibration-like parameters measured by the provider. 
For non-IBM devices (Rigetti/IQM via AWS Braket), we ingest JSON snapshots and
normalize qubit identifiers into a contiguous index set so that selection and routing logic is
consistent across devices. When live Braket metadata is available, device topology
and properties can also be accessed through the Braket console or the \texttt{GetDevice} API.

\paragraph{Topology summaries: connectivity and diameter.}
Routing overhead is driven primarily by the mismatch between the circuit's interaction graph and the
device coupling graph. Rather than using a high-dimensional graph embedding, we summarize the active
coupling subgraph by two interpretable scalars: average degree (connectivity) and graph diameter (swap
pressure proxy). These are computed on the induced subgraph of \emph{available} qubits and, when the
graph is disconnected, on the largest connected component. The use of
diameter/connectivity is motivated by the transpiler mechanism: during the routing stage, SWAP gates
are inserted to satisfy coupling constraints, and overhead increases when shortest-path distances are
large.

\paragraph{Problem-complexity feature.}
The bandit context includes a scalar \emph{Hamiltonian complexity} feature that summarizes the
structural difficulty of the current QUBO. For a knapsack QUBO with $m$ binary variables, let
$E_Q := \{(p,q): p<q,\; Q_{pq}\neq 0\}$ denote the set of nonzero quadratic couplings and let
$|E_Q|$ be its cardinality. We define
\begin{equation}
h \;:=\; \mathrm{clip}_{[1,12]}\!\left( 1 + 11\cdot \frac{|E_Q|}{m(m-1)/2} \right),
\label{eq:ham_complexity}
\end{equation}
so $h$ increases with coupling density (from nearly separable to dense interactions).

\paragraph{Error summary: blended average error.}
For bandit context, we compress heterogeneous calibration fields into a single hardware error signal.
In bandit training, we compute a blended error proxy that mixes average single-qubit error, readout
error, and two-qubit error (when available), ensuring JSON traces remain distinguishable even when some
providers omit specific calibration fields. This blended error enters the context as
\texttt{hardware\_avg\_error}. In online evaluation, the same field is computed from the current
descriptor whenever possible, with conservative defaults otherwise.

\paragraph{Context vector (GlobalState).}
Given a knapsack subproblem of logical width $n$ and a candidate device descriptor, we instantiate a
\texttt{GlobalState} vector

\[
\begin{aligned}
x = \big[ & 1, 0.1\cdot\texttt{complexity}, 100\cdot\texttt{avg\_error}, \\
& 0.5\cdot\texttt{connectivity},  0.1\cdot\texttt{diameter},  0.05\cdot n \big]
\end{aligned}
\]

which is the normalized feature map used by LinUCB. The normalization constants are
chosen to keep all features $\mathcal{O}(1)$ for typical ranges (e.g., errors on the order of $10^{-2}$
and problem sizes on the order of $10$--$60$ qubits), improving numerical stability of the ridge
regression updates. This results in a compact six-dimensional context that captures:
(i) subproblem difficulty/scale, (ii) expected noise level, and (iii) expected routing pressure.

\paragraph{Freshness and drift handling.}
Device calibrations drift over time. To encourage robustness, our trace-driven bandit training samples
\texttt{GlobalState} vectors from a pool of device traces and injects controlled multiplicative drift
into the error signal (Gaussian perturbation with clipping), approximating day-to-day calibration
changes. In online evaluation, descriptors are refreshed when possible (e.g., IBM
backend properties at runtime), and the multi-QPU layer falls back to alternate devices or local
simulation when a device becomes unavailable.

\subsection{Action space and feasibility screening (gate-budget constrained selection)}
\label{sec:bandit_actions_screening}

Our configuration policy selects from a discrete action space that factors into three orthogonal 
design choices: placement, entanglement pattern, and circuit depth, and then applies an explicit feasibility
screen before any hardware submission. This ``rank-then-filter'' design keeps the policy simple while
making compilation risk transparent and controllable.

\paragraph{Action space (27 arms).}
We define the arm set as the Cartesian product

\[
\mathcal{A}=\{\textsf{dense},\textsf{quality},\textsf{random}\}\times
\{\textsf{linear},\textsf{circular},\textsf{full}\}\times\{1,2,3\},
\]

yielding 27 configurations. Each arm fixes (i) a placement rule that chooses a physical-qubit subset,
(ii) an entanglement map that specifies which two-qubit interactions are attempted per layer, and (iii)
an ansatz depth. The bandit ranks these arms by an upper-confidence score and proposes the
best arm for the current context.

\paragraph{Why feasibility screening is necessary.}
Even for fixed logical width, the compiled circuit can become infeasible on a given device because
routing introduces SWAPs when two-qubit interactions are not supported by the coupling graph. This is a
first-order effect of compilation: in Qiskit’s transpilation pipeline, the \emph{layout} stage chooses a
mapping from logical to physical qubits, and the \emph{routing} stage inserts SWAPs to satisfy hardware
connectivity constraints. Consequently, depth and two-qubit gate counts can
inflate sharply on sparse, high-diameter topologies.

\paragraph{Gate-budget estimator and constraint.}
To prevent wasted submissions (and to enable consistent cross-backend evaluation), we enforce an explicit
gate-budget constraint. For each candidate arm $a=(p,e,d)\in\mathcal{A}$ (placement $p$, entanglement
pattern $e$, and depth $d$) and context $(n,\mathrm{conn},\mathrm{diam})$, we estimate the compiled gate
count as

\begin{equation}
\begin{split}
\widehat{G}(a; n,\mathrm{conn},\mathrm{diam}) = \bigg\lceil & n(d+1) \\
& + \mathrm{pairs}(n,e)\cdot d \\
& \cdot \mathrm{swap}(p,e;\mathrm{conn},\mathrm{diam}) \cdot \gamma \bigg\rceil
\end{split}
\label{eq:gate_estimator}
\end{equation}

where $\mathrm{pairs}(n,e)\in\{n-1,\,n,\,n(n-1)/2\}$ for linear/circular/full entanglement,
$\gamma>1$ is a compilation inflation factor, and $\mathrm{swap}(\cdot)$ is a routing inflation term
based on a topology proxy $\mathrm{avg\_hops}=\max\{1,\mathrm{diam}/\max(\mathrm{conn},1)\}$. In
particular, $\mathrm{swap}(\cdot)$ increases with $\mathrm{avg\_hops}$ and is larger for dense
entanglement patterns and for placement strategies that are less aligned with the backend connectivity.
We declare an arm safe if $\widehat{G}(a; n,\mathrm{conn},\mathrm{diam})\le G_{\max}$.
When a hardware descriptor is unavailable, we apply the same rule using conservative fallback topology
defaults.

\paragraph{Rank-then-filter selection with override.}
Given a ranked list of arms from LinUCB, we select the highest-ranked arm that passes the gate-budget
screen. If the top-ranked arm violates the budget, we override it with the best safe alternative; if all
arms violate the budget, we fall back to a fixed safe default (dense/linear with shallow depth). This
logic iterates through ranked
actions and returns the first safe candidate, logging an explicit ``gate-budget override'' when the
top-ranked arm is rejected. This produces a constrained contextual bandit decision rule
that is easy to audit: exploration remains bandit-driven, while constraint satisfaction is deterministic.

\paragraph{Placement details.}
The placement component enforces strict availability constraints and avoids dead qubits by selecting 
from nodes flagged as available in the descriptor graph. The \textsf{dense} strategy grows a compact 
connected cluster to reduce routing distances; \textsf{quality} chooses qubits with high coherence/low
readout error; and \textsf{random} provides exploration and a stress test for robustness. These
strategies are evaluated on JSON and IBM-derived descriptors using unit tests to ensure that selected 
layouts do not include dead qubits and respect basic connectivity properties.

\paragraph{Integration with online multi-QPU execution.}
The screened configuration is applied immediately in the per-vehicle knapsack solve: the system 
selects an
available QPU (respecting provider time windows, runtime headroom, and failover rules), 
queries the bandit,
screens the candidate configuration, and executes the resulting circuit (or falls back to local 
noisy simulation
on repeated failures/timeouts). This end-to-end orchestration is implemented in our workflow.

\subsection{Bandit training protocol and trace-driven robustness}
\label{sec:bandit_training}

We train the configuration policy offline using a trace-driven simulator and then deploy it online in
the multi-QPU solver. This ``train on traces, act online'' approach is motivated by two practical
constraints: (i) real QPU time is scarce and costly, and (ii) calibration drift makes a single static
configuration brittle across days and devices. Contextual bandits are a natural fit in this regime
because they can learn efficiently from sequential feedback and can also be evaluated and improved
using logged interaction data \cite{li2010contextual}.

\paragraph{Training data: backend trace pool.}
We construct a pool of backend traces capturing heterogeneous device properties, including coupling
topologies and calibration-like error proxies. Traces are loaded from JSON descriptors (for non-IBM
devices) and from IBM backend snapshots when available. Each trace is
converted into a \texttt{HardwareDescriptor} and summarized into a \texttt{GlobalState} context vector
(Section~\ref{sec:hardware_descriptors}). To model calibration drift, we inject
multiplicative noise into the hardware error signal (Gaussian perturbation with clipping), producing a
family of plausible contexts around each trace. This is consistent with empirical
observations that variability across qubits/edges and time-dependent drift can materially affect
execution outcomes.

\paragraph{Trace-driven reward model.}
During bandit training, we avoid executing circuits on hardware. Instead, each episode samples a
backend trace and a synthetic Hamiltonian-complexity level, then simulates an execution outcome for
the selected arm. The simulator produces a bounded proxy reward $r_{\text{proxy}}\in[0,10]$:

\begin{equation}
\begin{split}
r_{\text{proxy}} = \;& \mathbb{I}\{\widehat{G}\le G_{\max}\} \cdot \\
& \mathrm{clip}_{[0,10]}\big( 10\cdot F_{\text{eff}}(a,x)\cdot \\
& X_{\text{expr}}(a,h) - P_{\text{swap}}(a,x) \big)
\end{split}
\label{eq:proxy_reward}
\end{equation}

where $F_{\text{eff}}$ is an effective-fidelity proxy derived from trace error features, $X_{\text{expr}}$
is an expressibility proxy increasing with depth and entanglement richness, and $P_{\text{swap}}$
penalizes routing overhead as a function of topology (connectivity/diameter) and placement choice.
Actions predicted to violate the gate budget receive zero reward via the feasibility filter
(Section~\ref{sec:bandit_actions_screening}). This proxy reward is used only for offline policy learning;
all reported solution quality and cost metrics in Section~\ref{sec:bandit_results} are execution-realized.

\paragraph{Bandit updates and logging.}
We train a LinUCB model with per-arm ridge regressors, updating the chosen arm after each episode and
maintaining $A_a^{-1}$ via Sherman--Morrison rank-1 updates for computational efficiency
\cite{li2010contextual}. 

\paragraph{Offline sanity checks and online deployment.}
The trace-driven training loop provides two forms of validation prior to online deployment:
(i) convergence of average reward and stabilization of arm selection frequencies, and
(ii) robustness of the chosen arm distribution under drift-augmented traces.
In online use, the learned bandit is consulted at each knapsack solve, but decisions are still subject
to deterministic feasibility screening and multi-QPU failover (timeouts, backend unavailability), so
the system remains safe under distribution shift. This design aligns with the
standard UCB principle of balancing exploration and exploitation under bounded rewards,
while recognizing that physical execution introduces hard constraints and operational failures that
must be handled deterministically.

\subsection{Multi-QPU orchestration and online execution loop}
\label{sec:multi_qpu_orchestration}

Our hardware-aware execution layer operates in two nested loops: an \emph{outer} multi-QPU selection 
loop that chooses which backend to use (and handles availability/failover), and an \emph{inner} 
configuration loop that selects a screened circuit configuration via the contextual bandit (Sections~\ref{sec:bandit_formulation}--\ref{sec:bandit_training}). 
This section describes the online orchestration logic and how configuration decisions are embedded 
into the knapsack-solve calls.

\paragraph{Backend discovery and availability.}
We treat the set of candidate QPUs as time-varying. For IBM Quantum, available backends and their
properties can be retrieved programmatically, enabling device discovery and refresh of calibration-like
metadata. For AWS Braket, devices and their status/metadata can be retrieved via the 
\texttt{GetDevice} API (or equivalent SDK/CLI methods), which provides a stable interface for device 
availability and device capabilities across providers. In our implementation, backend descriptors 
are refreshed when possible, and the system maintains a fallback ordering (alternate QPUs, then a 
local noisy simulator) when a device becomes unavailable or repeatedly fails.

\paragraph{Scheduling policy and time-window constraints.}
The multi-QPU manager maintains a pool of candidate backends with operational constraints such as:
provider-specific access windows, a per-episode wall-clock budget, and reliability scores estimated 
from recent failures. For each knapsack call, the manager selects a backend that can accommodate the 
required logical width and has sufficient runtime headroom. 
This is implemented by \texttt{MultiQPUManager}, which tracks per-device availability, 
per-device success/failure counts, and time-window gating.
When multiple backends are eligible, selection is based on a simple priority rule that prefers 
devices with
(i) larger effective capacity for the required width and (ii) higher recent success rate, while still 
allowing exploration across devices for robustness.

\paragraph{Where the bandit is called in the execution path.}
Once a backend is selected, the solver constructs the \texttt{GlobalState} context 
(Section~\ref{sec:hardware_descriptors}), consults the LinUCB bandit to obtain a ranked list of 
configuration arms, and applies gate-budget screening to select the highest-ranked safe configuration 
(Section~\ref{sec:bandit_actions_screening}). The selected configuration determines:
(i) a physical-qubit subset (placement), (ii) a logical entanglement map, and (iii) circuit depth. 
This selection is performed for each knapsack subproblem invocation and is therefore adaptive across 
LR iterations and across heterogeneous backends.

\paragraph{Parallel dispatch across per-vehicle subproblems.}
Lagrangian decomposition yields one knapsack subproblem per vehicle. These subproblems are 
independent conditional on multipliers and can therefore be dispatched in parallel across CPU 
threads and/or across multiple QPUs. Our implementation uses a thread pool to run multiple 
per-vehicle solves concurrently, with backend selection and bandit configuration performed 
independently per subproblem. This parallelization is essential in the hybrid setting because 
quantum evaluations are expensive; parallel dispatch increases throughput and reduces wall-clock 
time per LR iteration when multiple devices are available.

\paragraph{Failure handling and deterministic fallbacks.}
Physical execution can fail due to queue delays, transient backend errors, or unexpected compilation 
issues. The orchestration layer therefore implements deterministic failover: if a submission times 
out or a device returns an error, the system retries on an alternate backend when available; after a 
bounded number of failures, it falls back to a local noisy simulator to 
ensure progress and preserve reproducibility. Importantly, the bandit decision remains \emph{screened} by the feasibility filter, so failures
are treated as operational events rather than as training signals; bandit updates are performed in the trace-driven
training regime (Section~\ref{sec:bandit_training}).

\paragraph{Audit logging.}
To support circuit-level reporting (Section~\ref{sec:metrics}) and reproducibility, we log for each 
knapsack call:
backend chosen, configuration arm (placement/entanglement/depth), whether an override occurred due to 
gate-budget screening, runtime and failure codes, and (when available) compilation statistics such 
as depth and gate counts.
These compilation statistics are obtained from the transpiled circuit, where routing algorithms 
insert SWAPs to map two-qubit interactions to the coupling graph. This produces a complete audit trail
linking hardware decisions to realized optimization utility.

\subsection{Reporting and diagnostic plots for the hardware-aware layer}
\label{app:bandit_reporting}

To make the hardware-aware execution layer auditable and reproducible, we report results that separate
(i) \emph{bandit learning behavior}, (ii) \emph{constraint/feasibility behavior}, and (iii) \emph{downstream
optimization utility}. This aligns with standard practice for contextual bandits, where learning is
typically evaluated by cumulative reward (or regret), action-selection frequencies, and constraint
violations under bounded rewards \cite{auer2002finite,li2010contextual}.

\paragraph{Bandit learning curves (offline training).}
Using the trace-driven training loop (Section~\ref{sec:bandit_training}), we plot:
(i) moving-average reward per episode (see Fig.~\ref{fig:bandit_learning}) ,
(ii) cumulative reward, and
(iii) arm-selection frequencies over time.

Because our training reward is bounded in $[0,10]$ and the policy is UCB-based,
these plots directly indicate whether exploration stabilizes and whether the policy concentrates on a
small subset of high-utility arms, as expected under UCB-style learning \cite{auer2002finite}.

\paragraph{Feasibility/override statistics (gate-budget screening).}
We report the rate at which the top-ranked arm is rejected by the gate-budget feasibility filter and
replaced by a safe alternative (``override rate''), as well as the overall gate-budget violation rate
under the final deployed policy. These quantities are logged explicitly by the online consult routine
and by the training simulator (actions exceeding the gate budget receive
zero reward). In addition, we include a size-scaling diagnostic
showing when configurations become unsafe as logical width increases, which helps interpret when
hardware constraints dominate and why shallow/linear configurations are selected more often at larger 
widths.

\paragraph{Circuit-level overhead metrics (depth, two-qubit gates, SWAPs).}
To connect configuration decisions to hardware costs, we record compilation-time circuit statistics
for the selected configuration on each backend: circuit depth, counts of two-qubit gates, and SWAP
counts. We report these metrics alongside reward because they
capture the primary mechanism by which connectivity induces overhead: the transpiler \emph{routing}
stage inserts SWAPs to satisfy coupling constraints, increasing depth and two-qubit gate counts. These statistics support the ``circuit-level feasibility and overhead'' results
reported in Section~\ref{sec:metrics}.

\paragraph{Downstream utility (subproblem and end-to-end impact).}
Finally, we quantify how hardware-aware configuration affects optimization utility. At the subproblem
level, we report the best-of-samples-after-repair score and feasible-yield statistics under the bandit
policy versus fixed baselines (e.g., always dense/linear at fixed depth). At the end-to-end level, we
report CVRP cost/gap after reconstruction using identical sampling budgets. Together, these
results provide a transparent linkage from (context $\rightarrow$ arm selection $\rightarrow$ compiled
overhead) to realized routing utility under heterogeneous backends.

\section{Experimental Design}
\label{sec:experiments}

This section specifies the details of our experimental protocol used to evaluate our approach. 

\subsection{Instances and data splits}
\label{sec:instances}

\paragraph{Test set (CVRPLIB).}
Our primary evaluation set consists of 30 benchmark CVRP instances drawn from multiple CVRPLIB 
families, covering a range of customer counts and fleet sizes.
These families are widely
used in the CVRP literature and provide heterogeneous geometric structure and demand patterns, while
retaining consistent file format, best-known-solution (BKS) tracking, and comparability across studies.
Across our selected set, the number of customers ranges from $n=13$ to $n=151$ and the fleet size ranges
from $K=2$ to $K=14$ (see Appendix~\ref{app:full_instance_tables} for the full list). This range spans
small and medium instances where quantum subproblems are easily executable and large-sized instances where routing
overhead and queueing effects become meaningful under real-device execution.

CVRPLIB is 
a standard repository of capacitated vehicle routing benchmarks with widely reported best-known 
(and often optimal) solutions, enabling direct comparison through optimality gaps \cite{cvrplib}. 
We treat these 30 instances as a held-out test set and do not use them for training.

\paragraph{Training set (synthetic CVRPLIB-style instances).}
To train the learned multiplier-update policy without overfitting to the small benchmark test set, 
we generate a synthetic training corpus designed to mimic the scale and basic structure of the 
CVRPLIB instances. Instances are produced in TSPLIB CVRP format by sampling 2D coordinates uniformly 
on a bounded grid and sampling customer demands from a fixed discrete range, with depot demand set 
to zero. Specifically, our generator selects the number of nodes uniformly from a prescribed 
range, sets capacity to a fixed constant, draws customer 
demands i.i.d.\ from $\{1,\dots,29\}$, and sets the number of vehicles to the minimum 
capacity-feasible value $k=\lceil \sum_i d_i / Q \rceil$. The resulting training instances provide 
diverse demand patterns and geometries while maintaining feasibility and ``tightness'' consistent 
with capacity-driven CVRP regimes.

\paragraph{Rationale.}
This split isolates two goals: (i) evaluate routing utility and generalization on standardized, 
widely used benchmarks (CVRPLIB), and (ii) learn multiplier control from a broader distribution of 
instances without leaking benchmark-specific structure into training. All reported test performance 
is computed exclusively on the 30 held-out CVRPLIB instances.

\subsection{Evaluation methods and ablation matrix}
\label{sec:methods_compared}


First, we report the primary end-to-end comparison, which evaluates the full pipeline against classical baselines and hybird variants. 
We then use a structured ablation matrix to isolate the contribution of individual components, namely 
decomposition and QUBO encoding into bounded-width knapsack subproblems,
learned multiplier updates in the LR outer loop, and
hardware-aware execution (contextual bandit + feasibility screening).
All variants share the same decoding/repair and route reconstruction procedures
(presented in Section~\ref{sec:primal_recovery}) and are evaluated on the held-out CVRPLIB test set described above \cite{cvrplib}.

\paragraph{Decomposition and QUBO encoding variants.}
We compare (a) a direct monolithic QUBO encoding baseline (used only for logical-width accounting) 
against
(b) the Fisher--Jaikumar assignment linearization followed by Lagrangian relaxation into per-vehicle
knapsack QUBOs \cite{fisher1981generalized}. For the knapsack capacity constraint we use the tilted
slack-free penalty as the default (Section~\ref{sec:knapsack_qubo}; Appendix~\ref{app:penalties}). 
Qubit-scaling results report
logical width (pre-compilation) to isolate modeling effects from hardware-specific compilation effects.

\paragraph{Multiplier-update controller (outer loop).}
To isolate the effect of multiplier control from hardware noise, we first evaluate three outer-loop 
update
rules using classical knapsack subproblem solves:
(i) classical subgradient ascent (stepsize-tuned baseline),
(ii) an expert-pretrained policy (behavior cloning), and
(iii) the same policy after PPO fine-tuning under the curriculum reward 
(Section~\ref{sec:training_protocol}).
PPO is used as the default fine-tuning method due to its stability and widespread adoption in noisy RL settings
\cite{schulman2017proximal}. These experiments establish whether learning improves convergence 
behavior and end-to-end
routing utility under a controlled subproblem solver.

\paragraph{Hardware-aware execution (contextual bandit) and baselines.}
For hardware-aware execution, we compare:
(i) the learned LinUCB configuration policy with gate-budget screening 
(Section~\ref{sec:bandit_formulation}) against
(ii) fixed configuration baselines (e.g., fixed placement/entanglement/depth). 
LinUCB provides a standard and computationally lightweight contextual bandit baseline with 
theoretical motivation
and strong empirical performance in large-scale online settings \cite{li2010contextual,auer2002finite}.
We report
both offline bandit diagnostics (reward curves, arm frequencies, override rates) and online impact 
on subproblem utility
and end-to-end routing performance.



\subsection{Budgets, stopping rules, and fairness of comparison}
\label{sec:budgets}

All methods are evaluated under fixed compute budgets to ensure fair comparison. Budgets are 
specified at three levels: (i) the LR outer-loop horizon, (ii) the sampling budget for each 
knapsack-QUBO evaluation, and (iii) hardware-execution constraints (timeouts and a gate-budget 
feasibility limit). We report wall-clock runtime as elapsed seconds measured around the full 
solver call; by definition, only differences between two timer calls are meaningful.

\paragraph{Outer-loop budget (LR horizon and early stopping).}
Each instance is solved for at most $T_{\max}$ Lagrangian iterations. At iteration $t$, the solver 
performs one knapsack-QUBO evaluation per vehicle (parallelized when possible) followed by 
decoding/repair and, at prescribed intervals, route reconstruction. We employ two standard stopping 
criteria:
(i) \emph{budget exhaustion} ($t=T_{\max}$), and
(ii) \emph{stagnation}, where the best reconstructed cost has not improved for a fixed patience 
window $P$ (measured in outer-loop iterations). The exact values of $(T_{\max},P)$ are fixed across 
all compared multiplier-update methods.

\paragraph{Subproblem sampling budget (shots and candidate selection).}
For each knapsack-QUBO evaluation, we allocate a fixed sampling budget of $S$ circuit repetitions 
(shots) per circuit evaluation and retain a fixed number of candidate samples for decoding/repair. 
This ensures that any improvement in feasible-yield or best-of-samples quality is attributable to 
the multiplier controller or configuration policy, rather than increased sampling. When Qiskit 
primitives are used, sampling corresponds to repeated circuit execution via the \texttt{Sampler} 
primitive, which returns samples from the circuit output register. In classical-only controller 
ablations (Section~\ref{sec:multiplier_results}), knapsack subproblems are solved 
exactly/approximately by the classical solver under the same outer-loop horizon.

\paragraph{Hardware feasibility budget (gate limit) and timeouts.}
For hardware-aware experiments (Section~\ref{sec:bandit_results}), we enforce a hard gate-budget 
limit $G_{\max}$ (default 20k) using the feasibility screen described in 
Section~\ref{sec:bandit_actions_screening}. This safeguard is motivated by the compilation 
mechanism: the transpiler’s layout and routing stages can introduce SWAP overhead to satisfy the 
device coupling graph, increasing depth and two-qubit gate counts. In addition, each backend 
submission is subject to a per-call timeout $\tau$; failed or timed-out executions trigger 
deterministic failover to alternate devices and, ultimately, to a local simulator. 
All configurations (bandit and baselines) are subject to the same gate-budget and timeout rules.

\paragraph{Fairness controls and fixed seeds.}
To make comparisons reproducible, we fix all configuration parameters (budgets, thresholds, 
penalty scales, and hardware screening limits) across methods, and we evaluate each test instance 
once under the fixed protocol (no multi-seed averaging in the reported tables). In the 
reproducibility package we provide the exact configuration files and command lines used to 
generate each table and figure.

\subsection{Backends and noise settings}
\label{sec:backends_noise}

We evaluate the proposed methods under three execution regimes: (i) classical subproblem solves 
(used to isolate the effect of multiplier control), (ii) noise-aware simulation (used to study 
sampling and noise sensitivity under controlled conditions), and (iii) heterogeneous-backend 
execution with a multi-QPU orchestration layer (used to assess hardware-aware configuration and 
failover behavior). The set of candidate backends includes IBM Quantum devices (accessed via Qiskit 
runtime metadata) and non-IBM devices accessed through JSON snapshots and AWS Braket metadata 
(Section~\ref{sec:hardware_descriptors}).

\paragraph{IBM Quantum backends and snapshots.}
For IBM devices, we obtain coupling topology and calibration-like parameters from Qiskit backend 
properties (e.g., $T_1$, $T_2$, readout error, and gate error estimates). When running in 
noise-aware simulation, we construct a device-derived noise model from these properties using 
Qiskit Aer’s noise utilities, which incorporate depolarizing gate errors, readout errors, 
and (optionally) thermal relaxation based on coherence times and gate durations. In addition, 
we use backend snapshot models (``fake backends'') when needed to emulate IBM systems without 
requiring live access; these snapshots are designed for testing and noisy simulation using real 
backend calibrations.

\paragraph{AWS Braket backends and metadata.}
For non-IBM devices accessed through AWS Braket, device capability and availability metadata are 
retrieved through the Braket device interface (e.g., the \texttt{GetDevice} API), which provides 
a stable programmatic method to query device properties and status. When fine-grained 
calibration-like information is not available uniformly across providers, we rely on our 
JSON-based descriptor interface (Section~\ref{sec:hardware_descriptors}) to standardize 
topology and error proxies across devices for bandit training and evaluation.

\paragraph{Noise-aware simulation definition.}
Noise-aware simulation is performed by executing circuits on a simulator equipped with a 
device-derived noise model. For IBM-based noise models, this corresponds to an Aer noise model 
constructed from backend properties as described above. For snapshot-based simulation, 
fake backends can be run via an Aer simulator with the snapshot noise model when Qiskit Aer is 
installed. In all cases, we treat the resulting measurement distribution as the source of 
stochasticity affecting (i) feasible-yield after decoding/repair and (ii) the best-of-samples 
outcome used by the outer loop (Section~\ref{sec:decode_rule}).

\paragraph{Real-hardware execution.}
For hardware experiments, we execute the knapsack-QUBO circuits on gate-based QPUs accessed through
IBM Quantum (via Qiskit Runtime) and AWS Braket (Rigetti and IQM devices), with cloud simulators and
a local simulator as fallbacks. On IBM, circuits are submitted through the Runtime Sampler primitive
(SamplerV2), which returns measurement counts from executed circuits. On Braket,
circuits are submitted through the device interface, and results are retrieved
as measurement counts upon task completion. Because QPU availability is
time-varying and queue delays can be significant, we enforce provider-specific availability windows,
per-call timeouts, and deterministic failover to alternate devices and finally to local simulation;
all such events are logged (Section~\ref{sec:multi_qpu_orchestration}).

\paragraph{Execution constraints: basis gates, compilation, and routing.}
All gate-based executions are compiled to the backend’s native basis and coupling constraints via 
the transpiler. Routing overhead is introduced by the transpiler routing stage (SWAP insertion) 
when the logical interaction pattern is not directly supported by the coupling graph; this 
motivates the circuit-level overhead metrics and the explicit gate-budget screening used by 
the hardware-aware policy. For cross-backend evaluation, we log both the logical width 
(pre-compilation) and the transpiled circuit statistics (depth and gate counts) whenever available.

\paragraph{Failover and fallback.}
In heterogeneous execution, physical backends can fail due to queue delays or transient errors. 
Our multi-QPU manager therefore implements deterministic failover to alternate devices and a 
final fallback to local simulation to ensure that the LR outer loop remains well-defined and 
that evaluation is reproducible (Section~\ref{sec:multi_qpu_orchestration}).

\subsection{Metrics and aggregation}
\label{sec:metric_aggregation}

We report metrics at three levels: circuit, subproblem, and end-to-end, and we distinguish the
execution regime for each run using an explicit tag
$\texttt{mode}\in\{\textsf{classical},\textsf{noisy-sim},\textsf{hardware}\}$.
This separation prevents conflating algorithmic effects (multiplier control) with execution effects
(noise, compilation overhead, or queue delays) and supports transparent ablations.

\paragraph{End-to-end CVRP metrics.}
For each instance, we report:
(i) the best reconstructed routing cost $C^{\star}$ produced within the fixed budget,
(ii) feasibility indicators (all customers served exactly once and capacity respected),
and (iii) the optimality gap relative to the instance best-known (often optimal) solution (BKS),
\begin{equation}
\mathrm{Gap}(\%) \;=\; 100\cdot\frac{C^{\star}-\mathrm{BKS}}{\mathrm{BKS}}.
\end{equation}
BKS values are read from the instance metadata when available. Runtime is reported as wall-clock 
elapsed time (seconds) for the \emph{entire} solve episode, measured around the complete solver 
call (including LR iterations, subproblem solves, repair, and routing). When quantum execution 
is used, we additionally report the time split into total solve time and time spent in quantum 
subproblem execution (``VQE time'') as logged by the multi-QPU evaluator.
\footnote{In the multi-QPU evaluator, \texttt{solve\_time} is measured around
the full evaluation loop, and \texttt{vqe\_time} accumulates time spent inside quantum subproblem 
calls.}

\paragraph{Subproblem-level metrics (knapsack QUBOs).}
For each per-vehicle knapsack call we record:
(i) logical width (number of selection variables; and, for slack-bit encodings, total logical width),
(ii) feasibility yield: the fraction of measurement samples that satisfy the capacity inequality 
prior to global assignment repair,
and (iii) the best-of-samples-after-repair outcome used by the outer loop 
(Section~\ref{sec:decode_rule}).
For noisy-simulator and hardware modes, these statistics quantify whether the chosen penalty 
encoding and circuit configuration produce usable candidates under finite-shot sampling.

\paragraph{Circuit-level overhead and feasibility.}
For hardware-aware execution experiments, we report compilation- and routing-driven overhead 
statistics for the selected configuration:
circuit depth, counts of two-qubit gates, and SWAP counts when available. These metrics are 
motivated by the compilation mechanism: in Qiskit's prebuilt transpiler pipeline, the 
\emph{routing} stage inserts SWAPs to satisfy coupling constraints, inflating depth and 
two-qubit gate counts. We also report the gate-budget 
screening outcomes (pass/override rates) used by the constrained bandit policy.

\paragraph{Bandit diagnostics.}
To assess the hardware-aware layer independently of downstream routing, we report:
(i) moving-average reward and cumulative reward during offline bandit training,
(ii) arm-selection frequencies, and
(iii) override rates induced by feasibility screening (Section~\ref{sec:bandit_actions_screening}).
This is consistent with standard practice for contextual bandits, where performance is 
evaluated by accumulated reward (or regret proxies) and action-selection behavior under 
bounded rewards \cite{auer2002finite,li2010contextual}.

\paragraph{Aggregation and reporting convention.}
Each test instance is evaluated once under the fixed protocol. We therefore summarize results across 
the 30-instance test set using both \emph{means} and \emph{medians} (reported alongside per-instance 
tables in the appendix) to reduce sensitivity to outliers, following common recommendations for 
reporting computational experiments in heuristic settings. Full per-instance results are provided 
in the appendix.

\section{Experimental Results}
\label{sec:results}


We now report results in the same order as the experimental design described in Section~\ref{sec:experiments}.
First, we present the primary end-to-end comparison of the full pipeline against baselines
(Section~\ref{sec:end_to_end_results}). We then report ablations that isolate the contribution of
the decomposition/encoding layer via logical-width scalability (Section~\ref{sec:scaling_results}),
the learned multiplier controller under classical subproblem solves (Section~\ref{sec:multiplier_results}),
and the hardware-aware execution layer under heterogeneous backends (Section~\ref{sec:bandit_results}).


\subsection{End-to-end routing performance}
\label{sec:end_to_end_results}

\begin{figure*}[ht]
  \centering
  \includegraphics[width=\textwidth]{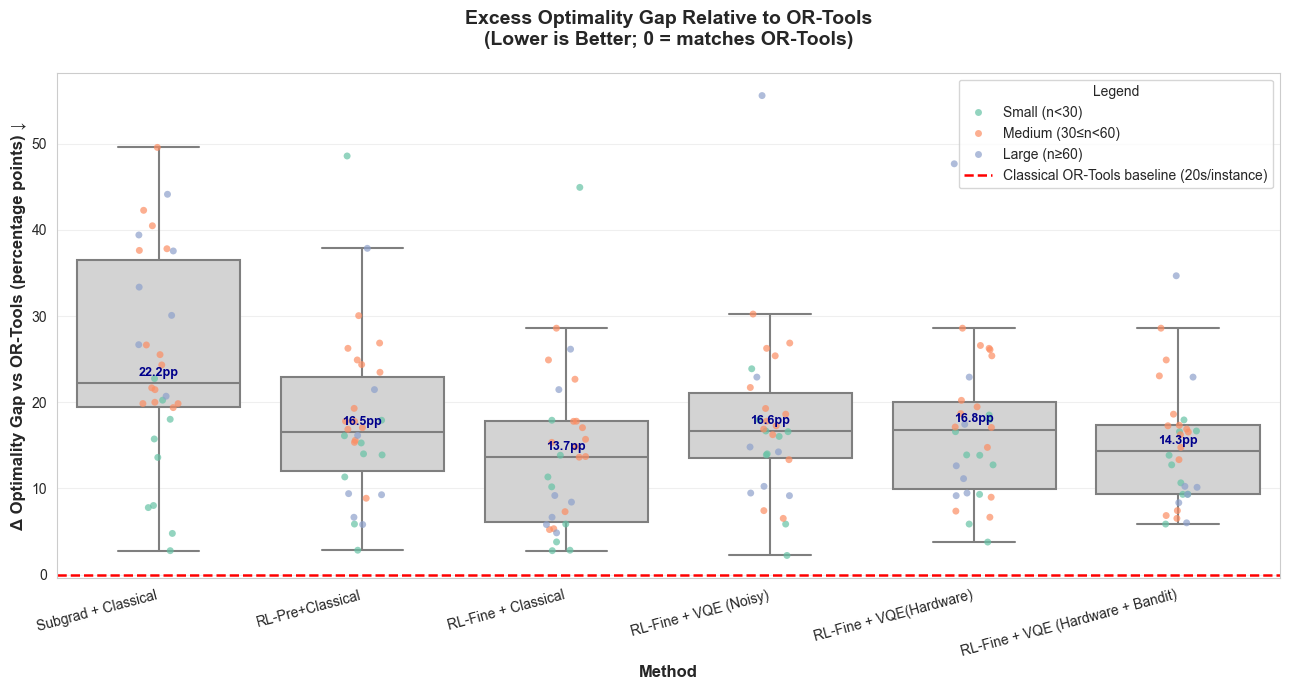}
  \caption{Excess optimality gap relative to OR-Tools (20s/instance).
For each instance $i$ and method $m$, $\Delta\mathrm{gap}_m(i)=\mathrm{gap}_m(i)-\mathrm{gap}_{\textsc{ORTools}}(i)$
(percentage points), where $\mathrm{gap}$ is computed end-to-end after decoding/repair and route
reconstruction. Lower is better; the red dashed line at $0$ denotes parity with OR-Tools. Gray boxes
show distributions across test instances and colored points denote size buckets.}
\label{fig:delta_gap_vs_ortools}
  \label{fig:all_gap_comparison}
\end{figure*}

Figure~\ref{fig:delta_gap_vs_ortools} reports \emph{excess} optimality gap relative to a classical
OR-Tools baseline (20s per instance). For each instance $i$ and method $m$, we plot
$\Delta\mathrm{gap}_m(i) = \mathrm{gap}_m(i) - \mathrm{gap}_{\textsc{ORTools}}(i)$ in percentage points,
where $\mathrm{gap}(\cdot)$ is the end-to-end CVRP gap to the best-known solution after decoding/repair
and route reconstruction. The dashed horizontal line at $0$ indicates parity with OR-Tools; positive
values indicate worse performance than the classical baseline. Across all methods, the distributions
remain above zero, reflecting the strength of OR-Tools under the fixed budget. Within our hybrid
pipeline variants, hardware-aware configuration improves performance: the bandit-enabled hardware
setting reduces the median excess gap compared with hardware execution without bandit, consistent with
the intended mechanism that selecting circuit families, depths, and qubit subsets to reduce routing
overhead and avoid gate-budget violations increases the utility of each expensive quantum evaluation.
Point colors indicate instance-size buckets, showing that larger instances tend to exhibit larger
excess gaps under all methods.

\subsection{Logical-width scalability and encoding ablations}
\label{sec:scaling_results}

A central motivation for our decomposition pipeline is \emph{logical-width} control: direct QUBO
encodings of CVRP quickly require thousands of binary variables (and hence logical qubits),
whereas Fisher--Jaikumar (FJ) linearization followed by Lagrangian relaxation (LR) yields a stream of
\emph{quantum-sized} subproblems, one knapsack-type QUBO per vehicle, whose width is governed by the
candidate-set size rather than the full customer set. We measure logical width as the number of
binary decision variables required by the QUBO before any hardware-specific embedding or routing, so
that the comparison isolates formulation-level scaling.

\begin{figure}[H]
   \centering
   \includegraphics[width=\linewidth]{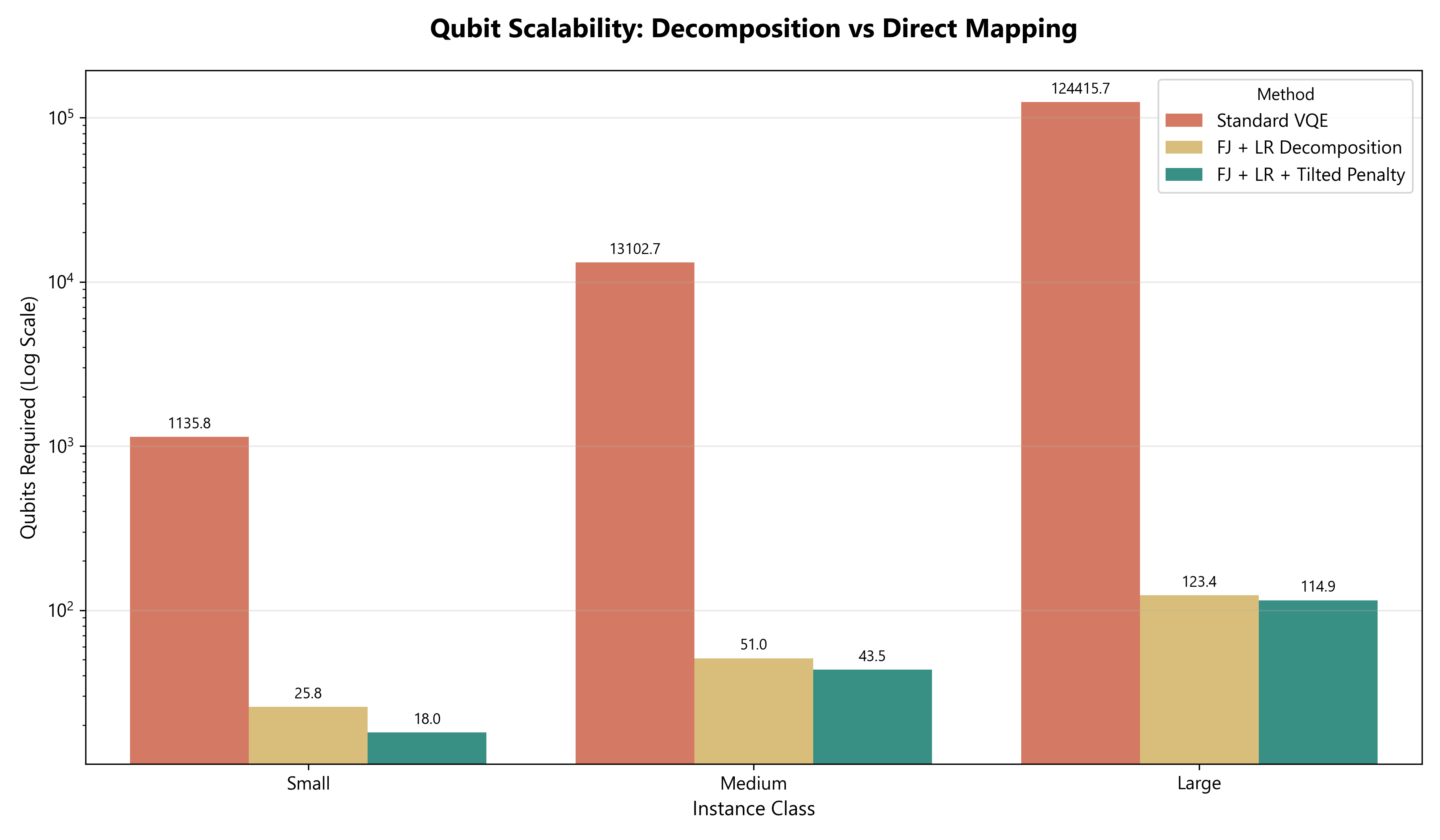}
   \caption{Logical-width (qubit) scaling of direct CVRP QUBO encoding versus Fisher--Jaikumar 
   (FJ) + Lagrangian relaxation (LR) decomposition into per-vehicle knapsack QUBOs. Bars report 
   required logical qubits (log scale) for three instance classes. Decomposition reduces logical 
   width by orders of magnitude by replacing a monolithic encoding with bounded-width subproblems; 
   the slack-free tilted penalty further reduces width by avoiding auxiliary slack bits.} 
   \label{fig:qubit_scalability_bar} 
\end{figure}

Figure~\ref{fig:qubit_scalability_bar} summarizes this effect across three CVRP instance classes.
Direct CVRP encodings scale superlinearly with problem size due to the coupling between assignment
and routing decisions in a monolithic formulation, while FJ+LR replaces the global encoding with
bounded-width per-vehicle subproblems, producing orders-of-magnitude reductions in logical qubits.
In addition to decomposition, we ablate the \emph{inequality-penalty encoding}: a conventional
slack-based encoding introduces auxiliary bits to represent one-sided capacity constraints, whereas
our \emph{slack-free tilted} penalty avoids these slack variables and further reduces logical width
at fixed instance size, as reflected in the bar comparisons.

Appendix~\ref{app:penalties} provides complementary scaling evidence at the instance-level.
Figure~\ref{fig:qubit_scalability_scatter} plots logical width versus $N$ over individual instances,
confirming that FJ+LR yields slowly growing (effectively bounded-width) knapsack QUBOs, while direct
CVRP QUBO encodings grow rapidly with $N$. 

\paragraph*{Scalability beyond 150 customers.}
We evaluate up to 150 customers because the study targets near-term QPU constraints (routing overhead, queueing latency, and execution failures). 
For larger instances, the primary bottleneck shifts from logical width to orchestration cost, candidate-restriction
quality, and backend availability. Since the number of subproblem evaluations grows with the problem
size, scaling to several hundred or thousand customers will depend less on QUBO width itself than on
sample efficiency, scheduling across heterogeneous resources, and stronger candidate-selection and
reconstruction procedures. Evaluating these effects on larger CVRPLIB benchmark families is beyond our scope.

\subsection{Learned multiplier controller results}
\label{sec:multiplier_results}

We first isolate the effect of multiplier control by evaluating three update rules using 
\emph{classical}
knapsack subproblem solves: (i) subgradient ascent with a tuned stepsize schedule, 
(ii) an expert-pretrained
policy (behavior cloning), and (iii) the same policy after PPO fine-tuning under the curriculum reward
(Section~\ref{sec:training_protocol}). This experiment removes quantum execution noise and compilation
variability so that differences primarily reflect outer-loop control behavior rather than backend 
effects.
Consistent with recommended practice for reporting heuristic computational experiments, we summarize 
results
across instances using both means and medians, and we provide a full per-instance breakdown in the 
appendix.

\begin{figure*}
    \centering
    \includegraphics[width=\textwidth]{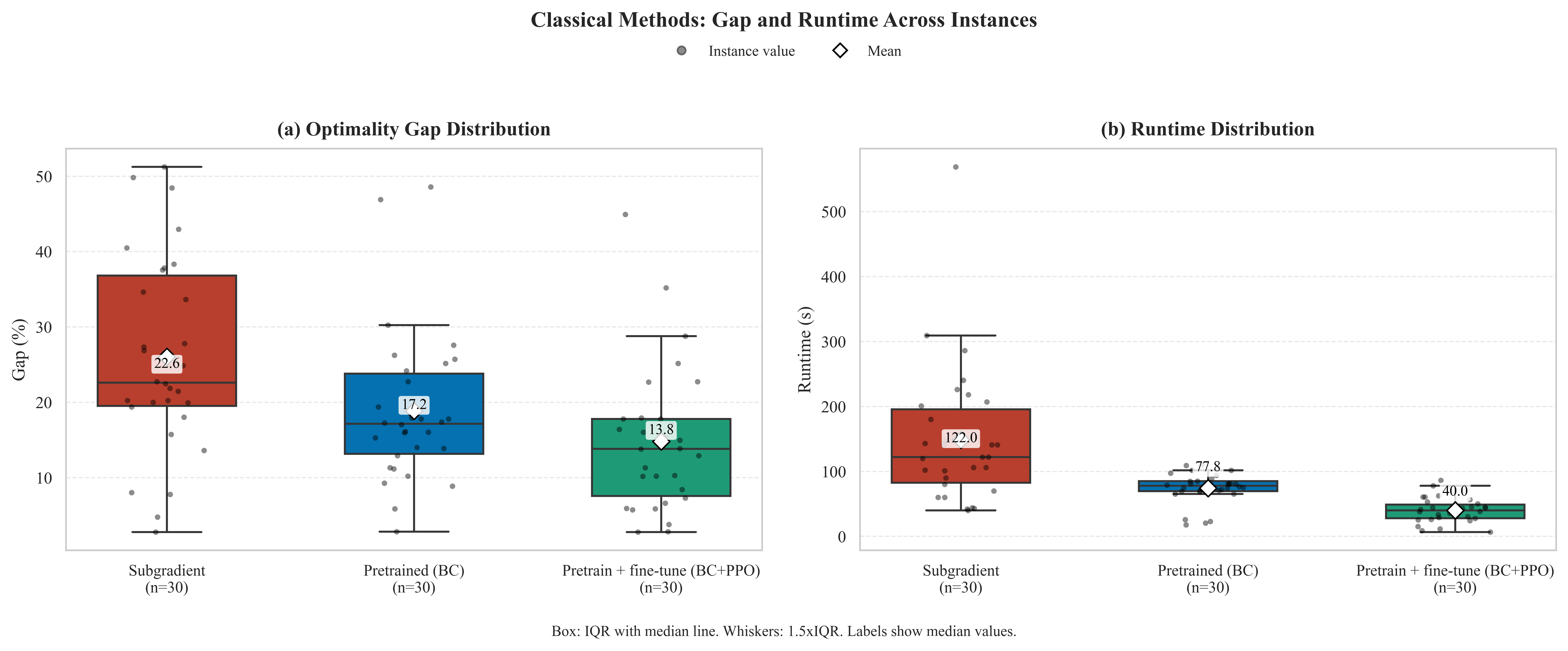}
    \caption{Multiplier-control ablation under \emph{classical} knapsack subproblem solves (30 CVRPLIB instances).
(a) Distribution of end-to-end CVRP optimality gaps (\%) after decoding/repair and route reconstruction (lower is better).
(b) Distribution of wall-clock runtime (seconds) for a full LR episode, including LR iterations, subproblem solves, repair,
and routing. Boxes show median and interquartile range (IQR) with 1.5$\times$IQR whiskers; points are individual instances.
Learned multiplier control reduces both median gap and time-to-solution relative to subgradient updates.}
    \label{fig:multiplier_classical_gap_runtime}
\end{figure*}

\paragraph{Metrics.}
For each instance, we report (i) the end-to-end CVRP optimality gap (\%) after decoding/repair and route
reconstruction, and (ii) wall-clock runtime (seconds) for a full LR episode. 
Runtime is measured as elapsed
time around the complete solver call, so it includes LR iterations,
subproblem solves, repair, and routing. See Figure~\ref{fig:multiplier_classical_gap_runtime} for 
the distribution of these metrics across the 30 CVRPLIB instances.

\paragraph{Summary across 30 CVRPLIB instances.}
Table~\ref{tab:multiplier_summary} shows that learning improves both routing quality and time-to-solution.
Relative to subgradient updates, the pretrained controller reduces mean gap from 26.03\% to 
18.86\% while
roughly halving mean runtime; PPO fine-tuning further reduces mean gap to 14.82\% and mean runtime 
to 40.07s.
The full per-instance results are reported in Appendix~\ref{app:extended_tables}.

\begin{table*}[t]
\centering
\caption{Multiplier-controller comparison under classical subproblem solves (30 CVRPLIB instances). 
We report end-to-end CVRP optimality gap (\%) after decoding/repair and route reconstruction, 
and wall-clock runtime per instance (seconds) for the full LR episode measured.}
\label{tab:multiplier_summary}
\begin{tabular}{lS[table-format=2.2]S[table-format=2.2]S[table-format=3.2]S[table-format=3.2]}
\hline
\textbf{Method} & {\textbf{Mean Gap (\%)}} & {\textbf{Median Gap (\%)}} & {\textbf{Mean Time (s)}} & {\textbf{Median Time (s)}} \\
\hline
Subgradient & 26.03 & 22.61 & 148.61 & 122.04 \\
Pretrained & 18.86 & 17.15 & 74.18 & 77.80 \\
Pretrain + fine-tune & 14.82 & 13.82 & 40.07 & 40.05 \\
\hline
\end{tabular}
\end{table*}

\paragraph{Interpretation and limitations.}
These results indicate that learned multiplier updates improve the quality of recovered primal
solutions and reduce time-to-solution under a fixed evaluation protocol, even before introducing
sampling noise. The same evaluation pipeline is then carried forward to quantum runs and 
hardware-aware
execution in Sections~\ref{sec:bandit_results}--\ref{app:full_instance_tables}. For hardware 
experiments, the
multi-QPU evaluator additionally logs total runtime and quantum-runtime split, to separate 
orchestration overhead from QPU execution time.

\subsection{Hardware-aware execution results}
\label{sec:bandit_results}

We now evaluate the effectiveness of the hardware-aware execution layer (presented in Section~\ref{sec:hardware_aware}). The results
focus on four aspects: offline bandit behavior, the frequency with which gate-budget screening becomes binding, execution cost and robustness across heterogeneous backends, and the resulting end-to-end
routing impact under noisy simulation and real-hardware execution.


\subsubsection{Offline bandit behavior}
\label{sec:bandit_offline_behavior}

\begin{figure}[H]
  \centering
  \includegraphics[width=\linewidth]{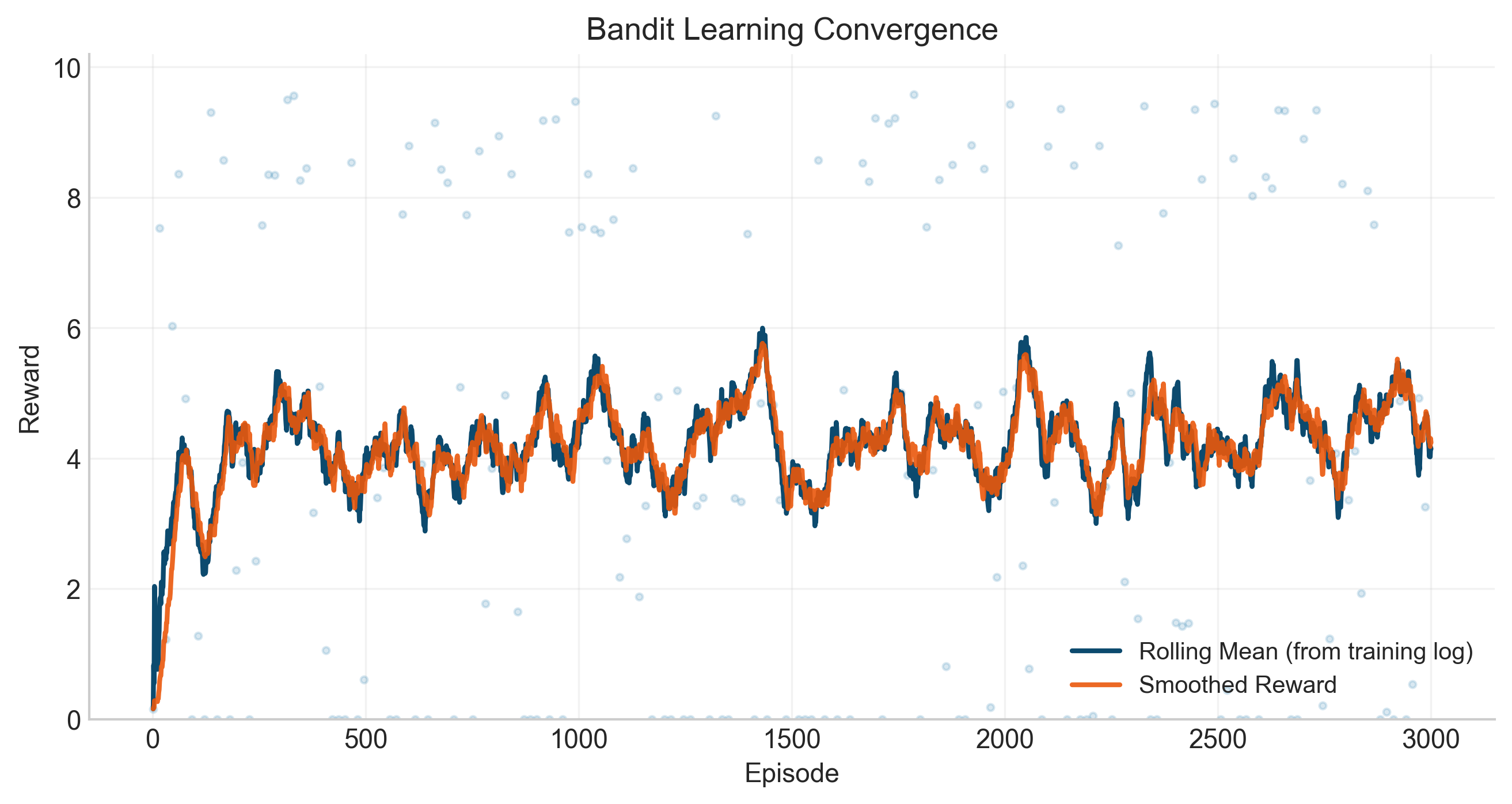}
  \caption{Offline bandit learning curve under trace-driven training. Points show per-episode bounded reward
(in $[0,10]$); the rolling-mean and smoothed traces summarize the learning trend. 
}\label{fig:bandit_learning}
\end{figure}

\begin{figure}[H]
  \centering
  \includegraphics[width=\linewidth]{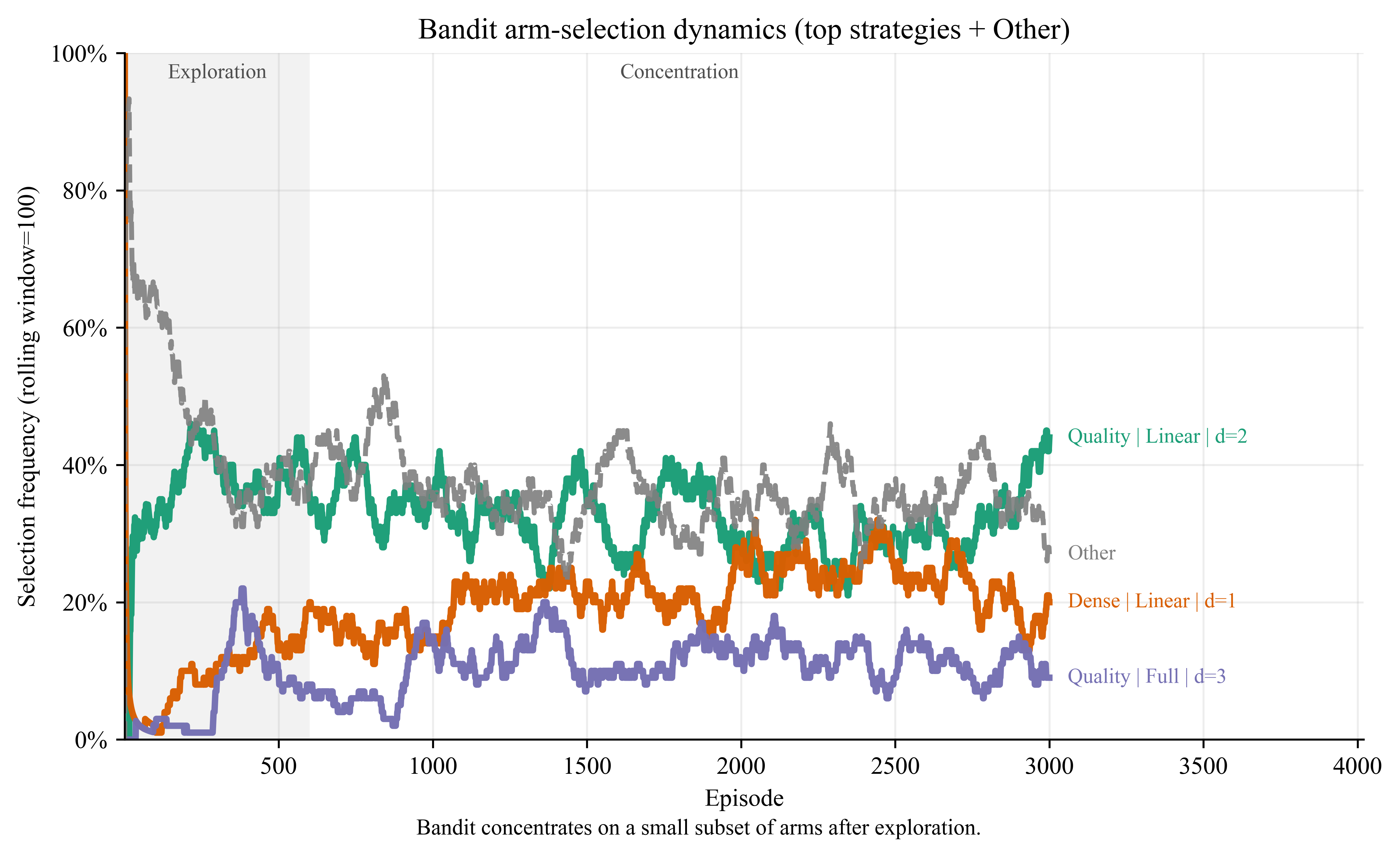}
  \caption{Bandit arm-selection dynamics during trace-driven offline training. Curves report rolling-window
(window=100) selection frequencies for the most frequently selected configuration arms; remaining arms are
aggregated as \emph{Other}. The policy transitions from broad exploration to concentration on a small subset
of high-utility configurations, consistent with UCB-style learning under bounded rewards.}
  \label{fig:bandit_arm_dist}
\end{figure}

Figure~\ref{fig:bandit_learning} summarizes the trace-driven offline training (Section~\ref{sec:bandit_training}). The bounded execution-proxy reward rises rapidly from a cold
start and then plateaus, indicating that LinUCB identifies a stable set of high-utility configuration
arms from the joint hardware/problem context (Section~\ref{sec:bandit_formulation}) rather than
continuing to explore broadly. Figure~\ref{fig:bandit_arm_dist} corroborates this behavior: after an
initial exploration phase, arm-selection frequencies concentrate on a small subset of configurations,
consistent with UCB-style learning under bounded rewards.

\subsubsection{Constraint pressure: gate-budget risk versus logical width}
\label{sec:gate_budget_risk}

\begin{figure}[H]
  \centering
  \includegraphics[width=\linewidth]{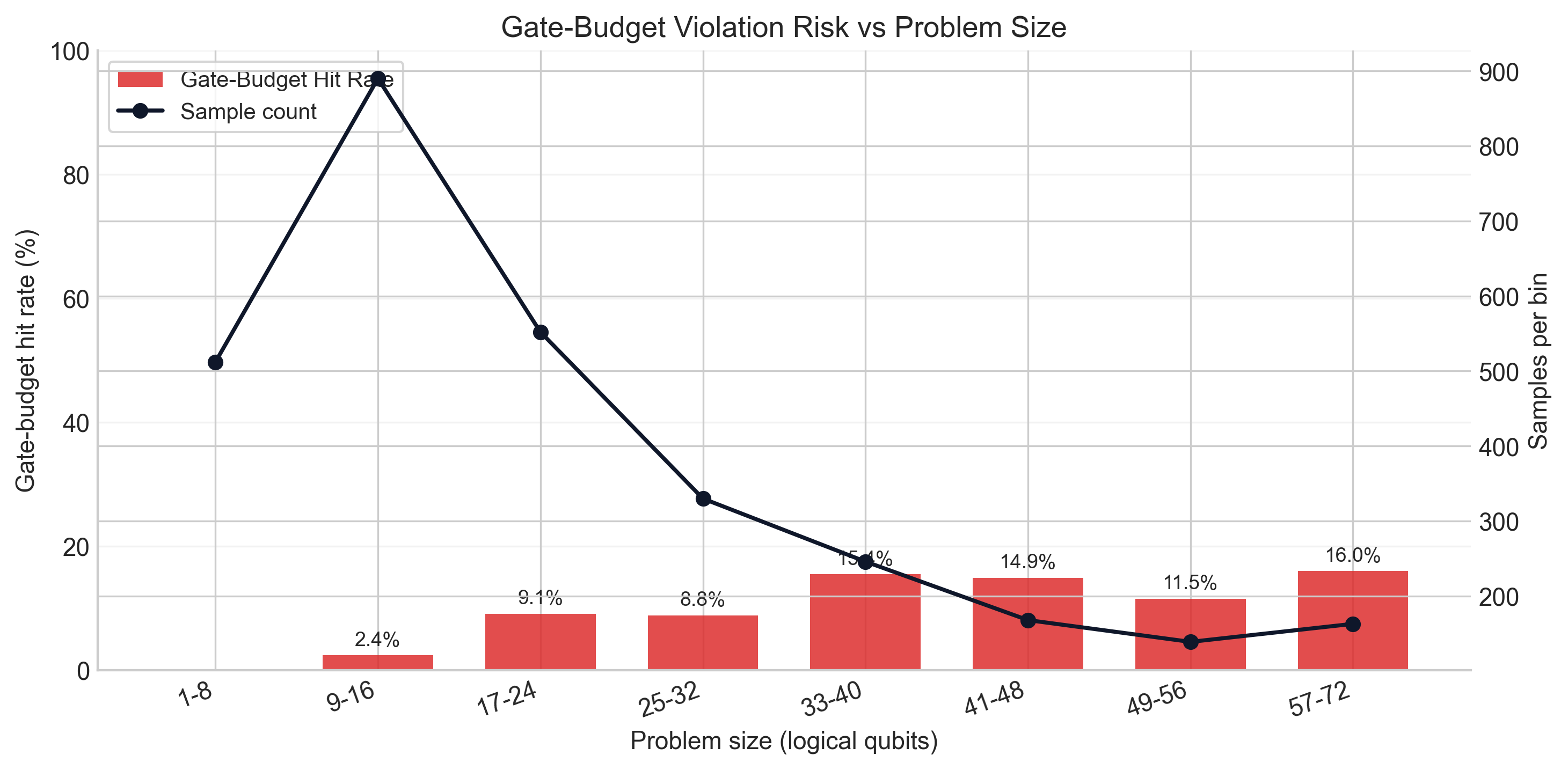}
  \caption{Gate-budget hit rate by logical-width bin.}
  \label{fig:gate_budget_risk}
  \vspace{-1ex}
  {\footnotesize \textit{Note.} Bars show the fraction of consult calls in which the top-ranked arm is rejected because $\widehat{G}>G_{\max}$; the line shows the number of logged consult calls per bin (right axis).}
\end{figure}

Figure~\ref{fig:gate_budget_risk} reports the empirical \emph{gate-budget hit rate} as a function of
logical width, defined as the fraction of consult calls in which the bandit’s top-ranked arm is
rejected by feasibility screening (because $\widehat{G} > G_{\max}$) and the selector falls back to
the best-ranked feasible alternative. The hit rate is negligible at small widths and rises to the
$\approx 9$--$16\%$ range beyond moderate widths, consistent with routing overhead becoming a binding
constraint as compiled circuits grow. This supports the constrained decision rule in
Section~\ref{sec:bandit_actions_screening}: the bandit provides a utility ranking, while deterministic
screening enforces executability. Importantly, nonzero hit rates also translate into outer-loop cost
pressure, since rejected top arms induce re-selection and typically correspond to higher expected
routing overhead and latency. The secondary axis in Figure~\ref{fig:gate_budget_risk} reports the
number of logged consult calls per width bin.

\subsubsection{Outer-loop sample efficiency and robustness}
\label{sec:outer_loop_efficiency}

Because quantum evaluations are costly and subject to queueing and execution failures, we additionally
evaluate the hardware-aware layer using outer-loop efficiency and robustness metrics: (i) quantum
evaluation budget (total transpiled circuit executions / quantum jobs), (ii) wall-clock cost, and
(iii) robustness measured via job-level latency and submission failures. Table~\ref{tab:bucket_summary}
reports bucketed summaries by instance size and execution variant, including total wall-clock time,
total circuit count, average job latency, and resulting end-to-end gap to the BKS.

\begin{table*}[t]
\centering
\caption{Outer-loop sample-efficiency and robustness summaries, bucketed by instance size and execution
variant. ``Total time'' is end-to-end wall-clock time for the full LR episode; ``Total circuits'' is the
number of transpiled circuit executions submitted; ``Avg job latency'' is the per-job turnaround time
(including queueing). Entries are reported as median [min, max] over instances within each bucket.
``Gap to BKS'' is computed after decoding/repair and route reconstruction.}
\label{tab:bucket_summary}
\begin{tabular}{lccccc}
\hline
\textbf{Source} & \textbf{Size bucket} & \textbf{Total time (sec)} & \textbf{Total circuits} &
\textbf{Avg job latency (sec)} & \textbf{Gap to BKS (\%)} \\
\hline
Multi-QPU (noise-aware) & 10--20   & 386.3 [13.9, 8363.0]    &  20.0 [20.0, 30.0]   & 24.1 [0.1, 84.4]  & 13.89 [13.29, 18.21] \\
Multi-QPU (noise-aware) & 21--50   &  4829.5 [3107.6, 8280.9]     & 100.0 [80.0, 100.0]  &  161.0 [119.2, 241.1]   & 18.68 [14.76, 26.26] \\
Multi-QPU (noise-aware) & 51--100  &  9767.3 [6187.5, 36009.9] & 140.0 [98.0, 140.0]  &  65.1 [54.3, 312.7]  & 17.92 [11.90, 24.32] \\
Multi-QPU (baseline)    & 10--20   &  8856.9 [4456.3, 9406.9]  &  40.0 [40.0, 60.0]   & 138.2 [111.4, 147.6] & 13.85 [12.74, 13.89] \\
Multi-QPU (baseline)    & 21--50   &  7625.4 [2775.9, 9282.7]  &  98.0 [80.0, 105.0]  &  55.8 [33.3, 110.1]  & 19.07 [13.31, 26.80] \\
Multi-QPU (baseline)    & 51--100  &  4401.9 [2080.5, 8650.1]  & 140.0 [95.0, 170.0]  &  29.8 [17.8, 50.2]   & 17.01 [12.77, 23.15] \\
Multi-QPU (baseline)    & 101+     &  2617.3 [2617.3, 2617.3]  &  70.0 [70.0, 70.0]   &  37.4 [37.4, 37.4]   & 56.71 [56.71, 56.71] \\
Single-QPU              & 10--20   &  6931.6 [3879.2, 12581.4] &  40.0 [35.0, 70.0]   & 124.8 [110.4, 162.4] & 13.32 [12.39, 21.95] \\
Single-QPU              & 21--50   &  8377.6 [2103.2, 10702.0] &  77.5 [18.8, 80.0]   & 107.7 [26.8, 133.2]  & 22.16 [11.41, 27.71] \\
\hline
\end{tabular}
\end{table*}

\begin{figure}[h]
  \centering
  \includegraphics[width=\linewidth]{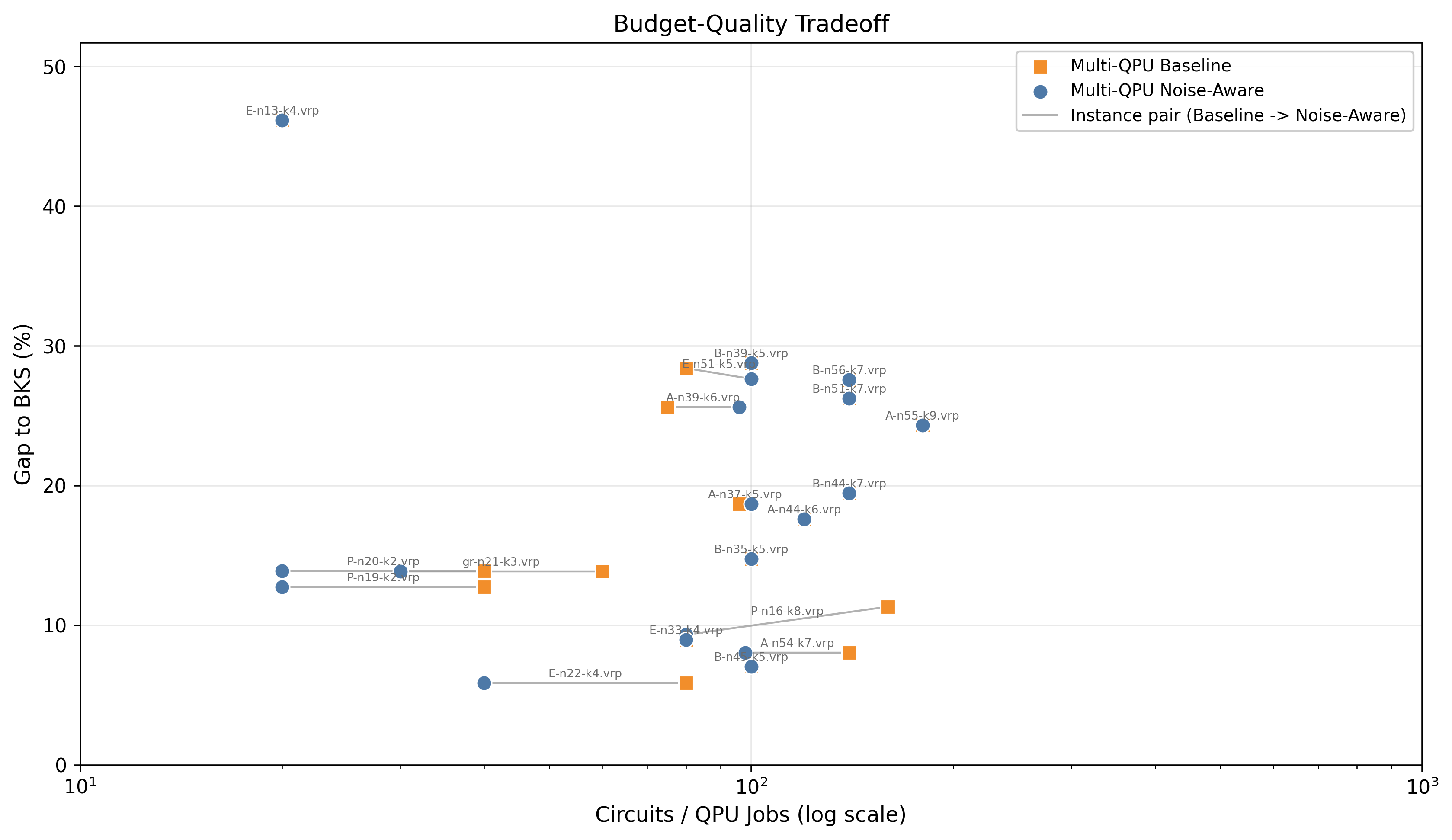}
  \caption{Budget--quality tradeoff (gap to BKS versus quantum-evaluation budget) for multi-QPU baseline and noise-aware runs.}
  \label{fig:tradeoff_circuits}
  \vspace{-1ex}
  {\footnotesize \textit{Note.} Quantum budget is the total number of transpiled circuit executions / QPU jobs (log scale); gray segments connect matched instances (baseline $\rightarrow$ noise-aware).}
\end{figure}

Figure~\ref{fig:tradeoff_circuits} reports the budget--quality tradeoff by plotting the end-to-end
gap to the best-known solution (BKS) against the quantum-evaluation budget, measured as the total
number of transpiled circuit executions / QPU jobs (log scale). Gray segments connect matched
instance runs (baseline $\rightarrow$ noise-aware) to highlight per-instance changes under a fixed
benchmark instance.

\subsubsection{End-to-end impact under noisy simulation and hardware}
\label{sec:bandit_end_to_end}

\begin{figure}[h]
  \centering
  \includegraphics[width=\linewidth]{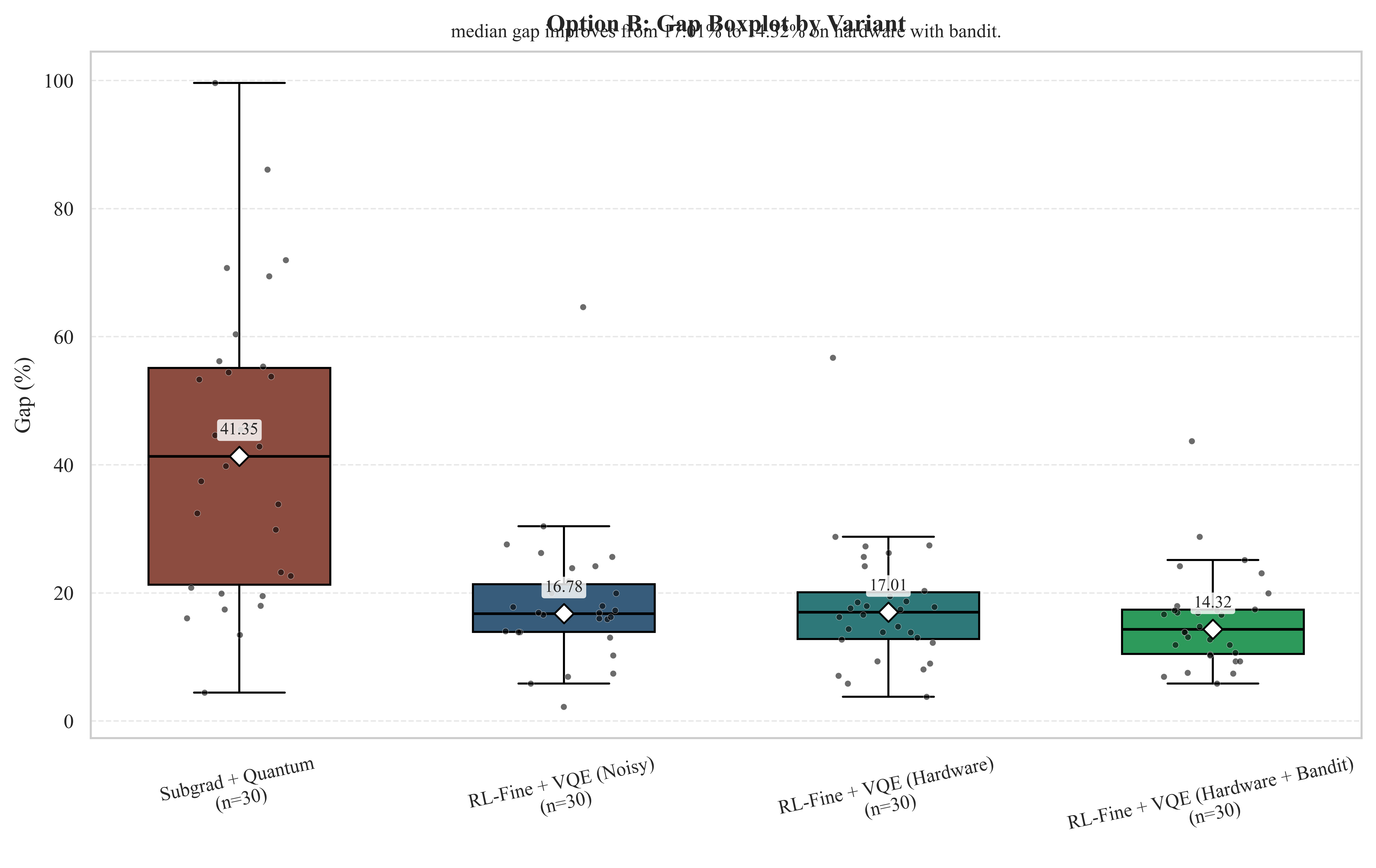}
  \caption{End-to-end CVRP optimality gap (\%) after decoding/repair and route reconstruction, comparing execution variants under learned multiplier control.}
  \label{fig:gap_comparison}
\end{figure}

Finally, we evaluate the end-to-end routing impact of hardware-aware configuration.
Figure~\ref{fig:gap_comparison} compares CVRP optimality gaps (to BKS) under three execution settings
for the same learned multiplier controller: (i) noisy simulation without bandit-based configuration,
(ii) real-hardware execution without bandit-based configuration, and (iii) real-hardware execution
with bandit-based configuration. The bandit-enabled setting yields a lower median gap than hardware
execution without the bandit, indicating that hardware-aware configuration improves the utility of
quantum evaluations within the hybrid loop.

\section{Discussion}
\label{sec:discussion}

The experimental results support three main conclusions. First, the decomposition achieves a substantial reduction in
logical width relative to direct CVRP encodings, making repeated subproblem evaluation feasible within
near-term hardware limits. Second, learned multiplier control materially improves both routing quality and
runtime relative to classical subgradient updates under fixed budgets. Third, hardware-aware execution
provides additional gains in some regimes, but those gains depend on the balance between improved execution
quality and increased orchestration cost.

A first observation comes from the multiplier-control ablation under classical subproblem solves. Relative to
subgradient updates, the full learned controller (pretraining followed by fine-tuning) reduces the mean
end-to-end CVRP gap from 26.03\% to 14.82\% and the median gap from 22.61\% to 13.82\%. It also reduces mean
runtime from 148.61s to 40.07s and median runtime from 122.04s to 40.05s. Since these improvements are observed
before introducing quantum execution noise, they provide evidence that the learned policy improves the outer-loop
update process itself. In particular, the policy appears to produce more effective multiplier adjustments than
standard subgradient rules for this decomposition.

The comparison between the pretrained and fine-tuned variants is also informative. Pretraining alone improves
over subgradient updates, reducing the mean gap to 18.86\% and the median gap to 17.15\%, while reducing mean
runtime to 74.18s. Fine-tuning yields a further improvement in both gap and runtime. This pattern supports the
two-stage training design: the expert-guided phase provides a strong initialization, and the reinforcement-learning
phase improves performance with respect to the end-to-end routing objective.

A second observation concerns the hardware-aware execution layer. In the end-to-end quantum comparison, the
best-performing quantum variant is the bandit-enabled hardware setting. Across the 30-instance test set, the
mean gap decreases from 17.83\% for RL + HW to 15.72\% for RL + HW + Bandit, and the median gap decreases from
17.02\% to 14.32\%. The noisy-simulator variant performs worse than the bandit-enabled hardware setting as well,
with mean and median gaps of 18.54\% and 16.78\%, respectively. These improvements are smaller than those observed
for learned multiplier control, but they are consistent with the intended role of the bandit: improving the quality
of quantum evaluations by selecting more suitable execution configurations.

At the same time, the hardware results show that hardware-aware execution is not uniformly beneficial across all
instance sizes. In the 21--50 bucket, the noise-aware multi-QPU setting improves the median gap slightly, from
19.07\% to 18.68\%, relative to the baseline multi-QPU setting, but at the cost of substantially higher total
time and average job latency. In the 51--100 bucket, the noise-aware setting is worse in median gap
(17.92\% versus 17.01\%) and also incurs higher total time. These results show that the hardware-aware layer
introduces a quality--cost tradeoff rather than a guaranteed improvement. Its benefit depends on whether the gain
in execution quality is large enough to offset additional orchestration and latency costs.

This interpretation is supported by the feasibility-screening statistics. The gate-budget hit rate is negligible
at small logical widths, but rises to approximately 9--16\% at larger widths. As the subproblems grow, the
highest-ranked arm proposed by the bandit is therefore more likely to violate feasibility constraints and be
replaced by an alternative configuration. This indicates that feasibility screening becomes increasingly important
as circuit width and execution cost increase, and that the constrained action-selection mechanism is operationally
relevant rather than merely a design safeguard.

The comparison against OR-Tools should also be interpreted carefully. The results do not show that the hybrid
quantum pipeline surpasses a strong classical routing method in absolute route quality under modest computational
budgets. The aggregate gaps remain above the OR-Tools baseline, which is expected given the additional overhead
associated with decomposition, repeated subproblem solution, decoding, repair, and noisy execution. The more
relevant conclusion is that, within this hybrid setting, learned multiplier control and hardware-aware execution
produce measurable improvements over simpler baselines. This is the appropriate level of empirical claim for the
present study.

A broader implication of the results is that the main scalability challenge has shifted. The decomposition
addresses the immediate issue of qubit width, but it does not remove the optimization difficulty; instead,
it changes its form. Once the subproblems are small enough to execute, performance depends increasingly on
outer-loop sample efficiency, the informativeness of the resulting quantum evaluations, and the overhead
required to obtain them. This suggests that, beyond logical-width reduction, further progress will depend on
improving orchestration, reconstruction quality, and evaluation efficiency.

\section{Conclusion}
\label{sec:conclusion}

Our hybrid quantum framework for CVRP integrates (i) an OR-based Lagrangian decomposition into bounded-width knapsack subproblems, (ii) a learned multiplier-update controller (expert pretraining + RL fine-tuning), and (iii) a hardware-aware execution policy based on constrained contextual bandits. Across the CVRPLIB benchmarks studied, the combined system improves end-to-end routing performance over classical subgradient control and static execution under matched evaluation budgets, while maintaining substantially better logical-width scaling than direct CVRP QUBO encodings.

The main empirical takeaway is systems-level: hybrid quantum routing performance is governed by the interaction between decomposition, dual control, and execution policy. Gains do not arise reliably from any single component in isolation; decomposition must be paired with stable multiplier updates and with execution choices that preserve the utility of each expensive quantum call.

Several limitations remain. (1) Primal recovery can dominate final quality: better subproblem samples do not always translate into better routes, and reconstruction choices can account for a large fraction of performance on some families. (2) The hardware study covers a limited set of devices, calibrations, and budget regimes; broader coverage is needed to draw stronger robustness conclusions. (3) The bandit action space is intentionally constrained for feasibility and sample efficiency, leaving richer controls—e.g., adaptive shot allocation, continuous depth selection, and coordinated multi-job scheduling—unexplored. (4) All comparisons are budget-dependent and should be interpreted under the fixed experimental protocols rather than as hardware-independent rankings.

These observations suggest four directions. First, tighter coupling of multiplier control and execution control to trade off dual progress against quantum cost. Second, stronger cross-family generalization via domain randomization over instance and hardware descriptors (or meta-learning). Third, improved primal recovery, including learning-augmented or exact postprocessing that better exploits knapsack structure while preserving feasibility. Fourth, uncertainty-aware control (e.g., CVaR-style objectives or robust bandits) that accounts for latency variation and job-failure risk in addition to expected solution quality.

More broadly, our results support a pragmatic path for near-term quantum routing: progress is more likely to come from decomposition-based, control-aware pipelines than from monolithic encodings. The near-term objective is not to solve large CVRP instances directly on quantum hardware, but to extract reliable optimization progress from bounded-width subproblems under noisy, heterogeneous execution conditions—a perspective that should remain relevant as hardware evolves.
\clearpage
\bibliographystyle{apsrev4-2}
\bibliography{reference} 

\appendix
\onecolumngrid

\section{Penalty encodings for knapsack capacity constraints}
\label{app:penalties}

\subsection{Quantum knapsack encoding in LR-decomposed CVRP: Taylor vs.\ Tilted penalties}
\label{app:penalties:taylor_vs_tilt}

\paragraph{Context (LR-decomposed per-vehicle knapsack).}
Under the Lagrangian relaxation of the Fisher--Jaikumar assignment surrogate, 
the relaxed problem decomposes into one subproblem per vehicle $k$ 
(Section~\ref{sec:lagrangian_relaxation}). Each vehicle solves a 0--1 
knapsack-structured selection problem of the form
\begin{align}
\min_{y\in\{0,1\}^{n}} \quad & \sum_{i=1}^{n} \tilde a_i\, y_i \label{eq:app_knap_obj}\\
\text{s.t.}\quad & \sum_{i=1}^{n} w_i\, y_i \le C, \label{eq:app_knap_cap}
\end{align}
where $y_i=1$ indicates selecting customer $i$ for the vehicle, $w_i$ is the demand/weight, 
$C$ is the vehicle capacity, and $\tilde a_i$ denotes the LR-adjusted cost (reduced cost) for 
that vehicle (we suppress the vehicle index for readability).

\paragraph{Quadratic unconstrained objective.}
Variational quantum routines operate on unconstrained Ising/QUBO objectives 
\cite{lucas2014ising,farhi2014quantum,peruzzo2014variational}. Hence we convert 
\eqref{eq:app_knap_obj}--\eqref{eq:app_knap_cap} into a QUBO by adding a quadratic penalty 
for violating the capacity inequality. 

Define

\begin{equation}
W(y) := \sum_{i=1}^{n} w_i y_i,\qquad
r(y) := W(y) - C,\qquad
t(y) := C - W(y) = -r(y).
\end{equation}
We seek an unconstrained energy of the form
\begin{equation}
E(y) \;=\; \sum_{i=1}^{n} \tilde a_i\, y_i \;+\; \Pi(y),
\qquad y\in\{0,1\}^n,
\label{eq:app_energy}
\end{equation}

where $\Pi(y)$ is a quadratic penalization term. Below we document two encodings used in our 
implementation and experiments: a Taylor-style quadratic penalty baseline and a tilted 
quadratic penalty (default) that introduces explicit \emph{headroom} below capacity.

\subsubsection{Taylor penalty encoding}
\label{app:penalties:taylor}

\paragraph{Definition.}
The Taylor encoding constructs a quadratic penalty in the slack $t(y)=C-W(y)$ using a 
second-order expansion:
\begin{equation}
\Pi_{\mathrm{Taylor}}(y)
\;:=\;
\alpha\left(1 - t(y) + \tfrac{1}{2} t(y)^2\right),
\label{eq:app_taylor_penalty_def}
\end{equation}
where $\alpha>0$ controls penalty scale. The resulting energy is
\begin{equation}
E_{\mathrm{Taylor}}(y)
=
\sum_{i=1}^{n} \tilde a_i\, y_i
+
\alpha\left(1 - (C-W(y)) + \tfrac{1}{2}(C-W(y))^2\right).
\label{eq:app_taylor_energy}
\end{equation}

\paragraph{QUBO expansion.}
Using $y_i^2=y_i$ and
\begin{equation}
W(y)^2
=
\left(\sum_i w_i y_i\right)^2
=
\sum_i w_i^2 y_i
+
2\sum_{i<j} w_i w_j y_i y_j,
\end{equation}
we can write \eqref{eq:app_taylor_energy} (up to an additive constant) as
\begin{equation}
E_{\mathrm{Taylor}}(y)
=
\sum_{i=1}^{n} a_i^{\mathrm{Taylor}}\, y_i
+
\sum_{i<j} b_{ij}^{\mathrm{Taylor}}\, y_i y_j
+\mathrm{const},
\label{eq:app_taylor_qubo}
\end{equation}
with
\begin{align}
b_{ij}^{\mathrm{Taylor}} &= \alpha\, w_i w_j,\\
a_i^{\mathrm{Taylor}} &= \tilde a_i + \alpha(1-C) w_i + \tfrac{1}{2}\alpha w_i^2.
\end{align}

\paragraph{Interpretation.}
Ignoring constants, $\alpha(-t+\tfrac12 t^2)=\tfrac{\alpha}{2}(t-1)^2-\tfrac{\alpha}{2}$, 
so this penalty is minimized at $t^\star=1$, i.e., $W^\star=C-1$. Thus, Taylor encoding 
implicitly prefers near-capacity loading. This can be beneficial in classical settings but 
may be brittle under approximate sampling: near-feasible samples can flip to slight 
infeasibility, reducing the feasible yield after decoding/repair.

\subsubsection{Tilted quadratic penalty encoding}
\label{app:penalties:tilt}

\paragraph{Definition.}
Our default encoding uses a quadratic penalty in the residual $r(y)=W(y)-C$ with an added 
linear ``tilt'' term:
\begin{equation}
\Pi_{\mathrm{Tilt}}(y)
\;:=\;
\rho\left(r(y)^2 + s\, r(y)\right),
\label{eq:app_tilt_penalty_def}
\end{equation}
where $\rho>0$ sets penalty strength and $s\ge 0$ controls the degree of headroom encouraged 
below capacity. The resulting energy is
\begin{equation}
E_{\mathrm{Tilt}}(y)
=
\sum_{i=1}^{n} \tilde a_i\, y_i
+
\rho\left((W(y)-C)^2 + s(W(y)-C)\right).
\label{eq:app_tilt_energy}
\end{equation}
This is closely related in spirit to ``unbalanced penalization'' approaches that avoid slack 
variables and bias the energy landscape to improve feasibility behavior under approximate 
quantum optimization \cite{montanez2024unbalanced}.

\paragraph{QUBO expansion.}
Expanding \eqref{eq:app_tilt_energy} yields (up to constants)
\begin{equation}
E_{\mathrm{Tilt}}(y)
=
\sum_{i=1}^{n} a_i^{\mathrm{Tilt}}\, y_i
+
\sum_{i<j} b_{ij}^{\mathrm{Tilt}}\, y_i y_j
+\mathrm{const},
\label{eq:app_tilt_qubo}
\end{equation}
with
\begin{align}
b_{ij}^{\mathrm{Tilt}} &= 2\rho\, w_i w_j,\\
a_i^{\mathrm{Tilt}} &= \tilde a_i + \rho w_i^2 + \rho(s-2C) w_i.
\end{align}

\paragraph{Headroom window.}
The penalty term is $\rho(r^2+s r)=\rho\,r(r+s)$. Its minimizer is $r^\star=-s/2$, i.e.,
\begin{equation}
W^\star = C - \tfrac{s}{2}.
\end{equation}
Moreover, the factor $r(r+s)\le 0$ for $r\in[-s,0]$, which corresponds to a preferred band 
$W\in[C-s,C]$. Hence the tilt explicitly encourages near-capacity but \emph{under}-capacity 
selections with a controllable headroom margin. In sampling-based subproblem solves, this 
reduces the probability that high-quality candidates are slightly infeasible and improves 
feasible yield after decoding/repair.

\subsubsection{Pros/cons and recommended default}
\label{app:penalties:recommendation}

\paragraph{Taylor encoding.}
\textbf{Pros:} simple (single parameter $\alpha$) and yields a smooth quadratic landscape. \\
\textbf{Cons:} implicitly enforces a ``capacity-hugging'' target ($W\approx C-1$) rather than 
expressing a one-sided inequality, and can be sensitive to scaling as LR-adjusted costs $\tilde 
a_i$ evolve across outer-loop iterations.

\paragraph{Tilted encoding.}
\textbf{Pros:} provides explicit, interpretable headroom control via $s$, avoids slack 
variables (reducing width), and is better aligned with approximate/noisy sampling because it 
biases candidates toward feasibility. \\
\textbf{Cons:} introduces an additional parameter $s$ beyond the penalty scale $\rho$; 
if $s$ is set too large, it can bias toward overly conservative loads.

\begin{figure}[t]
  \centering
  \includegraphics[width=\linewidth]{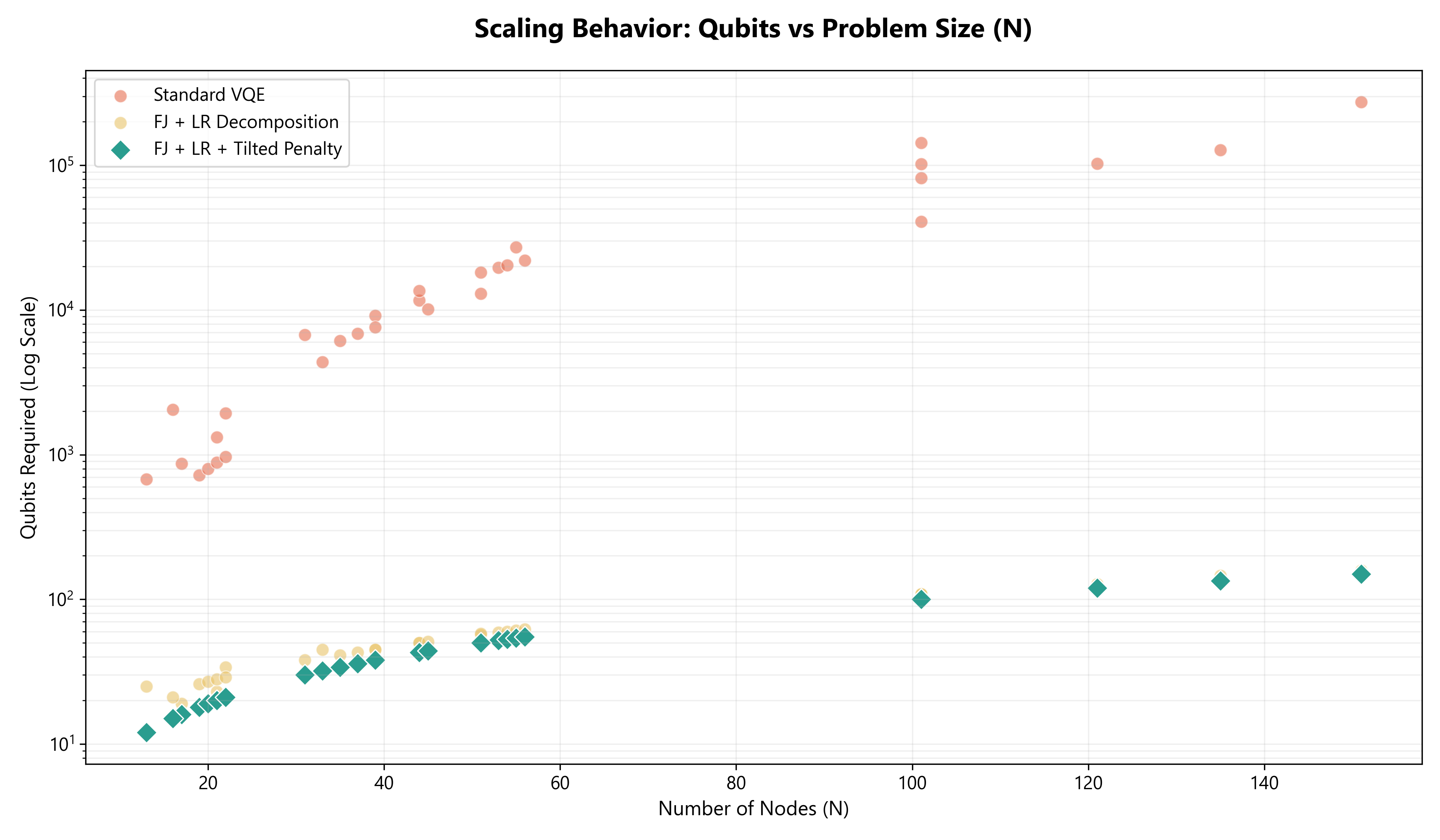}
  \caption{Instance-level scaling behavior of logical qubits versus problem size $N$. Decomposition (FJ+LR) yields bounded-width knapsack QUBOs whose width grows slowly with $N$ (driven by candidate-set size), while a direct CVRP QUBO encoding grows superlinearly.}
  \label{fig:qubit_scalability_scatter}
\end{figure}

Instance-level scaling trends are consistent with this summary view; Fig.~\ref{fig:qubit_scalability_scatter}
reports a per-instance scatter of logical width versus $N$.

\subsection{Micro-benchmark under noisy simulation: feasibility and optimality yield}
\label{app:penalties:microbench}

To motivate the choice of tilted penalization in the LR--CVRP pipeline, we perform a controlled 
micro-benchmark on CVRP instances. We generated random CVRP instances, and used our framework
of decomposition and LR to extract the corresponding knapsack subproblems. We then compile 
and solve the corresponding QUBOs under a noisy simulator using two penalty encodings 
(Taylor vs.\ Tilted) and two variational objectives (standard VQE vs.\ CVaR-VQE). 
For each method, we report (i) the \emph{feasibility rate}, defined as the fraction of 
measurement samples that satisfy the capacity inequality, and (ii) the \emph{success rate}, 
defined as the fraction of runs in which the best decoded sample attains the exact optimal value.

\begin{figure}[t]
  \centering
  \includegraphics[width=\linewidth]{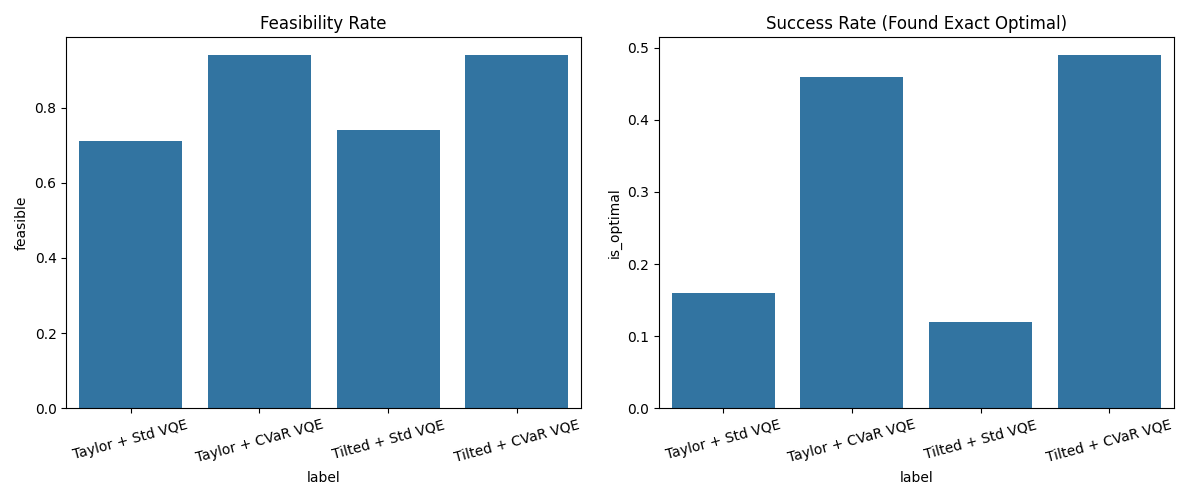}
  \caption{Noisy-simulator micro-benchmark on random instances comparing penalty 
  encodings and variational objectives. \emph{Left:} feasibility rate (fraction of samples 
  satisfying capacity). \emph{Right:} success rate (fraction of runs where the best decoded 
  sample achieves the exact optimum). Tilted penalization maintains feasible yield 
  relative to the Taylor baseline and, when combined with CVaR-VQE, attains the highest 
  optimal-solution recovery rate.}
  \label{fig:tilt_microbench}
\end{figure}

Figure~\ref{fig:tilt_microbench} summarizes the results. The tilted encoding maintains feasibility yield relative to the Taylor baseline. 
Moreover, combining the tilted encoding with CVaR-based 
optimization yields the highest success rate of recovering exact optimal solutions, aligning 
with prior evidence that CVaR objectives can improve robustness of variational optimization on 
noisy hardware \cite{barkoutsos2020improving,sharma2025comparative}. These trends support using the tilted penalty as the 
default encoding in the LR--CVRP subproblem solver, while retaining the Taylor encoding as an 
ablation baseline.

\section{Assignment repair pseudo-code}
\label{app:repair_pseudocode}

This appendix provides pseudo-code for the deterministic repair map used in 
Section~\ref{sec:assignment_repair}. The goal is to convert per-vehicle tentative 
selections (obtained from sampled knapsack bitstrings) into a feasible generalized 
assignment $\hat y$ satisfying: (i) each customer is assigned exactly once and (ii) 
each vehicle capacity is not exceeded. The design follows standard ``Lagrangian heuristic''
practice in which dual subproblem solutions are complemented by explicit primal recovery 
\cite{fisher1981lagrangian} and uses building blocks commonly employed in GAP heuristics 
\cite{cattrysse1992survey,osman1995heuristics}.

\paragraph{Inputs and outputs.}
Let $V=\{1,\dots,n\}$ be customers, $K$ vehicles, $d_i$ demands, and $Q$ capacity 
(identical capacities for simplicity).
For each vehicle $k$, a candidate sample provides a selection set $S_k\subseteq V$.
Let $c_{ik}$ denote the assignment surrogate used for repair scoring (either $a_{ik}$ or 
$\tilde a_{ik}(\lambda)$; we treat this choice as a parameter).

The repair returns either:
(i) a feasible assignment $\hat y\in\{0,1\}^{n\times K}$ and vehicle loads 
$L_k=\sum_i d_i \hat y_{ik}$, or
(ii) an infeasibility flag if completion fails.

\begin{figure}[H]
\caption{Repair$(\{S_k\}_{k=1}^K)$: conflict resolution + completion}
\label{alg:repair}
\begin{algorithmic}[1]
\Require Customer set $V$, vehicles $k=1,\dots,K$, demands $d_i$, capacity $Q$, 
candidate sets $S_k$, costs $c_{ik}$
\Ensure Feasible assignment $\hat y$ or \textsc{Infeasible}

\State Initialize $\hat y_{ik}\gets 0$ for all $i,k$; loads $L_k\gets 0$.
\State $\texttt{owners}(i)\gets \{k: i\in S_k\}$ for all $i\in V$.

\Comment{\textbf{Stage 1: Conflict resolution (duplicates)}}
\ForAll{$i\in V$ with $|\texttt{owners}(i)|\ge 1$}
    \State Choose $k^\star \in \texttt{owners}(i)$ minimizing $c_{ik}$ (tie-break 
    deterministically).
    \If{$L_{k^\star}+d_i \le Q$}
        \State Set $\hat y_{ik^\star}\gets 1$; $L_{k^\star}\gets L_{k^\star}+d_i$.
    \Else
        \State \textbf{continue} (defer $i$ to completion stage)
    \EndIf
\EndFor

\Comment{\textbf{Stage 2: Completion (unassigned customers)}}
\State $U\gets \{i\in V: \sum_{k=1}^K \hat y_{ik}=0\}$.
\ForAll{$i\in U$ (process in increasing $\min_k c_{ik}$ or decreasing $d_i$)}
    \State $\mathcal{K}(i)\gets \{k: L_k+d_i\le Q\}$.
    \If{$\mathcal{K}(i)\neq \emptyset$}
        \State Choose $k^\star \in \mathcal{K}(i)$ minimizing $c_{ik}$.
        \State Set $\hat y_{ik^\star}\gets 1$; $L_{k^\star}\gets L_{k^\star}+d_i$.
    \Else
        \Comment{Attempt local exchange to free capacity}
        \State Try to find $(k, j)$ with $\hat y_{jk}=1$ such that $L_k - d_j + d_i \le Q$
        \State \hspace{2em}and $\Delta = c_{ik} - c_{jk}$ is minimal (or negative if available).
        \If{such $(k,j)$ found}
            \State Set $\hat y_{jk}\gets 0$; $L_k\gets L_k-d_j$.
            \State Set $\hat y_{ik}\gets 1$; $L_k\gets L_k+d_i$.
            \State Add $j$ back to $U$ (to be reassigned later).
        \Else
            \State \Return \textsc{Infeasible}
        \EndIf
    \EndIf
\EndFor

\State \Return $\hat y$, $\{L_k\}_{k=1}^K$
\end{algorithmic}
\end{figure}

\paragraph{Remarks.}
(1) The exchange step can be extended to a bounded-depth ($2$-swap) search; we keep it 
lightweight to preserve runtime predictability.
(2) Repair scoring can use either surrogate assignment costs $a_{ik}$ or LR-adjusted 
reduced costs $\tilde a_{ik}(\lambda)$; we use the latter when the goal is to preserve 
LR structure.
(3) The repair is applied per sampled candidate and paired with the best-of-samples 
selection rule in \eqref{eq:best_of_samples}.

\section{Route reconstruction details}
\label{app:reconstruction_details}

This appendix specifies the reconstruction procedure referenced in 
Section~\ref{sec:reconstruction}. Given a feasible assignment $\hat y$, we form 
per-vehicle customer clusters $V_k=\{i\in V: \hat y_{ik}=1\}$ and then construct a 
depot-starting route for each cluster. This is a standard ``cluster-first, route-second'' 
strategy within the broader vehicle-routing literature \cite{toth2002vehicle}.

\subsection{Cluster-level route construction}
\label{app:recon:construction}

For each vehicle $k$:
\begin{enumerate}
\item Construct an initial tour on $\{0\}\cup V_k$ using a fast heuristic 
(e.g., nearest-neighbor or savings-based initialization).
\item Apply a fixed budget of local improvements (e.g., 2-opt and node relabeling moves) 
to reduce travel cost.
\item Output the improved tour as the route for vehicle $k$.
\end{enumerate}
Because capacity feasibility is enforced at the assignment level, each cluster 
route is capacity-feasible by construction; reconstruction failures are therefore 
rare and typically arise only from degenerate edge cases (e.g., empty cluster handling 
or numerical issues), which are handled deterministically in the implementation.

\subsection{Cost evaluation and feasibility flags}
\label{app:recon:evaluation}

The reconstruction module returns:
(i) the set of routes $\mathcal{R}=\{R_k\}_{k=1}^K$,
(ii) the total routing cost $\mathrm{Cost}(\mathcal{R})=\sum_k \sum_{(i,j)\in R_k} c_{ij}$, 
and
(iii) feasibility flags:
\begin{itemize}
\item \textbf{Assignment-feasible}: all customers assigned exactly once and all vehicle 
loads $\le Q$.
\item \textbf{Route-feasible}: a valid depot-starting tour exists for each nonempty cluster.
\end{itemize}
In the main experiments, we report end-to-end feasibility as the conjunction of 
assignment-feasible and route-feasible.

\subsection{Determinism and reproducibility}
\label{app:recon:repro}

To ensure that differences in end-to-end performance are attributable to subproblem 
generation and quantum sampling (rather than reconstruction randomness), reconstruction 
is implemented deterministically given $\hat y$ and the distance matrix. Any heuristic 
that uses random choices (e.g., randomized nearest-neighbor) is run with fixed seeds.

\section{Training and model hyperparameters}
\label{app:training_hparams}

This appendix reports the hyperparameters used for (i) expert-guided pretraining and 
(ii) PPO fine-tuning of the multiplier-update policy, as well as the architectural 
settings of the default policy (\texttt{DiagonalPrecondPolicy}).

\subsection{Policy architecture (DiagonalPrecondPolicy)}
\label{app:training_hparams:policy}

Table~\ref{tab:policy_hparams} summarizes the default graph-policy architecture used in our two-phase training pipeline.

\begin{table}[H]
\centering
\caption{Default policy hyperparameters (\texttt{DiagonalPrecondPolicy}).}
\label{tab:policy_hparams}
\begin{tabular}{ll}
\hline
\textbf{Parameter} & \textbf{Value} \\
\hline
Node feature dimension (\texttt{input\_dim}) & 14 \\
GNN hidden dimension (\texttt{hidden\_dim}) & 256 \\
Number of GATv2 layers (\texttt{num\_layers}) & 4 \\
Attention heads per layer (\texttt{heads}) & 8 \\
Dropout & 0.10 \\
Edge feature dimension (\texttt{edge\_dim}) & 1 \\
Actor mean parameterization & $\mu_i = \alpha_i g_i + \beta_i$ \\
Scale head nonlinearity & $\alpha_i=\mathrm{softplus}(\cdot)+0.01$ \\
Initial log-std (\texttt{log\_std}) & $-1.5$ \\
Value head pooling & global mean pooling \\
\hline
\end{tabular}
\end{table}

\subsection{Environment and multiplier bounds}
\label{app:training_hparams:env}

We clamp multipliers to a bounded interval for numerical stability and to prevent extreme dual oscillations. The default bounds and horizon used by the curriculum environment are:
\begin{itemize}
\item Maximum outer-loop steps per episode: $200$.
\item Multiplier bounds: $\lambda_{\min}=-400$, $\lambda_{\max}=800$.
\item Frequency of primal evaluation: every $2$ iterations.
\item Frequency of cost evaluation: every iteration.
\end{itemize}

\subsection{Two-stage training schedule}
\label{app:training_hparams:training}

Table~\ref{tab:training_hparams} summarizes the training schedule and optimizer settings. ``BC'' denotes the expert-guided pretraining phase; ``BC+RL'' denotes the PPO fine-tuning phase with an annealed imitation weight.

\begin{table}[t]
\centering
\caption{Training hyperparameters (default configuration).}
\label{tab:training_hparams}
\begin{tabular}{ll}
\hline
\textbf{Parameter} & \textbf{Value} \\
\hline
Discount factor $\gamma$ & 0.99 \\
GAE parameter $\lambda$ & 0.95 \\
BC epochs & 120 \\
DAgger iterations & 20 \\
Total BC+RL steps & 200{,}000 \\
Rollout steps per PPO update & 2048 \\
PPO epochs per update & 4 \\
Number of PPO mini-batches & 4 \\
PPO clip parameter $\epsilon$ & 0.20 \\
Entropy coefficient & 0.01 \\
Value loss coefficient & 0.50 \\
Max gradient norm & 0.50 \\
BC learning rate & $1\times 10^{-4}$ \\
RL learning rate (initial) & $3\times 10^{-4}$ \\
RL learning rate (minimum) & $1\times 10^{-6}$ \\
Imitation weight (start) & 0.30 \\
Imitation weight (end) & 0.05 \\
Imitation anneal steps & 100{,}000 \\
Evaluation episodes per checkpoint & 25 \\
Evaluation horizon (steps) & 200 \\
\hline
\end{tabular}
\end{table}

\subsection{Curriculum parameters}
\label{app:training_hparams:curriculum}

The curriculum transitions and phase-specific weights are fixed as follows:
\begin{itemize}
\item Phase 1$\rightarrow$2 transition when assignment ratio exceeds 0.65.
\item Phase 2$\rightarrow$3 transition when gap falls below 0.20.
\item Transition smoothing window: 30 iterations.
\item Phase weights: feasibility (phase 1) = 3.0, gap (phase 2) = 3.0.
\item Phase 3 weights: cost = 0.10, gap = 0.01, feasibility = 0.10.
\end{itemize}
Additional fine-tuning adjustments applied in the fine-tuning-only setting (e.g., stability rescaling of curriculum weights) are documented in the reproducibility package.

\section{Extended per-instance results for multiplier-control ablations}
\label{app:extended_tables}

This appendix reports the full per-instance breakdown (Tab.~\ref{tab:multiplier_per_instance}) underlying the summary statistics in
Section~\ref{sec:multiplier_results}. Each entry is the end-to-end CVRP optimality gap (\%)
after decoding/repair and route reconstruction under the \emph{classical} subproblem-solve protocol
described in Section~\ref{sec:multiplier_results}. By using classical knapsack solves, these results
isolate the effect of multiplier-update rules from quantum execution noise and compilation variability.

\begin{table}[h]
\centering
\caption{Per-instance optimality gaps (\%) and runtime (s) for multiplier-control ablations under classical knapsack subproblem solves. Lower is better.}
\label{tab:multiplier_per_instance}
\small
\setlength{\tabcolsep}{5pt}
\resizebox{\textwidth}{!}{%
\begin{tabular}{lrrrrrrrrrr} 
\hline
\textbf{Instance} & \textbf{n} & \textbf{K} & 
\multicolumn{2}{c}{\textbf{OR Tools}} &
\multicolumn{2}{c}{\textbf{Subgradient}} & 
\multicolumn{2}{c}{\textbf{Pretrained}} & 
\multicolumn{2}{c}{\textbf{Pretrain+fine-tune}} \\ 
\cline{4-5} \cline{6-7} \cline{8-9} \cline{10-11}
& & & \textbf{Opt gap (\%)} & \textbf{Time (s)} & \textbf{Opt gap (\%)} & \textbf{Time (s)} & \textbf{Opt gap (\%)} & \textbf{Time (s)} & \textbf{Opt gap (\%)} & \textbf{Time (s)} \\ 
\hline
E-n13-k4 & 13 & 4 & 0.00 & 20 & 20.24 & 70.02 & 48.58 & 22.9 & 44.94 & 6.6 \\
P-n16-k8 & 16 & 8 & 0.00 & 20 & 7.78 & 90.02 & 11.33 & 25.7 & 11.33 & 15.3 \\
gr-n17-k3 & 17 & 3 & 0.00 & 20 & 18.03 & 60.02 & 17.91 & 20.4 & 17.91 & 11.6 \\
P-n19-k2 & 19 & 2 & 0.00& 20 & 8.02 & 40.03 & 2.83 & 17.6 & 2.83 & 8.8 \\
P-n20-k2 & 20 & 2 & 0.00 & 20 & 15.74 & 43.04 & 13.89 & 69.4 & 2.78 & 26.2 \\
P-n21-k2 & 21 & 2 & 0.00 & 20 & 22.75 & 42.05 & 16.11 & 65.2 & 3.79 & 25.5 \\
gr-n21-k3 & 21 & 3 & 0.00 & 20 & 4.78 & 60.02 & 14.01 & 65.2 & 13.85 & 29.3 \\
E-n22-k4 & 22 & 4 & 0.00 & 20 & 13.60 & 80.02 & 5.87 & 68.8 & 5.87 & 30.3 \\
P-n22-k2 & 22 & 2 & 0.00 & 20 & 2.78 & 44.05 & 15.28 & 74.4 & 10.19 & 27.5 \\
A-n37-k5 & 37 & 5 & 0.00 & 20 & 37.82 & 218.06 & 17.79 & 70.7 & 17.79 & 24.1 \\
A-n39-k6 & 39 & 6 & 0.24 & 20 & 49.82 & 122.04 & 25.15 & 75.2 & 25.15 & 38.7 \\
A-n44-k6 & 44 & 6 & 0.53 & 20 & 24.87 & 120.03 & 17.40 & 74.4 & 15.37 & 43.1 \\
A-n53-k7 & 53 & 7 & 0.69 & 20 & 42.97 & 147.06 & 16.04 & 84.8 & 16.04 & 45.3 \\
A-n54-k7 & 54 & 7 & 0.68 & 20 & 27.34 & 143.05 & 24.16 & 80.4 & 5.91 & 44.4 \\
A-n55-k9 & 55 & 9 & 0.09 & 20 & 19.94 & 180.04 & 19.38 & 81.2 & 13.79 & 60.7 \\
B-n35-k5 & 35 & 5 & 0.00 & 20 & 21.47 & 101.04 & 17.80 & 76.6 & 17.80 & 35.8 \\
B-n39-k5 & 39 & 5 & 0.18 & 20 & 21.86 & 102.03 & 30.24 & 81.0 & 28.78 & 38.1 \\
B-n44-k7 & 44 & 7 & 0.00 & 20 & 40.48 & 141.04 & 17.05 & 81.0 & 17.05 & 44.0 \\
B-n45-k5 & 45 & 5 & 0.39 & 20 & 20.24 & 106.06 & 15.98 & 79.0 & 5.73 & 38.3 \\
B-n51-k7 & 51 & 7 & 0.00 & 20 & 19.38 & 141.05 & 26.26 & 80.6 & 22.67 & 41.4 \\
B-n56-k7 & 56 & 7 & 0.70 & 20 & 38.33 & 141.04 & 27.58 & 85.1 & 16.41 & 45.1 \\
E-n33-k4 & 33 & 4 & 0.00 & 20 & 20.00 & 122.04 & 8.86 & 72.0 & 7.31 & 33.6 \\
E-n51-k5 & 51 & 5 & 1.34 & 20 & 26.87 & 106.05 & 25.72 & 70.2 & 14.97 & 53.0 \\
E-n101-k8 & 101 & 8 & 1.10 & 20 & 27.78 & 207.12 & 17.26 & 93.9 & 10.28 & 86.3 \\
E-n101-k14 & 101 & 14 & 3.56  & 20 & 33.65 & 286.08 & 10.22 & 110.5 & 10.22 & 62.4 \\
F-n135-k7 & 135 & 7 & 9.03 & 20 & 48.45 & 568.71 & 46.90 & 97.3 & 35.20 & 77.9 \\
P-n101-k4 & 101 & 4 & 1.76 & 20 & 22.47 & 200.84 & 11.16 & 101.6 & 6.61 & 41.4 \\
M-n101-k10 & 101 & 10 &  0.00 & 20 & 37.56 & 226.11 & 9.27 & 109.2 & 8.41 & 56.5 \\
M-n121-k7 & 121 & 7 & 1.25 & 20 & 34.62 & 240.29 & 22.73 & 89.4 & 22.73 & 50.3 \\
M-n151-k12 & 151 & 12 & 7.09 & 20 & 51.23 & 309.16 & 12.91 & 101.6 & 12.91 & 60.5 \\
\hline
Mean Gap (\%) & & & 0.95 & 20 & 26.02 & 148.61 & 18.85 & 74.17 & 14.82 & 40.06 \\
Median Gap (\%) & & & 0.04 & 20 & 22.60 & 122.04 & 17.15 & 77.8 & 13.82 & 40.05 \\
\end{tabular}%
}
\end{table}

\paragraph{Notes.}
Optimality gaps are computed relative to the CVRPLIB best-known solution (BKS) for each instance.
All methods are evaluated under the same stopping criteria and reconstruction pipeline; differences
reflect only the multiplier-update rule.

\section{Full per-instance results for end-to-end experiments}
\label{app:results_end_to_end}

This appendix provides supporting results for the hardware-aware execution layer
(Section~\ref{sec:bandit_results}), including circuit-level compilation overhead, full
per-instance cost and quality outcomes, backend usage counts, and robustness diagnostics.

\subsection{Circuit-level overhead under selected configurations}
\label{sec:circuit_overhead_results}

To connect configuration choices to compilation overhead, we summarize transpiled circuit statistics
recorded under the configurations selected by the consult routine. Table~\ref{tab:circuit_overhead}
reports mean transpiled depth and mean total gate count as proxies for compiled circuit size,
aggregated by backend, along with the feasibility-screen estimator pass rate
$\Pr(\widehat{G}\le G_{\max})$. These summaries quantify how backend topology and configuration choice
translate into practical routing overhead and clarify when the gate-budget screen becomes active in
practice.

\begin{table}[H]
\centering
\caption{Circuit-level overhead under selected configurations across heterogeneous backends. We report mean transpiled depth and mean total gate count,
aggregated over logged decisions. ``Estimator pass rate'' is the fraction of cases satisfying
$\widehat{G}\le G_{\max}$ under the gate-budget screen (with $G_{\max}=20{,}000$).}
\label{tab:circuit_overhead}
\begin{tabular}{lrrrr}
\hline
\textbf{Backend} & \textbf{N (logs)} & \textbf{Mean depth} & \textbf{Mean total gates} & \textbf{Estimator pass rate} \\
\hline
IBM ibm\_marrakesh & 16 & 182.0 & 483.3 & 1.00 \\
IBM ibm\_fez       &  6 & 333.5 & 876.0 & 1.00 \\
IBM ibm\_torino    & 12 &  89.3 & 253.3 & 1.00 \\
Rigetti Ankaa-3    & 10 &  72.0 & 284.1 & 1.00 \\
IQM Garnet         &  5 &  42.0 & 275.2 & 1.00 \\
IQM Emerald        &  8 &  85.1 & 406.4 & 0.625 \\
\hline
\end{tabular}
\end{table}

\subsection{Full per-instance cost and quality results}
\label{app:full_instance_tables}

Table~\ref{tab:panel_a_full} reports per-instance outer-loop sample-efficiency metrics, including
quantum-evaluation budget, wall-clock time, and latency statistics, together with the resulting
end-to-end gap after decoding/repair and route reconstruction.

\begin{table}[t]
\centering
\caption{Per-instance quantum decomposition results across CVRP benchmarks. Metrics include circuit counts, runtime, QPU utilization, job latency, and optimality gap to best-known solution (BKS).}
\label{tab:panel_a_full}
\scriptsize
\setlength{\tabcolsep}{2.8pt}
\begin{tabular}{lrrllrrrrl}
\hline
\textbf{Instance} & \textbf{$n$} & \textbf{$K$} & \textbf{Source} & \textbf{Circuits} & \textbf{Time (s)} & \textbf{QPU (\%)} & \textbf{Latency (s)} & \textbf{Gap (\%)} & \textbf{Device Mix} \\
\hline
E-n13-k4.vrp & 12 & 4 & Multi-QPU (noise-aware) & 20 & 1867.8 & 100.0 & 93.4 & 46.15 & IBM:0 Rig:0 IQM:20 Sim:0 \\
E-n13-k4.vrp & 12 & 4 & Multi-QPU (baseline) & 20 & 2147.9 & 100.0 & 107.4 & 46.15 & IBM:12 Rig:4 IQM:3 Sim:1 \\
E-n13-k4.vrp & 12 & 4 & Single-QPU (Rigetti Ankaa-3) & 20 & 2147.9 & 100.0 & 107.4 & 46.15 & Rigetti Ankaa-3 \\
P-n16-k8.vrp & 15 & 8 & Multi-QPU (noise-aware) & 80 & 13114.5 & 100.0 & 163.9 & 9.33 & IBM:0 Rig:4 IQM:76 Sim:0 \\
P-n16-k8.vrp & 15 & 8 & Multi-QPU (baseline) & 160 & 22104.8 & 100.0 & 138.2 & 11.33 & IBM:93 Rig:35 IQM:24 Sim:8 \\
P-n16-k8.vrp & 15 & 8 & Single-QPU (IQM Emerald) & 160 & 22104.8 & 100.0 & 138.2 & 11.33 & IQM Emerald \\
gr-n17-k3.vrp & 16 & 3 & Multi-QPU (noise-aware) & 30 & 4347.4 & 100.0 & 144.9 & 17.95 & IBM:9 Rig:10 IQM:11 Sim:0 \\
P-n19-k2.vrp & 18 & 2 & Multi-QPU (noise-aware) & 20 & 6365.3 & 100.0 & 318.3 & 12.74 & IBM:0 Rig:12 IQM:8 Sim:0 \\
P-n19-k2.vrp & 18 & 2 & Multi-QPU (baseline) & 40 & 4456.3 & 100.0 & 111.4 & 12.74 & IBM:23 Rig:9 IQM:6 Sim:2 \\
P-n19-k2.vrp & 18 & 2 & Single-QPU (IQM Garnet) & 40 & 4456.3 & 100.0 & 111.4 & 12.74 & IQM Garnet \\
P-n20-k2.vrp & 19 & 2 & Multi-QPU (noise-aware) & 20 & 10196.5 & 100.0 & 509.8 & 13.89 & IBM:20 Rig:0 IQM:0 Sim:0 \\
P-n20-k2.vrp & 19 & 2 & Multi-QPU (baseline) & 40 & 9406.9 & 100.0 & 235.2 & 13.89 & IBM:23 Rig:9 IQM:6 Sim:2 \\
P-n20-k2.vrp & 19 & 2 & Single-QPU (IQM Garnet) & 40 & 9406.9 & 100.0 & 235.2 & 13.89 & IQM Garnet \\
P-n21-k2.vrp & 20 & 2 & Multi-QPU (noise-aware) & 20 & 398.6 & 100.0 & 19.9 & 18.48 & IBM:20 Rig:0 IQM:0 Sim:0 \\
gr-n21-k3.vrp & 20 & 3 & Multi-QPU (noise-aware) & 30 & 4829.5 & 100.0 & 161.0 & 13.85 & IBM:10 Rig:19 IQM:1 Sim:0 \\
gr-n21-k3.vrp & 20 & 3 & Multi-QPU (baseline) & 60 & 8856.9 & 100.0 & 147.6 & 13.85 & IBM:0 Rig:28 IQM:32 Sim:0 \\
E-n22-k4.vrp & 21 & 4 & Multi-QPU (noise-aware) & 40 & 3917.0 & 100.0 & 97.9 & 5.87 & IBM:0 Rig:40 IQM:0 Sim:0 \\
E-n22-k4.vrp & 21 & 4 & Multi-QPU (baseline) & 80 & 8573.0 & 100.0 & 107.2 & 5.87 & IBM:46 Rig:18 IQM:12 Sim:4 \\
E-n22-k4.vrp & 21 & 4 & Single-QPU (Rigetti Ankaa-3) & 80 & 8573.0 & 100.0 & 107.2 & 5.87 & Rigetti Ankaa-3 \\
P-n22-k2.vrp & 21 & 2 & Multi-QPU (noise-aware) & 16 & 386.3 & 100.0 & 24.1 & 18.52 & IBM:16 Rig:0 IQM:0 Sim:0 \\
A-n37-k5.vrp & 36 & 5 & Multi-QPU (noise-aware) & 100 & 8363.0 & 100.0 & 83.6 & 18.68 & IBM:48 Rig:47 IQM:0 Sim:5 \\
A-n37-k5.vrp & 36 & 5 & Multi-QPU (baseline) & 96 & 16517.8 & 100.0 & 172.1 & 18.68 & IBM:56 Rig:21 IQM:14 Sim:5 \\
A-n37-k5.vrp & 36 & 5 & Single-QPU (IQM Emerald) & 96 & 16517.8 & 100.0 & 172.0 & 18.68 & IQM Emerald \\
A-n39-k6.vrp & 38 & 6 & Multi-QPU (noise-aware) & 96 & 10863.0 & 100.0 & 113.2 & 25.63 & IBM:0 Rig:91 IQM:0 Sim:5 \\
A-n39-k6.vrp & 38 & 6 & Multi-QPU (baseline) & 75 & 8182.2 & 100.0 & 109.1 & 25.63 & IBM:44 Rig:16 IQM:11 Sim:4 \\
A-n39-k6.vrp & 38 & 6 & Single-QPU (Rigetti Ankaa-3) & 75 & 8182.2 & 100.0 & 108.1 & 25.63 & Rigetti Ankaa-3 \\
A-n44-k6.vrp & 43 & 6 & Multi-QPU (noise-aware) & 120 & 15376.7 & 100.0 & 128.1 & 17.61 & IBM:72 Rig:36 IQM:12 Sim:0 \\
A-n44-k6.vrp & 43 & 6 & Multi-QPU (baseline) & 120 & 13571.3 & 100.0 & 113.1 & 17.61 & IBM:70 Rig:26 IQM:18 Sim:6 \\
A-n53-k7.vrp & 52 & 7 & Multi-QPU (noise-aware) & 140 & 43778.4 & 100.0 & 312.7 & 17.92 & IBM:64 Rig:40 IQM:29 Sim:7 \\
A-n54-k7.vrp & 53 & 7 & Multi-QPU (noise-aware) & 98 & 36009.9 & 100.0 & 367.4 & 8.05 & IBM:43 Rig:55 IQM:0 Sim:0 \\
A-n54-k7.vrp & 53 & 7 & Multi-QPU (baseline) & 140 & 10048.3 & 100.0 & 71.8 & 8.05 & IBM:53 Rig:41 IQM:39 Sim:7 \\
A-n55-k9.vrp & 54 & 9 & Multi-QPU (noise-aware) & 180 & 9767.3 & 100.0 & 54.3 & 24.32 & IBM:180 Rig:0 IQM:0 Sim:0 \\
A-n55-k9.vrp & 54 & 9 & Multi-QPU (baseline) & 180 & 9720.1 & 100.0 & 54.0 & 24.32 & IBM:74 Rig:50 IQM:47 Sim:9 \\
B-n35-k5.vrp & 34 & 5 & Multi-QPU (noise-aware) & 100 & 8437.0 & 100.0 & 84.4 & 14.76 & IBM:100 Rig:0 IQM:0 Sim:0 \\
B-n35-k5.vrp & 34 & 5 & Multi-QPU (baseline) & 100 & 2775.1 & 100.0 & 27.7 & 14.76 & IBM:63 Rig:32 IQM:0 Sim:5 \\
B-n39-k5.vrp & 38 & 5 & Multi-QPU (noise-aware) & 100 & 4085.6 & 100.0 & 40.8 & 28.78 & IBM:95 Rig:0 IQM:0 Sim:5 \\
B-n39-k5.vrp & 38 & 5 & Multi-QPU (baseline) & 100 & 2901.8 & 100.0 & 29.0 & 28.78 & IBM:62 Rig:0 IQM:33 Sim:5 \\
B-n44-k7.vrp & 43 & 7 & Multi-QPU (noise-aware) & 140 & 13.9 & 92.0 & 0.1 & 19.47 & IBM:81 Rig:31 IQM:21 Sim:7 \\
B-n44-k7.vrp & 43 & 7 & Multi-QPU (baseline) & 140 & 7068.5 & 100.0 & 50.5 & 19.47 & IBM:74 Rig:11 IQM:48 Sim:7 \\
B-n45-k5.vrp & 44 & 5 & Multi-QPU (noise-aware) & 100 & 11.4 & 90.5 & 0.1 & 7.06 & IBM:58 Rig:22 IQM:15 Sim:5 \\
B-n45-k5.vrp & 44 & 5 & Multi-QPU (baseline) & 100 & 2588.6 & 100.0 & 25.9 & 7.06 & IBM:63 Rig:0 IQM:32 Sim:5 \\
B-n51-k7.vrp & 50 & 7 & Multi-QPU (noise-aware) & 140 & 17.7 & 91.5 & 0.1 & 26.26 & IBM:81 Rig:31 IQM:21 Sim:7 \\
B-n51-k7.vrp & 50 & 7 & Multi-QPU (baseline) & 140 & 8548.6 & 100.0 & 61.1 & 26.26 & IBM:63 Rig:46 IQM:24 Sim:7 \\
B-n56-k7.vrp & 55 & 7 & Multi-QPU (noise-aware) & 140 & 19.7 & 91.7 & 0.1 & 27.58 & IBM:81 Rig:31 IQM:21 Sim:7 \\
B-n56-k7.vrp & 55 & 7 & Multi-QPU (baseline) & 140 & 5440.3 & 100.0 & 38.9 & 27.58 & IBM:80 Rig:53 IQM:0 Sim:7 \\
E-n33-k4.vrp & 32 & 4 & Multi-QPU (noise-aware) & 80 & 7.1 & 83.6 & 0.1 & 8.98 & IBM:46 Rig:18 IQM:12 Sim:4 \\
E-n33-k4.vrp & 32 & 4 & Multi-QPU (baseline) & 80 & 2776.2 & 100.0 & 34.7 & 8.98 & IBM:51 Rig:0 IQM:25 Sim:4 \\
E-n33-k4.vrp & 32 & 4 & Single-QPU (Rigetti Ankaa-3) & 0 & 76.8 & 99.6 & 0.0 & 8.98 & Rigetti Ankaa-3 \\
E-n51-k5.vrp & 50 & 5 & Multi-QPU (noise-aware) & 100 & 14.4 & 88.7 & 0.1 & 27.64 & IBM:58 Rig:22 IQM:15 Sim:5 \\
E-n51-k5.vrp & 50 & 5 & Multi-QPU (baseline) & 80 & 11411.7 & 100.0 & 142.6 & 28.41 & IBM:46 Rig:18 IQM:12 Sim:4 \\
E-n51-k5.vrp & 50 & 5 & Single-QPU (Rigetti Ankaa-3) & 80 & 11411.7 & 100.0 & 141.5 & 28.41 & Rigetti Ankaa-3 \\
E-n101-k8.vrp & 100 & 8 & Multi-QPU (baseline) & 80 & 1652.8 & 100.0 & 20.7 & 12.23 & IBM:80 Rig:0 IQM:0 Sim:0 \\
E-n101-k14.vrp & 100 & 14 & Multi-QPU (noise-aware) & 95 & 6187.5 & 100.0 & 65.1 & 11.90 & IBM:95 Rig:0 IQM:0 Sim:0 \\
F-n135-k7.vrp & 134 & 7 & Multi-QPU (baseline) & 70 & 2617.3 & 100.0 & 37.4 & 56.71 & IBM:70 Rig:0 IQM:0 Sim:0 \\
P-n101-k4.vrp & 100 & 4 & Multi-QPU (baseline) & 40 & 323.2 & 99.9 & 8.1 & 14.39 & IBM:38 Rig:0 IQM:0 Sim:2 \\
M-n101-k10.vrp & 100 & 10 & Multi-QPU (baseline) & 200 & 3363.5 & 100.0 & 16.8 & 19.63 & IBM:190 Rig:0 IQM:0 Sim:10 \\
\hline
\end{tabular}
\end{table}

The main paper reports aggregate cost--quality trends and median gap improvements to keep the
hardware-aware results readable. This appendix provides the corresponding per-instance end-to-end
results for transparency and reproducibility. Table~\ref{tab:quantum_per_instance} reports the
optimality gap to the best-known solution (BKS) for each CVRP benchmark instance under (i) classical
subgradient multiplier updates with VQE subproblem solves on a noisy simulator, (ii) learned multiplier
control with the same noisy-simulator VQE primitive, and (iii)--(iv) learned control executed on real
hardware with and without bandit-based configuration. Together with the robustness and tradeoff
diagnostics in Appendix~\ref{app:results_end_to_end}, these results clarify where improvements arise and
under what execution conditions they persist.

\begin{table}[t]
\centering
\caption{Per-instance end-to-end optimality gaps (\%) to the best-known solution (BKS) across controller and execution variants.}
\label{tab:quantum_per_instance}
\scriptsize
\setlength{\tabcolsep}{3.5pt}
\begin{tabular}{lrrrrrrrr}
\hline
\textbf{Instance} & \textbf{$n$} & \textbf{$K$} & \textbf{BKS} &
\textbf{OR Tools } &
\textbf{SG+VQE (Noisy)} &
\textbf{RL+VQE (Noisy)} &
\textbf{RL+VQE (HW)} &
\textbf{RL+VQE (HW+Bandit)} \\
\hline
E-n13-k4 & 13 & 4 &  247 & 0.00 &17.41 & 23.89 & 9.31 & 9.31 \\
P-n16-k8 & 16 & 8 & 450 & 0.00 & 4.44 & 2.22 & 3.78 & 9.33 \\
gr-n17-k3 & 17 & 3 & 2685 & 0.00 & 29.87 & 17.95 & 17.95 & 17.95 \\
P-n19-k2 & 19 & 2 & 212 & 0.00 & 16.04 & 16.04 & 12.74 & 12.74 \\
P-n20-k2 & 20 & 2 & 216 & 0.00 & 19.91 & 13.89 & 13.89 & 10.65 \\
P-n21-k2 & 21 & 2 & 211 & 0.00 & 23.22 & 16.59 & 16.59 & 16.59 \\
gr-n21-k3 & 21 & 3 & 3704 & 0.00 & 13.47 & 14.01 & 13.85 & 13.85 \\
E-n22-k4 & 22 & 4 & 375 & 0.00 & 53.33 & 5.87 & 5.87 & 5.87 \\
P-n22-k2 & 22 & 2 & 216 & 0.00 & 20.83 & 16.67 & 18.52 & 16.67 \\
A-n37-k5 & 37 & 5 & 669 & 0.00 & 32.44 & 16.89 & 18.68 & 16.89 \\
A-n39-k6 & 39 & 6 & 831 & 0.02 & 37.42 & 25.63 & 25.63 & 25.15 \\
A-n44-k6 & 44 & 6 & 937 & 0.53 & 45.89 & 13.87 & 17.61 & 13.87 \\
A-n53-k7 & 53 & 7 & 1010 & 0.69 & 54.46 & 16.93 & 17.82 & 16.93 \\
A-n54-k7 & 54 & 7 & 1167 & 0.68 & 33.85 & 19.97 & 8.05 & 7.54 \\
A-n55-k9 & 55 & 9 & 1073 & 0.09 & 42.87 & 21.81 & 20.32 & 17.43 \\
B-n35-k5 & 35 & 5 & 955 & 0.00 & 18.01 & 17.80 & 14.76 & 14.76 \\
B-n39-k5 & 39 & 5 & 549 & 0.18 & 99.64 & 30.42 & 28.78 & 28.78 \\
B-n44-k7 & 44 & 7 & 909 & 0.00 & 60.40 & 17.27 & 19.47 & 17.27 \\
B-n45-k5 & 45 & 5 & 751 & 0.39 & 56.19 & 6.92 & 7.06 & 6.92 \\
B-n51-k7 & 51 & 7 & 1032 & 0.00 & 39.83 & 26.26 & 26.26 & 23.06 \\
B-n56-k7 & 56 & 7 & 707 & 0.70 & 69.45 & 27.58 & 27.30 & 17.26 \\
E-n33-k4 & 33 & 4 & 835 & 0.00 & 19.52 & 7.43 & 8.98 & 7.43 \\
E-n51-k5 & 51 & 5 & 521 & 1.34 & 22.65 & 19.96 & 27.45 & 19.96 \\
E-n101-k8 & 101 & 8 & 817 & 1.10 & 45.53 & 15.91 & 12.24 & 10.40 \\
E-n101-k14 & 101 & 14 & 1067 & 3.56 & 44.61 & 13.03 & 13.03 & 11.90 \\
F-n135-k7 & 135 & 7 & 1162 & 9.03 & 53.79 & 64.63 & 56.71 & 43.72 \\
P-n101-k4 & 101 & 4 & 681 & 1.76 & 55.36 & 16.01 & 14.39 & 11.89 \\
M-n101-k10 & 101 & 10 & 820 & 0.00 & 70.73 & 10.24 & 17.44 & 10.24 \\
M-n121-k7 & 121 & 7 & 1034 & 1.25 & 71.95 & 24.18 & 24.18 & 24.18 \\
M-n151-k12 & 151 & 12 & 1015 & 7.09 & 86.11& 16.26 & 16.26 & 13.10 \\
\hline
Mean gap & & & & 0.95 & 41.97 & 18.54 & 17.83 & 15.72 \\
Median gap & & & & 0.05 & 41.35 & 16.78 & 17.02 & 14.32 \\
\end{tabular}

{\footnotesize \textit{Note.} SG = classical subgradient multiplier updates; RL = learned multiplier updates; VQE = variational quantum eigensolver used for knapsack subproblems; Noisy = noisy quantum simulator; HW = real QPU execution; Bandit = hardware-aware configuration selection.}
\end{table}

\subsection{Backend usage and orchestration behavior}
\label{app:backend_usage}

\begin{figure}[h]
  \centering
  \includegraphics[width=0.75\linewidth]{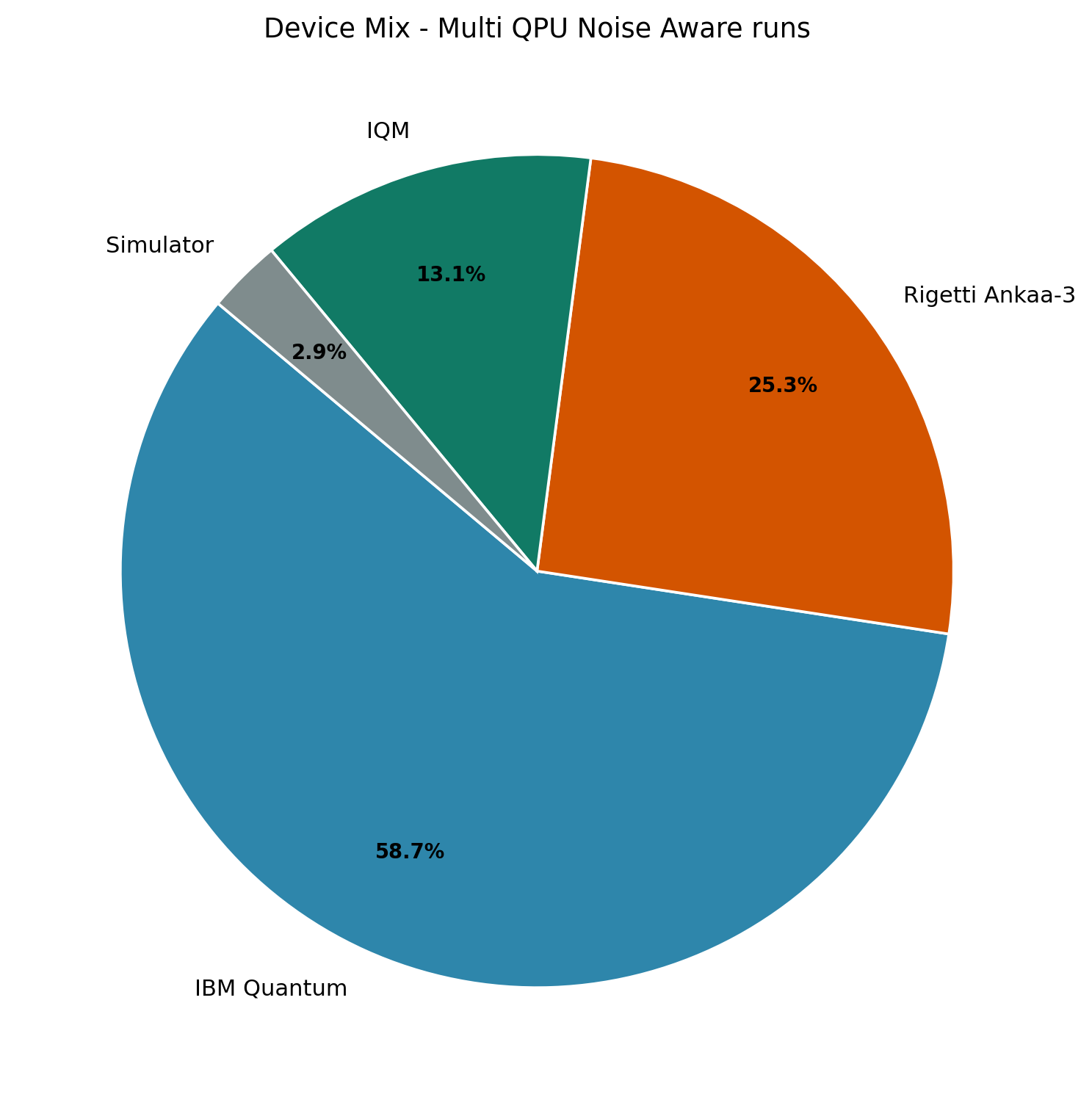}
  \caption{Fraction of completed quantum jobs executed on each backend in multi-QPU noise-aware runs.}
  \label{fig:qpu_mix}
  \vspace{-1ex}
  {\footnotesize \textit{Note.} Shares are computed over completed jobs; ``Simulator'' denotes jobs executed on the noisy simulator when hardware execution is not selected or not available.}
\end{figure}

For multi-QPU runs, we record which backend executes each submitted quantum job and summarize the
resulting allocation across device families. Figure~\ref{fig:qpu_mix} reports the aggregate
\emph{device mix} for the multi-QPU noise-aware setting, measured as the fraction of completed
quantum jobs executed on each backend. This diagnostic verifies that the orchestration layer
distributes work across heterogeneous resources rather than relying on a single provider.

\end{document}